\documentclass[aps,preprint,showpacs]{revtex4}
\usepackage{amsmath}
\usepackage{pstricks}
\usepackage{color}
\usepackage{graphicx}

\newcommand{\be}{\begin{equation}}
\newcommand{\ee}{\end{equation}}
\newcommand{\bea}{\begin{eqnarray}}
\newcommand{\eea}{\end{eqnarray}} 
\newcommand{\ba}{\begin{array}}
\newcommand{\ea}{\end{array}}

\newcommand{\bb}{\bibitem}

\begin{document}
\begin{flushright}
\phantom{a}
\vspace{-3cm}\large
YITP-35\\
\normalsize
\end{flushright}

\title{\bf Modern finite-size criticality: Dirichlet and Neumann boundary conditions}
\author{Messias V. S. Santos\footnote{e-mail:messiasvilbert@df.ufpe.br}, Jos\'e B. da Silva Jr.\footnote{e-mail:jborba@petrobras.com.br},}
\affiliation{{\it Laborat\'orio de F\'\i sica Te\'orica e Computacional, Departamento de F\'\i sica, Universidade Federal de Pernambuco, 
50670-901, Recife, PE, Brazil}}
\author{Marcelo M. Leite\footnote{e-mail:mleite@insti.physics.sunysb.edu}}
\affiliation{{\it C N Yang Institute for Theoretical Physics, State University of New York, Stony Brook, NY 11794-3840, USA}}
\vspace{0.2cm}
\begin{abstract}
{\it Finite-size critical systems defined on a parallel plate geometry of finite extent along one single ($z$) direction with 
Dirichlet and Neumann boundary conditions at $z=0,L$ are analyzed in momentum space. We 
introduce a modified representation for the discrete eigenfunctions in a renormalized one-particle irreducible vertex part ($1PI$) 
scalar field-theoretic framework using either massless or massive fields. The appearance of multiplicities in the 
Feynman rules to construct diagrams due to this choice of representation of the basis functions is discussed along with the modified 
normalization conditions. For nonvanishing external quasi-momenta, Dirichlet and Neumann boundary conditions are shown to be unified 
within a single formalism. We examine the dimensional crossover regimes for these and show a correspondence with those from antiperiodic 
and periodic boundary conditions. It is demonstrated that finite-size effects for Dirichlet and Neumann boundary conditions 
do not require surface fields necessarily but are implemented nontrivially from the Feynman rules involving only bulk terms in the 
Lagrangian. As an application, the critical exponents $\eta$ and $\nu$ are evaluated at least up to two-loop level through 
diagrammatic means. We show that the critical indices are the same as those from the bulk (infinite) system irrespective of the 
boundary conditions.}      
\end{abstract}
\vspace{1cm}

\maketitle

\newpage
\section{Introduction}
\par Fields confined in a certain space region have especial properties whose study is worthwhile. Focusing entirely on conventional critical 
behavior, they correspond to the order parameters of certain systems undergoing a second order phase transition \cite{Wil,amit}. The order 
parameter has well defined values at the boundary surfaces. This can be implemented through boundary conditions, {\it e. g.}, periodic ($PBC$), antiperiodic ($ABC$), Dirichlet ($DBC$), Neumann ($NBC$), etc. and/or 
via external surface fields. 
\par A layered parallel (hyper)plate geometry is defined by $(d-1)$-dimensional (hyper)planes of infinite extent whose limiting plates are 
located, say, at $z=0, L$ ($z$ is the space direction perpendicular to the ``planes''). We can either vary the distance between the 
boundaries or keep $L$ fixed and study universal amounts for the confined system. The change of fundamental quantities, like energy, with 
the variation on $L$ originates the Casimir effect \cite{Casimir,BorUp,Gamba,Farina}, in which long range forces arise in response to the 
confinement of fluctuations. A similar effect in confined geometries of this type is the thinning of $^{4}He$ near the superfluid transition 
due to a force causing the change of the surface free energy in the critical fluctuations whenever the order parameter vanishes ($DBC$) 
on both boundary surfaces \cite{Garcia}. 
\par By maintaining $L$ fixed, however, the critical behavior of the order parameter in thin films could manifest finite-size corrections. 
$PBC$ resembles closely experimental confined systems as in flows from glassy materials 
above and below the jamming transition in comparison with the unconfined behavior, whereas $DBC$ for the same system indicates a surface 
phenomenon \cite{Goyon}. Nevertheless, $DBC$ and $PBC$ take into account purely finite-size contributions in the experimental discussion of 
confined $^{4}He$ \cite{Getal} as well as other fluids \cite{SMMO}, between parallel plates. These corrections also take  place in  
the measurements of spatial correlation of current critical points in an open billiards system \cite{Hohmann} in describing the fields 
of turbulence, sound waves and acoustics, among others. On the other hand, Neumann boundary conditions govern the transition bulk-surface 
when the critical behavior deviates from the ($N,d$) to the $(N,d-1)$ universality class and a nontrivial mixing of finite-size and surface 
effects was argued to take place \cite{Di}. Moreover, $DBC$ and $NBC$ have been investigated on a wide class of finite-size models, like 
amorphous solid suffering plastic deformation in a certain class of depinning models \cite{Sand}, dynamics of order reconstruction in confined 
nematic liquid crystals \cite{Zhou}, determination of free energy corrections in the confined spherical model \cite{Chamati}, etc.
\par Phenomenological finite-size scaling theory states that close to the bulk critical temperature, the variable 
$\frac{L}{\xi}$ ($\xi$ is the bulk correlation length) measures the deviation from the bulk critical behavior \cite{F,FB,B,PF,P}. According 
to some authors \cite{F,FB,B}, the description of the finite system is not limited to the values $\frac{L}{\xi}>1$: they suggested that the 
finite-size critical exponents should be identical as those from the bulk (infinite) system. 
\par The description of finite-size critical systems using momentum space renormalization group field-theoretic methods has basic aspects 
which are simple to grasp. First, for parallel plate geometries the boundary conditions on the plates are 
implemented through the bare free propagator. Second, the typical length $L$ separating the boundaries 
can be included in the Feynman rules. For $PBC$ and $ABC$, the momentum along the 
finite-size direction turns into the quasi-momentum and the integral along the $z$ direction gets transformed to an infinite sum 
(Nemirovsky-Freed ($NF$) method \cite{NF}). Motivated by the $NF$ Green's functions formalism, a one-particle 
irreducible ($1PI$) vertex part framework was designed recently to the treatment of finite-size systems subject to $PBC$ and $ABC$ 
using either massive or massless fields. Within this finite-size technique, scaling theory holds rigorously in the whole 
region $0 < \frac{L}{\xi} \leq \infty$ in agreement with previous claims \cite{B,PF}. 
Explicit computations at higher order were performed corroborating that the finiteness corrections are not sufficient to modify the bulk 
critical exponents \cite{BL}. A different piece of folklore exists in the literature regarding 
$DBC$ and $NBC$: they represent free surfaces and are appropriate to describe finite-size plus surface effects. Indeed one can introduce 
surface fields in conjumination with bulk fields for this boundary conditions breaking translation invariance in this way \cite{Nami}. If 
we allow only bulk fields, ruling out ordering surface fields associated with boundary plates, is that possible 
to renormalize solely bulk fields subject to those boundary conditions which do not produce any surface contributions, but with manifest 
translation invariance breaking? Can a framework 
for $DBC$ and $NBC$ be devised to explore the full finite-size region $0 < \frac{L}{\xi} \leq \infty$? If we achieve consistency with the 
results obtained from $PBC$ and $ABC$, the emerging description represents a modern finite-size scaling regime. Is there any simple relation 
with the associated unconfined system (the ``bulk'' criticality)? How the decreasing of the confinement region rules over those 
situations? How the presence or absence of surface fields alters the criticality? 
\par In this work we build up an {\it ab initio} renormalized one-particle irreducible vertex part formalism in momentum space for 
Dirichlet and Neumann boundary conditions in order to calculate perturbatively universal critical properties including {\it only} bulk 
fields. For sake of comparison with previous one-loop results, we first employ massive fields. We modify the Feynman 
rules with respect to the previous approach using sines and cosines as basis functions \cite{NF} due to our decomposition of them 
in terms of exponentials only. The Feynman diagrams obtained from those rules can be expressed in terms of integrals 
(and summations) identical to the $PBC$ and $ABC$ cases, but contain extra ``nondiagonal'' terms in which there are not as many 
summations as there are integrals over the $(d-1)$-dimensional transverse space. 
\par This new feature will permit us to prove that each diagram is composed of two parts: the first one, where {\it the momenta as well as 
quasimomenta are conserved in every diagram order by order in perturbation theory} and the second, substantiated by the presence of 
the ``nondiagonal''  terms which break the translation invariance (violation of quasi-momentum conservation) along the finite-size 
direction. These terms are important for they make a clear-cut distinction in comparison with the simplicity obtained from $PBC$ and 
$ABC$ results. They look like surface contributions at first sight albeit they are purely finite-size corrections and perhaps this is 
the mathematical origin of the folklore above mentioned. We will show that they do not contribute to the leading singularities in 
dimensional regularization. We introduce new Feynman rules for vertex parts including composite operators which can be renormalized 
multiplicatively. We demonstrate that for a certain quasi-momentum distribution of external legs not belonging to composite 
fields, the external quasi-momentum of the insertion of the composite operator admits more than one combination of the external 
quasi-momenta of the other usual external legs (not associated to composite operators) and should be properly taking 
into account in the set of rules order by order in perturbation theory.
\par We discuss the dimensional crossover criteria and show the consistency with the previous results from $PBC$ and $ABC$ in \cite{BL}. 
Staying away from the problematic region where the $\epsilon$-expansion ceases to give meaningful results, all loop integrals considered 
will be shown to have the general structure made out of bulk plus finite-size terms with the latter depending on the boundary conditions. 
Within the finite-size plus bulk regime, we choose to renormalize the field theory with zero external momenta and nonvanishing external 
quasimomenta in order to unify Neumann and Dirichlet boundary conditions in a single framework. Afterwards, the unifying formalism of 
massless fields using nonvanishing external momenta and quasi-momenta for $DBC$ and $NBC$ is presented along with a discussion of the 
dimensional crossover criterion where the finite-size regime starts to give meaningless results. We prove its equivalence with the massive 
case. As an application we compute the critical exponents $\eta$ and $\nu$ using diagrammatic means. We show that the universal results are 
independent of the boundary condition.      
\par In Sec. II we discuss how the discrete eigenfunctions corresponding to $NBC$ and $DBC$ are expressed in 
terms of exponentials and the consequent modification of the Feynman rules.  We construct the tensor couplings for all the primitively 
divergent vertex parts which can be renormalized multiplicatively and demonstrate the emergence of multiplicities with respect to the 
bulk (infinite) theory due to the exponential representation of the basis functions. We show that $DBC$ possess diagram multiplicities 
identical to $NBC$ whenever both theories are defined at nonvanishing external quasi-momenta. In this picture we compute explicitly some 
sample Feynman diagrams. We conclude by presenting the unified set of all diagrams which are going to be relevant in our discussion. The 
arguments are valid for both formulations of massive and massless fields. 
\par We discuss the renormalization of massive fields in Sec. III. We start with a large number of diagrams and achieve the reduction to 
a smaller set of graphs owing to the nontrivial cancellation of the mass insertions. This feature includes both the diagonal and 
nondiagonal terms of tadpole diagrams. This argument is decisive in proving that the theory can be renormalized without making any reference 
to {\it ad hoc} surface fields. Normalization conditions are defined for the primitively divergent vertex parts in order to assure 
the finiteness of all vertex parts that can be renormalized multiplicatively. We briefly discuss how the flow in the mass scale affects the 
renormalized vertex parts in essentially the same way as in the bulk case. The limit $L \rightarrow \infty$ is shown to retrieve the bulk result, 
whereas the $L\rightarrow 0$ limit marks the onset of the dimensional crossover which invalidates $\epsilon$-expansion results. We also list 
the solution for the higher loop diagrams from Appendix A which will be required in the determination of critical exponents in Sec. IV. The 
dependence of the renormalization constants/functions on the boundary conditions disappears in the final expression for the exponents.   
\par Section V contains the explicit discussion of massless fields and their multiplicative renormalization for both boundary 
conditions in the unified description. The examination of how mass insertions are cancelled is explained in two different ways in order 
to get a minimal number of diagrams to work with in the determination of critical exponents. An in-depth discussion on the validity of the 
finite-size regime with the approach to the bulk criticality as well as to the dimensional crossover regime is presented. We give a brief 
description of the solution to the higher-order massless diagrams by writing down their expressions and point out their similarity with $PBC$ and 
$ABC$ arguments for the massless fields. In Sec. VI we compute the critical exponents in the massless approach.
\par In Sec. VII we discuss our results. The dimensional crossover regime is analyzed explicitly by focusing on the one-loop correction 
to the bulk case of the four-point vertex part. We perform a comparison of these regimes with those from $PBC$ and $ABC$ previously 
studied. We vary the correlation length and establish that even in the massive theory there are regions for fixed, finite $\xi$ where 
$\frac{L}{\xi} < 1$ and the $\epsilon$-expansion results are still valid.  
\par Section VIII displays the conclusions and future directions within the formalism introduced in the present work. In addition, we point 
out how these new aspects can be adapted to tackle the problem of competing systems. 
\par In Appendix A we compute higher order massive integrals. We decided not to give a detailed account of the solution of massless integrals 
of this problem in another appendix for the same reason. The reader is advised to consult Appendix A and Ref. \cite{BL} for grasping the details.  
\section{Modified $NF$ approach to Feynman rules with $DBC$ and $NBC$}
\subsection{Review of $NF$ approach}
\par First we will discuss briefly the field theory setting introduced by Nemirovsky and Freed $(NF)$ \cite{NF} for 
constructing Feynman diagrams in momentum space with the boundary conditions of interest in the present work. We are going to restrict 
ourselves to the situation where no external surface fields are allowed.
\par The bare Lagrangian (free energy) density is composed by scalar fields with $O(N)$ symmetry, defined on the volume enclosed by the 
two limiting $(d-1)$-dimensional parallel hyperplates located at $z=0,L$ (bulk fields). It is given by: 
\begin{equation}\label{1}
\mathcal{L} = \frac{1}{2}
|\bigtriangledown \phi_{0}|^{2} + \frac{1}{2} \mu_{0}^{2}\phi_0^{2} + \frac{1}{4!}\lambda_0(\phi_0^{2})^{2} ,
\end{equation}
where $\phi_{0}$, $\mu_{0}$ and $\lambda_{0}$ are the bare order parameter, mass 
($\mu_{0}^{2}= t_{0}$ is the bare reduced temperature proportional to $\frac{T-T_{C}}{T_{C}}$) and coupling 
constant, respectively \cite{amit,BLZ1,BLZ2}. The $O(N)$ symmetry of the bare order parameter means that it is a vector 
of $N$ components $((\phi_{0}^{2})^{2}= (\phi_{01}^{2} +...+\phi_{0N}^{2})^{2}$). The additional indices related to the $O(N)$ symmetry 
of the fields were not written since they are identical to the infinite system. The space directions split in two distinct sets 
denoted by $\vec{\rho}$ representing the coordinates along the $(d-1)$-dimensional subspace parallel 
to the plates and the perpendicular $z$ axis, collectively represented in the vector form as $x=(\vec{\rho},z)$. The space is layered and 
filled with parallel plates in the region between  $z=0$ and $z=L$. 
The field satisfies $\phi_{0}(z=0)= \phi_{0}(z=L)=0$ for Dirichlet 
boundary conditions $(DBC)$, whereas $\frac{\partial \phi_{0}}{\partial z}(z=0)= \frac{\partial \phi_{0}}{\partial z}(z=L)$ 
for Neumann boundary conditions $(NBC)$. 
\par The order parameter can be related to its Fourier modes in momentum space through 
$\phi_{0}(x)= {\underset{j}{\sum}} \int d^{d-1}k exp(i\vec{k}.\vec{\rho}) 
u_{j}(z) \phi_{0j}(\vec{k})$, where $\vec{k}$ is the momentum vector characterizing the $(d-1)$-dimensional space. The basis 
functions $u_{j}(z)$ satisfy the differential equation $-\frac{d^{2} u_{j}(z)}{dz^{2}}= 
\kappa_{j}^{2} u_{j}(z)$, $\kappa_{j}$ being the quasi-momentum along the $z$-direction. The 
eigenfunctions are orthonormalized according to 
${\underset{j}{\sum}} u_{j}(z)u_{j}(z')= \delta(z-z')$ and 
$\int_{0}^{L} dz u_{j}(z)u^{*}_{j'}(z)= \delta_{j,j'}$. Here  
$\kappa_{j}= \frac{\pi j}{L}\equiv \tilde{\sigma}j$, where $j=1, 2,...$ for $DBC$ and $j=0, 1, 2,...,$ for $NBC$ 
($\tilde{\sigma}= \frac{\pi}{L}$). We can attach a
label for each boundary condition such that $\tau =-1$ corresponds to DBC and $\tau=1$ to NBC. The free 
bare massive propagator ($\mu_{0}^{2} \neq 0$) in momentum space for either 
boundary condition is given by the expression 
$G_{0j}^{(\tau)}(k,j) = \frac{1}{k^{2} + \tilde{\sigma}^{2}j^{2} 
+ \mu_{0}^{2}}$. Although not explicitly mentioned so far, it is obvious that the eigenfunctions depend upon the boundary 
conditions. We will suppress their relation with the boundary conditions but can retrieve this dependence whenever it is 
appropriate. 
\par A generic Feynman diagram involves the product 
of many bare propagators $G_{0j}^{(\tau)}$ contracted with interaction vertices. The amazing fact about the structure of 
the finite-size to take effect on the field theory is that it is implemented as an internal symmetry. For example, beside the standard 
tensorial couplings of the infinite theory corresponding to the $N$-component order parameter, each momentum line (propagator) 
must be multiplied by $S^{(\tau)}_{j_{1}j_{2}} = \int_{0}^{L} dz u_{j_{1}}(z)u_{j_{2}}(z)$ and the $\phi^{4}$ vertices 
are multipled by the tensor $S^{(\tau)}_{j_{1}j_{2}j_{3}j_{4}}= \int_{0}^{L} 
dz u_{j_{1}}(z)u_{j_{2}}(z)u_{j_{3}}(z)u_{j_{4}}(z)$. Forgetting about the composite operators for the time being, this is nothing but the 
enhanced internal symmetry representation of the direct product $O(N) \times (fs)$, where the $(fs)$ symmetry is represented by the 
tensors  $S^{(\tau)}_{j_{1}j_{2}}$ and $S^{(\tau)}_{j_{1}j_{2}j_{3}j_{4}}$. The basis functions can be written as 
$u_{j}^{(\tau=-1)}(z)= \Bigl(\frac{2}{L}\Bigr)^{\frac{1}{2}} sin(\kappa_{j} z)$ for $DBC$, whereas for 
$NBC$ we have the nonzero mode as in $DBC$ $u_{j}^{(\tau=1)}(z)= \Bigl(\frac{2}{L}\Bigr)^{\frac{1}{2}} cos(\kappa_{j} z)$ 
($j=1,2...$) as well as $u_{0}^{(\tau=1)}= \Bigl(\frac{1}{L}\Bigr)^{\frac{1}{2}}$. The Feynman rules for vertex and propagators are 
given in \cite{NF}. 
\par In the computation of Feynman diagrams as the set of integrals in $(d-1)$ dimensions in conjunction with infinite 
summations, it is difficult to establish a direct comparison with periodic ($PBC$) and antiperiodic boundary conditions 
($ABC$). The reason is that $j \geq 0$ for $DBC$ and $NBC$, but varies in the interval ($-\infty,\infty$) por 
$PBC$ and $ABC$. The simplification which took place for $PBC$ and $ABC$ when we could compute integrals with 
all external quasi-momentum set to zero no longer occurs for $DBC$. In any serious attempt to unify the framework 
for $DBC$ and $PBC$, we should figure out how to compute graphs with nonvanishing external quasi-momentum for those 
conditions.  
\par Fortunately, at an external quasi-momentum symmetric point, we can compute the diagrams provided some 
modifications are introduced in the moding of the label $j$ and additional trivial orthonormality properties. Next, 
we shall  introduce new notation in order to give a unified description of the Dirichlet and 
Neumann problems for nonzero external quasi-momentum. 

\subsection{Exponential representation and unification of the Feynman rules for nonvanishing external quasi-momentum}
\subsubsection{One-loop diagrams for the two- and four-point vertex functions}
\par The construction of Feynman diagrams for $DBC$ (sine) and $NBC$ (cosine) takes into account  
solely the internal structure provided by the tensors $S^{(\tau)}_{j_{1}j_{2}}$, $S^{(\tau)}_{j_{1}j_{2}j_{3}j_{4}}$. The 
$O(N)$ underlying symmetry appears in exactly the same form as in the infinite $L$ limit and we shall simply attach 
the symmetry factor to each diagram under consideration. However, we are going to discuss explicitly how 
the combination of products of the finite size tensors shows up in particular one-loop graphs. We 
wish to express them in terms of summations varying in the range $(-\infty, \infty)$ and $(d-1)$-dimensional momentum 
integrals. 
\par Let us write the basis functions as
\begin{equation}\label{2}
u_j^{(\tau=-1)}(z)=
\left(\frac{2}{L}\right)^{\frac{1}{2}}\left[\frac{1}{2i}\left(e^{i\tilde{\sigma} j z}-e^{-i\tilde{\sigma} j z}\right)\right]
\;\;\;\;\; (\textit{DBC}),
\end{equation}
\begin{equation}\label{3}
u_j^{(\tau=1)}(z)=
\left(\frac{2}{L}\right)^{\frac{1}{2}}\left[\frac{1}{2}\left(e^{i\tilde{\sigma} j z}+e^{-i\tilde{\sigma} j z}\right)\right]
\;\;\;\;\;\;\;\; (\textit{NBC}).
\end{equation}
\par We can compute the tensors in this new representation. For instance, using the definition of $S^{(\tau)}_{j_{1} j_{2}}$ 
and restricting ourselves only to positive values ($j_1, j_2 \in Z_+^*$) we achieve the following unified form
\begin{equation}\label{4}
S^{(\tau)}_{j_{1} j_{2}} = \frac{(-1)^{\frac{1 - \tau}{2}}}{2L}\int_{0}^{L} dz[e^{i\tilde{\sigma} (j_{1}+j_{2}) z}+e^{-i\tilde{\sigma} (j_{1}+j_{2}) z} 
+\tau e^{i\tilde{\sigma} (j_{1}-j_{2}) z}+ \tau e^{-i\tilde{\sigma} (j_{1}-j_{2}) z}],
\end{equation}
which by the change $z \rightarrow -z$ in the second and fourth terms turns out to be given by:
\begin{equation}\label{5}
S^{(\tau)}_{j_1 j_2}=(-1)^{\frac{1 - \tau}{2}}[\tau \delta_{j_1 - j_2,0} + \delta_{j_1 + j_2,0}] \equiv (-1)^{\frac{1 - \tau}{2}}[\tau \delta(j_1 - j_2) 
+ \delta(j_1 + j_2)] .
\end{equation}
The new notation for the Kronecker's delta is going to be very useful in what follows. We have to be careful 
with the moding since $NBC$ allows $j_{1},j_{2}=0$ in the construction of a generic Feynman diagram: in the contraction 
of these tensors, this value of $j$ can occur and all components of the finite-size tensors should be computed 
(as the internal quasi-momentum indices are summed). For all practical purposes in the present paper, 
the result for the two-index tensor is simply $S^{(\tau)}_{j_1 j_2}= \delta(j_1 - j_2)$.   
\par It is a simple task to evaluate the tensor $S^{(\tau)}_{j_1 j_2 j_3 j_4}$ with the arguments at hand. We find
\begin{eqnarray}\label{6}
S^{(\tau)}_{j_1 j_2 j_3 j_4} &=& \frac{1}{2L}\Bigl[\delta(j_1+j_2+j_3+j_4) +\tau \delta(j_1-j_2+j_3+j_4) 
+ \tau \delta(j_1+j_2-j_3+j_4) \nonumber\\ 
&& \;\;\;\;\; +\; \tau\delta(j_1+j_2+j_3-j_4)+ \delta(j_1-j_2-j_3+j_4)+\delta(j_1-j_2+j_3-j_4) \nonumber \\
&& \;\;\;\;\; + \; \delta(j_1+j_2-j_3-j_4)+\tau\delta(j_1-j_2-j_3-j_4)\Bigr]. 
\end{eqnarray}.  
\par In $PBC$ and $ABC$ the quasi-momentum conservation is represented by the condition 
$S^{(\tau)}_{j_1 j_2 j_3 j_4}= \frac{1}{L} \delta(j_1+j_2+j_3+j_4)$, whereas $DBC$ and $NBC$ in the above expression 
possess the quasi-momentum conservation as well as additional contributions as expected from the intrinsic nontrivial nature of those boundary conditions. 
\par The last expression can be further simplified utilizing the notation $\tilde{j}_{\pm}=j_{1} \pm j_{2}$, 
$j_{\pm}=j_{3} \pm j_{4}$ which yields
\begin{equation}\label{7}
S^{(\tau)}_{j_1 j_2 j_3 j_4}=\frac{1}{2L} \sum_{\substack{\alpha_1=\pm 1 \\ \alpha_2=\pm 1 \\ \beta=\pm 1 }} (\alpha_1 \alpha_2)^{\frac{1 - \tau}{2}}\delta(\tilde{j}_{\alpha_1} + \beta j_{\alpha_2}).
\end{equation}
\par Let us consider the components of the finite-size tensors when $j=0$. In the simplest case of the two-index 
tensor, it is easy to show that zero modes do not mix with nonzero modes, 
which can be expressed in a compact form as $S^{(\tau)}_{0 j}=\delta(j) \delta^{\tau 1}$. When we analyze the four-index 
tensor, the only difference with respect to the components with all nonvanishing subscripts is different normalization 
factors. It is straightforward to prove that $S^{(\tau)}_{0 j_1 j_2 j_3}=\frac{1}{\sqrt{2}L} [\delta(j_1+j_2+j_3)
+\delta(j_1-j_2+j_3)+\delta(j_1+j_2-j_3)+\delta(j_1-j_2-j_3)] \delta^{\tau 1}$, 
$S^{(\tau)}_{0 0 j_1 j_2}=\frac{1}{L}[\delta(j_1+j_2)+\delta(j_1-j_2)]\delta^{\tau 1}$ and 
$ S^{(\tau)}_{0 0 0 j_1}=\frac{1}{L} \delta(j_{1}) \delta^{\tau 1}$. It is 
important to mention that neither $S^{(\tau)}_{0 0 j_1 j_2}$ nor $S^{(\tau)}_{0 j_1 j_2 j_3}$ are equivalent to $NF$ 
notation. The above tensor is similar but has different components when compared with the $NF$ counterpart (obtained in terms 
of sines and cosines) as can be verified in Eq. (A6) from the second paper in Ref. \cite{NF}. Therefore, the components of the four-index tensor 
in the exponential representation enlarges the possibilities for the finite-size indices.
\par These elements will suffice to our construction of vertex parts diagrams not including composite fields yet. Before 
working out explicitly the one-loop graphs, let us introduce the notation 
$\tilde{S}^{(\tau)}_{j_1 j_2 j_3 j_4}= 2 \pi S^{(\tau)}_{j_1 j_2 j_3 j_4}$ and from now on we are going to construct the 
integrals associated with Feynman diagrams with this modified tensor. Each loop integral will be represented by a 
$(d-1)$-dimensional integral multiplied by summations involving products of $S^{(\tau)}_{j_1 j_2}$ and 
$\tilde{S}^{(\tau)}_{j_1 j_2 j_3 j_4}$. This has the virtue of producing the metric factor $\tilde{\sigma}$ multiplying 
each loop integral.
\par In practice, we have to use the tensor structure to build up the diagrams perturbatively. The coupling constant factors will be 
omitted in all diagrams to be discussed, but will be retrieved during the discussion of the diagrammatic expansion of each primitively 
divergent vertex part. Within a $1PI$ vertex part framework, those primitively divergent are the two-, four- and two-point with 
one composite operator vertex parts, which are represented by $\Gamma^{(2)}$, $\Gamma^{(4)}$ and $\Gamma^{(2,1)}$, respectively. Whenever we 
do not mention explicitly the external quasi-momentum index in a certain graph, these modes are arbitrary. For instance, a generic 
graph from $\Gamma^{(2)}$ has external quasi-momenta labeled by $i_{1}$ and $i_{2}$, $\Gamma^{(4)}$ has associated 
external quasi-momenta $i_{1},i_{2},i_{3}$ and $i_{4}$ and so on. On the other hand, we are going to attach them to the particular diagram 
whenever we choose a particular fixed distribution of external quasi-momenta modes.
 \par As a first sample computation, consider the one-loop ``tadpole'' contribution to $\Gamma^{(2)}$. Its diagrammatic expression reads
\begin{eqnarray}\label{8}
\parbox{10mm}{\includegraphics[scale=0.32]{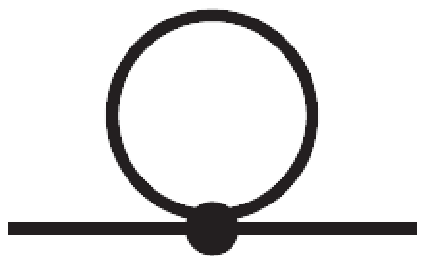}}\quad &=& \frac{(N+2)}{3}\overset{\infty}{\underset{l_{1}, l_{2} \geq 0}{\sum}} \tilde{S}^{(\tau)}_{i_1 i_2 l_1 l_2} S^{(\tau)}_{l_1 l_2} \int d^{d-1}q G_{0}(q,l_{1}).
\end{eqnarray}    
Let us consider explicitly Neumann boundary conditions ($\tau=1$). The summation splits in the contributions of zero and 
nonzero modes as:
\begin{eqnarray}\label{9}
&& \overset{\infty}{\underset{l_{1}, l_{2} \geq 0}{\sum}} \tilde{S}^{(\tau)}_{i_1 i_2 l_1 l_2} S^{(\tau)}_{l_1 l_2} \int d^{d-1}k G_{0}(q,l_{1}) = 
\tilde{S}^{(1)}_{i_{1} i_{2} 00} S^{(1)}_{00} \int d^{d-1}k G_{0}(q,0) + 2\overset{\infty}{\underset{l=1}{\sum}} \tilde{S}^{(1)}_{i_{1} i_{2} l0} S^{(1)}_{l0} \nonumber\\
&&\int d^{d-1}k G_{0}(q,l) + \overset{\infty}{\underset{l_{1}, l_{2} =1}{\sum}} \tilde{S}^{(1)}_{i_1 i_2 l_1 l_2} S^{(1)}_{l_1 l_2} \int d^{d-1}k G_{0}(q,l_{1}). 
\end{eqnarray} 
Note that $G_{0}(q,l)=G_{0}(q,-l)$. The term proportional to $S^{(1)}_{l0}$ vanishes. Replacing explicitly the values of the tensors 
already derived and defining $i_{\pm}= i_{1} \pm i_{2}$, we find:
\begin{eqnarray}\label{10}
&& \overset{\infty}{\underset{l_{1}, l_{2} \geq 0}{\sum}} \tilde{S}^{(\tau)}_{i_1 i_2 l_1 l_2} S^{(\tau)}_{l_1 l_2} \int d^{d-1}k G_{0}(q,l_{1}) = 
\tilde{\sigma} \int d^{d-1}k \Bigl[ 2(\delta(i_{+}) +\delta(i_{-}))G_{0}(q,0) + \overset{\infty}{\underset{l=1}{\sum}}(\delta(i_{+} + 2l)\nonumber\\ 
&& + \delta(i_{+} -2l) + 2\delta(i_{+}) +  \delta(i_{-} + 2l) 
+ \delta(i_{-} -2l) + 2\delta(i_{-}))G_{0}(q,l)\Bigr]. 
\end{eqnarray} 
Next, the identities 
\begin{subequations}\label{11}
\begin{eqnarray}
&& \overset{\infty}{\underset{l\neq 0, \; l= -\infty}{\sum}} \delta(i_{\pm} + 2l)G_{0}(q,l)= 
\overset{\infty}{\underset{l= -\infty}{\sum}} \delta(i_{\pm} + 2l)G_{0}(q,l) - \delta(i_{\pm})G_{0}(q,0),\label{11a}\\
&& \delta(i_{\pm})G(q,0) + 2 \delta(i_{\pm}) \overset{\infty}{\underset{l=1}{\sum}} G_{0}(q,l)=  \delta(i_{\pm}) 
\overset{\infty}{\underset{l= -\infty}{\sum}} G_{0}(q,l),\label{11b}
\end{eqnarray}
\end{subequations}
will be useful in order to achieve our goal of transforming summations with a bounded lower limit into 
those with unlimited negative values for the index $l$. After some manipulations the final expression can 
be rewritten as 
\begin{eqnarray}\label{12}
\parbox{10mm}{\includegraphics[scale=0.32]{fig1DN.eps}}\quad &=& \tilde{\sigma} 
\frac{(N+2)}{3} \int d^{d-1}q \overset{\infty}{\underset{l= -\infty}{\sum}} \frac{[\delta(i_{-}) + \delta(i_{+}) 
+ \delta(i_{+} + 2l) + \delta(i_{-} - 2l)]}{q^{2} + \tilde{\sigma}^{2} l^{2} + \mu_{0}^{2}}.
\end{eqnarray}    
\par For $DBC$ boundary conditions we have instead
\begin{eqnarray}\label{13}
\parbox{10mm}{\includegraphics[scale=0.32]{fig1DN.eps}}\quad &=& \tilde{\sigma} \frac{(N+2)}{3}\overset{\infty}{\underset{l_{1}, l_{2}=1}{\sum}} \tilde{S}^{(-1)}_{i_1 i_2 l_1 l_2} S^{(-1)}_{l_1 l_2} \int d^{d-1}k G_{0}(k,l_{1}).
\end{eqnarray}    
With $l_{1},l_{2} > 0$, $S^{(-1)}_{l_{1},l_{2}}= \delta(l_{1}-l_{2})$. We can perform one summation, say over $l_{2}$ and perform the 
change of index $l_{1}\rightarrow l$ in the remaining summation. Focusing only on the summation and replacing the expression 
for $\tilde{S}^{(-1)}_{i_{1} i_{2} ll}$ we get to 
\begin{eqnarray}\label{14}
&& \overset{\infty}{\underset{l=1}{\sum}} \tilde{S}^{(-1)}_{i_{1} i_{2} ll} \int d^{d-1}k G_{0}(k,l)= \tilde{\sigma} 
\overset{\infty}{\underset{l=1}{\sum}} [-2\delta(i_{+}) + \delta(i_{+} + 2l) + \delta(i_{+} - 2l) + 2\delta(i_{-}) - \delta(i_{-} + 2l) \nonumber\\ 
&& - \delta(i_{-} - 2l)] \int d^{d-1}k G_{0}(k,l).
\end{eqnarray} 
From the identity $\overset{\infty}{\underset{l=1}{\sum}} (\delta(i_{\pm} + 2l) + \delta(i_{\pm} - 2l))G_{0}(q,l)= 
\overset{\infty}{\underset{l= -\infty}{\sum}} \delta(i_{\pm} + 2l)G_{0}(q,l)$ and by employing Eqs. (\ref{11}), 
we obtain
\begin{eqnarray}\label{15}
\parbox{10mm}{\includegraphics[scale=0.32]{fig1DN.eps}}\quad &=& \tilde{\sigma} 
\frac{(N+2)}{3} \int d^{d-1}q \overset{\infty}{\underset{l= -\infty}{\sum}} \frac{[\delta(i_{-}) - \delta(i_{+}) 
+ \delta(i_{+} + 2l) - \delta(i_{-} - 2l)]}{q^{2} + \tilde{\sigma}^{2} l^{2} + \mu_{0}^{2}}.
\end{eqnarray}   
Comparing the $DBC$ ($\tau=-1$) and $NBC$ ($\tau=1$) results, we can write down the expression corresponding to this graph 
in a unified fashion for these boundary conditions as
\begin{eqnarray}\label{16}
\parbox{10mm}{\includegraphics[scale=0.32]{fig1DN.eps}}\quad &=& \tilde{\sigma} 
\frac{(N+2)}{3} \int d^{d-1}q \overset{\infty}{\underset{l= -\infty}{\sum}} \frac{[\delta(i_{-}) + \tau \delta(i_{+}) 
+ \delta(i_{+} + 2l) + \tau \delta(i_{-} - 2l)]}{q^{2} + \tilde{\sigma}^{2} l^{2} + \mu_{0}^{2}}.
 \end{eqnarray}
The summations in the third and fourth term can be performed and we are left with
\begin{eqnarray}\label{17}
&& \parbox{10mm}{\includegraphics[scale=0.32]{fig1DN.eps}}\quad = \frac{(N+2)}{3}\Bigl[(\delta(i_{-})+ \tau \delta(i_{+})) 
D_{1}(\tilde{\sigma},\mu_{0}) + \tau \tilde{D}_{1}(\frac{i_{-}}{2},\tilde{\sigma},\mu_{0}) 
+ \tilde{D}_{1}(\frac{i_{+}}{2},\tilde{\sigma},\mu_{0})\Bigr], 
\end{eqnarray}
where
\begin{subequations}\label{18}
\begin{eqnarray}
&& D_{1}(\tilde{\sigma},\mu_{0}) = \tilde{\sigma} \overset{\infty}{\underset{l= -\infty}{\sum}} \int \frac{d^{d-1}q}{q^{2} + \tilde{\sigma}^{2} l^{2} + \mu_{0}^{2}}, \label{18a}\\
&& \tilde{D}_{1}(i,\tilde{\sigma},\mu_{0})= \tilde{\sigma} \int \frac{d^{d-1}q}{q^{2} + \tilde{\sigma}^{2} i^{2} + \mu_{0}^{2}}.\label{18b}
\end{eqnarray}
\end{subequations}
Although the first and second terms in the diagram expression have a clear counterpart in $PBC$ and $ABC$, the third and fourth 
terms are entirely new, since no summation appears there. They look like surface terms albeit coming purely from 
finite-size contributions: they reflect the breaking of translation invariance for $DBC$ and $NBC$ along the finite-size direction. In what 
follows diagrams of this type (``tadpoles'') will not be required in the present method as we are going to show in the 
discussion of normalization conditions.      
\par Consider the one-loop contribution of the four-point vertex part $\Gamma^{(4)}$. Its graph in the present setting is 
given by
\begin{eqnarray}\label{19}
&& \parbox{10mm}{\includegraphics[scale=0.32]{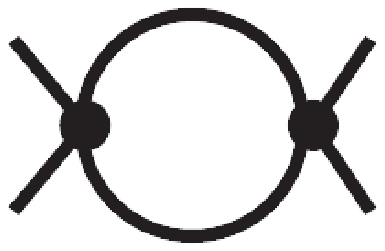}}\quad = \frac{(N+8)}{9} 
\overset{\infty}{\underset{l_{1},l_{2} \geq 0}{\sum}}\tilde{S}^{(\tau)}_{i_1 i_2 l_1 l_2} \tilde{S}^{(\tau)}_{l_1 l_2 i_3 i_4} \int d^{d-1}q G_{0}(q+k,l_{1}) G_{0}(q,l_{2}),
\end{eqnarray}
where $k$ is the external momentum, $i_{1}, i_{2}, i_{3}$ and $i_{4}$ are the external indices from quasi-momenta. We start 
with $\tau=1$ ($NBC$). Focusing only on the summmation, integrals and propagator products, we first decompose the whole thing 
in the following form
\begin{eqnarray}\label{20}
&& \overset{\infty}{\underset{l_{1},l_{2} \geq 0}{\sum}}\tilde{S}^{(\tau)}_{i_1 i_2 l_1 l_2} \tilde{S}^{(\tau)}_{l_1 l_2 i_3 i_4} \int d^{d-1}q G_{0}(q+k,l_{1}) G_{0}(q,l_{2})= \int d^{d-1}q \Bigl[\tilde{S}^{(1)}_{i_{1} i_{2} 0 0} \tilde{S}^{(1)}_{0 0 i_{3} i_{4}}G_{0}(q+k,0) \nonumber\\
&& G_{0}(q,0)  + \overset{\infty}{\underset{l_{1}=1}{\sum}}\tilde{S}^{(\tau)}_{i_{1} i_{2} l_{1}0} \tilde{S}^{(\tau)}_{l_{1} 0 i_{3} i_{4}} G_{0}(q+k,l_{1}) G_{0}(q,0) 
+ \overset{\infty}{\underset{l_{2}=1}{\sum}}\tilde{S}^{(\tau)}_{i_{1} i_{2} 0 l_{2}} \tilde{S}^{(\tau)}_{0 l_{2} i_{3} i_{4}} G_{0}(q+k,0) G_{0}(q,l_{2})\nonumber\\ 
&& + \overset{\infty}{\underset{l_{1},l_{2}=1}{\sum}}\tilde{S}^{(\tau)}_{i_1 i_2 l_1 l_2} \tilde{S}^{(\tau)}_{l_1 l_2 i_3 i_4} 
G_{0}(q+k,l_{1}) G_{0}(q,l_{2})\Bigr]. 
\end{eqnarray}   
Note that the single summations are actually equal.  For the sake of simplicity, the definitions 
$i_{\pm}=i_{1} \pm i_{2}$ and  $\tilde{i}_{\pm}=i_{3} \pm i_{4}$ will have their utility in what follows. Using the expressions 
for the tensors and the symmetry of the integrals yield
\begin{eqnarray}\label{21}
&&\overset{\infty}{\underset{l_{1},l_{2} \geq 0}{\sum}}\tilde{S}^{(\tau)}_{i_1 i_2 l_1 l_2} \tilde{S}^{(\tau)}_{l_1 l_2 i_3 i_4} \int d^{d-1}q G_{0}(q+k,l_{1}) G_{0}(q,l_{2})=4\tilde{\sigma}^{2} \int d^{d-1}q \Bigl([\delta(i_{+}) + \delta(i_{-})][\delta(\tilde{i}_{+})\nonumber\\
&& + \delta(\tilde{i}_{-})]G_{0}(q+k,0)G_{0}(q,0) 
+ \overset{\infty}{\underset{l=1}{\sum}} [\delta(i_{+} + l) + \delta(i_{+} - l) + \delta(i_{-} + l) + \delta(i_{-} - l)]
[\delta(\tilde{i}_{+}\nonumber\\ 
&& + l) + \delta(\tilde{i}_{+} - l) + \delta(\tilde{i}_{-} + l)+ \delta(\tilde{i}_{-}-l)]
G_{0}(q+k,l)G_{0}(q,0) 
+ \frac{1}{4}\overset{\infty}{\underset{l_{1},l_{2}=1}{\sum}}[\delta(i_{+}+l_{1}+l_{2}) \nonumber\\
&& + \delta(i_{+}-l_{1}+l_{2}) + \delta(i_{+}+l_{1}-l_{2})
+ \delta(i_{+}-l_{1}-l_{2}) + \delta(i_{-}+l_{1}+l_{2}) + \delta(i_{-}-l_{1}+l_{2})\nonumber\\ 
&& + \delta(i_{-}+l_{1}-l_{2}) + \delta(i_{-}-l_{1}-l_{2})][\delta(\tilde{i}_{+}+l_{1}+l_{2}) + \delta(\tilde{i}_{+}-l_{1}+l_{2}) + \delta(\tilde{i}_{+}+l_{1}-l_{2})\nonumber\\
&& + \delta(\tilde{i}_{+}-l_{1}-l_{2}) + \delta(\tilde{i}_{-}+l_{1}+l_{2})+ \delta(\tilde{i}_{-}-l_{1}+l_{2}) 
+ \delta(\tilde{i}_{-}+l_{1}-l_{2}) + \delta(\tilde{i}_{-} - l_{1} -l_{2})]\times \nonumber\\
&& G_{0}(q+k,l_{1})G_{0}(q,l_{2})\Bigr).
\end{eqnarray}
The summation involving the combination $[\delta(i_{\pm} + l) + \delta(i_{\pm} - l)]$ can be transformed into one whose range extends 
to negative integer values not including the zero. Indeed, the identity 
\begin{eqnarray}\label{22}
&& \overset{\infty}{\underset{l=1}{\sum}} 
[\delta(i_{\pm} + l) + \delta(i_{\pm} - l)][\delta(\tilde{i}_{+} + l) + \delta(\tilde{i}_{+} - l) + \delta(\tilde{i}_{-} + l)+ \delta(\tilde{i}_{-}-l)]= \overset{\infty}{\underset{l\neq0,
l=-\infty}{\sum}}\delta(i_{\pm}\nonumber\\
&&  + l)[\delta(\tilde{i}_{+}+ l) + \delta(\tilde{i}_{+} - l) + \delta(\tilde{i}_{-} + l)+ \delta(\tilde{i}_{-}-l)],
\end{eqnarray}
can be used so that  
\begin{eqnarray}\label{23} 
&& \int d^{d-1}q \overset{\infty}{\underset{l=1}{\sum}} [\delta(i_{\pm} + l) + \delta(i_{\pm} - l)][\delta(\tilde{i}_{+} + l) + \delta(\tilde{i}_{+} - l) + \delta(\tilde{i}_{-} + l)+ \delta(\tilde{i}_{-}-l)]G_{0}(q+k,l)\nonumber\\
&& G_{0}(q,0)= \int d^{d-1}q 
\overset{\infty}{\underset{l=-\infty}{\sum}}[\delta(i_{+} + l) + \delta(i_{-} - l)][\delta(\tilde{i}_{+} + l) 
+ \delta(\tilde{i}_{+} - l) + \delta(\tilde{i}_{-} + l)+ \delta(\tilde{i}_{-}-l)]\nonumber\\
&& G_{0}(q+k,l)G_{0}(q,0)- 2[\delta(i_{+}) 
+ \delta(i_{-})][\delta(\tilde{i}_{+}) + \delta(\tilde{i}_{-})]\int d^{d-1}q G_{0}(q+k,0)G_{0}(q,0).
\end{eqnarray}
Using these expressions 
inside the double summation and carrying out the computations for it in a similar way, we obtain the result in terms of 
summations with indices varying in the range $(-\infty,\infty)$. The single summation is 
cancelled by the zero mode of each index in the double summation and we are left with the following expression for this 
diagram
\begin{eqnarray}\label{24}
&& \parbox{10mm}{\includegraphics[scale=0.32]{fig2DN.eps}}\quad = \frac{(N+8)}{9} \tilde{\sigma}^{2} \overset{\infty}{\underset{l_{1},l_{2}=-\infty}{\sum}}[\delta(i_{+}+l_{1}+l_{2})+ \delta(i_{-}+l_{1}+l_{2})][\delta(\tilde{i}_{+}+l_{1}+l_{2}) + \delta(\tilde{i}_{+}-l_{1}\nonumber\\
&& +l_{2}) + \delta(\tilde{i}_{+}+l_{1}-l_{2})+ \delta(\tilde{i}_{+}-l_{1}-l_{2}) + \delta(\tilde{i}_{-}+l_{1}+l_{2})+ \delta(\tilde{i}_{-}-l_{1}+l_{2}) 
+ \delta(\tilde{i}_{-}+l_{1}-l_{2}) \nonumber\\
&& + \delta(\tilde{i}_{-} - l_{1} -l_{2})]\int d^{d-1}q G_{0}(q+k,l_{1})G_{0}(q,l_{2}).
\end{eqnarray}
This form can be further reduced in such a way that at maximum one summation 
for each loop graph takes place. We can perform the summations very easily and express the diagram in the simpler form
\begin{eqnarray}\label{25}
&& \parbox{10mm}{\includegraphics[scale=0.32]{fig2DN.eps}}\quad\;= \frac{(N+8)}{9} \tilde{\sigma}\Biggl\{[\delta(i_{+}+\tilde{i}_{+}) + \delta(i_{+}-\tilde{i}_{+})+\delta(i_{+}+\tilde{i}_{-}) + 
\delta(i_{+}-\tilde{i}_{-})]I_{2}(k,i_{+};\tilde{\sigma},\mu_{0}) \nonumber\\
&& + [\delta(i_{-}+\tilde{i}_{+}) + \delta(i_{-}-\tilde{i}_{+})+\delta(i_{-}+\tilde{i}_{-}) + \delta(i_{-}-\tilde{i}_{-})]I_{2}(k,i_{-};\tilde{\sigma},\mu_{0}) + \tilde{I}_{2}(k,\frac{i_{+}+\tilde{i}_{+}}{2},\frac{i_{+}-\tilde{i}_{+}}{2},\nonumber\\
&& \tilde{\sigma},\mu_{0})
+  \tilde{I}_{2}(k,\frac{i_{+}-\tilde{i}_{+}}{2},\frac{i_{+}+\tilde{i}_{+}}{2},\tilde{\sigma},\mu_{0})
+ \tilde{I}_{2}(k,\frac{i_{+}+\tilde{i}_{-}}{2},\frac{i_{+}-\tilde{i}_{-}}{2},\tilde{\sigma},\mu_{0})
+ \tilde{I}_{2}(k,\frac{i_{+}-\tilde{i}_{-}}{2},
\nonumber\\
&& \frac{i_{+}+\tilde{i}_{-}}{2},\tilde{\sigma},\mu_{0}) 
+ \tilde{I}_{2}(k,\frac{i_{-}+\tilde{i}_{+}}{2},\frac{i_{-}-\tilde{i}_{+}}{2},\tilde{\sigma},\mu_{0}) 
+ \tilde{I}_{2}(k,\frac{i_{-}-\tilde{i}_{+}}{2},\frac{i_{-}+\tilde{i}_{+}}{2},\tilde{\sigma},\mu_{0})
+ \tilde{I}_{2}(k,\nonumber\\
&& \frac{i_{-}+\tilde{i}_{-}}{2},\frac{i_{-}-\tilde{i}_{-}}{2},\tilde{\sigma},\mu_{0})
+ \tilde{I}_{2}(k,\frac{i_{-}-\tilde{i}_{-}}{2},\frac{i_{-}+\tilde{i}_{-}}{2},\tilde{\sigma},\mu_{0})\Biggr\}
\end{eqnarray} 
where 
\begin{subequations}\label{26}
\begin{eqnarray}
&& I_{2}(k,i,\tilde{\sigma},\mu_{0})= \tilde{\sigma} \overset{\infty}{\underset{l= -\infty}{\sum}} \int 
\frac{d^{d-1}q}{[q^{2} + \tilde{\sigma}^{2}l^{2}+ \mu_{0}^{2}][(q+k)^{2}+\tilde{\sigma}^{2}(l+i)^{2}+ \mu_{0}^{2}]},\label{26a}\\
&& \tilde{I}_{2}(k,i,j,\tilde{\sigma},\mu_{0})= \tilde{\sigma} \int \frac{d^{d-1}q}{[q^{2} + \tilde{\sigma}^{2}i^{2}+ \mu_{0}^{2}][(q+k)^{2}+\tilde{\sigma}^{2}j^{2}+ \mu_{0}^{2}]}.\label{26b}
\end{eqnarray}
\end{subequations}  
For $DBC$, the absence of zero modes in the four-index tensor makes the computation easier. Now the diagram reads:
\begin{eqnarray}\label{27}
\parbox{10mm}{\includegraphics[scale=0.32]{fig2DN.eps}}\quad &=& \frac{(N+8)}{9} 
\lambda_{0}^{2}\overset{\infty}{\underset{l_{1},l_{2}=1}{\sum}}\tilde{S}^{(-1)}_{i_1 i_2 l_1 l_2} \tilde{S}^{(-1)}_{l_1 l_2 i_3 i_4} \int d^{d-1}q G_{0}(q+k,l_{1}) G_{0}(q,l_{2}),
\end{eqnarray}
and we have to transform the summation to the range $(-\infty,\infty$) just as before after using the tensor components for 
$\tau=-1$. The unified form holding for both $\tau=-1$ ($DBC$) and $\tau=1$ ($NBC$) can be written as
\begin{eqnarray}\label{28}
&& \parbox{10mm}{\includegraphics[scale=0.32]{fig2DN.eps}}\quad\;= \frac{(N+8)}{9} 
\tilde{\sigma}\Biggl\{[\delta(i_{+}+\tilde{i}_{+}) + \delta(i_{+}-\tilde{i}_{+}) 
+ \tau(\delta(i_{+}+\tilde{i}_{-}) + \delta(i_{+}-\tilde{i}_{-}))]I_{2}(k,i_{+};\tilde{\sigma},\mu_{0}) \nonumber\\
&& + [\tau(\delta(i_{-}+\tilde{i}_{+}) + \delta(i_{-}-\tilde{i}_{+}))+\delta(i_{-}+\tilde{i}_{-}) 
+ \delta(i_{-}-\tilde{i}_{-})]I_{2}(k,i_{-};\tilde{\sigma},\mu_{0}) 
+  \tau \Bigl[\tilde{I}_{2}(k,\frac{i_{+}-\tilde{i}_{+}}{2},\nonumber\\
&& \frac{i_{+}+\tilde{i}_{+}}{2},\tilde{\sigma},\mu_{0})
+ \tilde{I}_{2}(k,\frac{i_{+}+\tilde{i}_{+}}{2},\frac{i_{+}-\tilde{i}_{+}}{2},\tilde{\sigma},\mu_{0})
+ \tilde{I}_{2}(k,\frac{i_{-}+\tilde{i}_{-}}{2},\frac{i_{-}-\tilde{i}_{-}}{2},\tilde{\sigma},\mu_{0}) 
+ \tilde{I}_{2}(k,\nonumber\\
&& \frac{i_{-}-\tilde{i}_{-}}{2},\frac{i_{-}+\tilde{i}_{-}}{2},\tilde{\sigma},\mu_{0})\Bigr]
+ \tilde{I}_{2}(k,\frac{i_{-}+\tilde{i}_{+}}{2},\frac{i_{-}-\tilde{i}_{+}}{2},\tilde{\sigma},\mu_{0})
+ \tilde{I}_{2}(k,\frac{i_{-}-\tilde{i}_{+}}{2},\frac{i_{-}+\tilde{i}_{+}}{2},\tilde{\sigma},\mu_{0})\nonumber\\
&& + \tilde{I}_{2}(k,\frac{i_{+}-\tilde{i}_{-}}{2},\frac{i_{+}+\tilde{i}_{-}}{2},\tilde{\sigma},\mu_{0}) 
+ \tilde{I}_{2}(k,\frac{i_{+} + \tilde{i}_{-}}{2},\frac{i_{+}-\tilde{i}_{-}}{2},\tilde{\sigma},\mu_{0})\Biggr\},
\end{eqnarray}    
with the integrals defined by Eqs. (\ref{26}). From this simple analysis, we conclude that 
the Feynman diagrams in the exponential representation involve more complicated objects like 
the ``nondiagonal'' integrals $\tilde{I}_{2}(k,i,j,\tilde{\sigma},\mu_{0})$. Consequently, each 
momentum loop integral in the finite system for $DBC$ and $NBC$ {\it cannot} be obtained from the 
infinite system through the substitution $\int d^{d}k \rightarrow \overset{\infty}{\underset{j=-\infty}{\sum}} \sigma \int d^{d-1}k$ as in $PBC$ and $ABC$. The higher the loop graph, the lengthier is the computation of the contractions of 
$\tilde{S}^{(\tau)}_{j_{1}j_{2}j_{3}j_{4}}$ giving the particular diagram. Nevertheless, the same structure is preserved as in the 
simple one-loop examples just worked out. We postpone the presentation of all relevant higher loop diagrams after our next task, namely, 
the discussion of vertex parts that can be renormalized multiplicatively and include composite operators.  

\subsubsection{The vertex $\Gamma^{(2,1)}$: tree-level and one-loop graph} 
\par The most nontrivial feature of the construction proposed in the present work manifests itself in the diagrams of the composed field. 
The reason for that are the basis functions of the composite fields associated to the $\Gamma^{(2,1)}$ vertex part (and its descendents 
that can be renormalized multiplicatively), which consists of one single type for Dirichlet and Neumann boundary conditions. 
\par Applying the operator $(\bigtriangledown^{2} - \mu_{0}^{2})$ to the composite field $(\phi_{0}^{(\tau)}(x))^{2}$ and using the field 
equations $(\bigtriangledown^{2} - \mu_{0}^{2})\phi_{0}^{(\tau)}(x)=0$, we find
\begin{equation}\label{29}
(\bigtriangledown^{2} - \mu_{0}^{2})(\phi_{0}^{(\tau)}(x))^{2}= 2 (\bigtriangledown \phi_{0}^{(\tau)}(x))^{2}.
\end{equation}
The composite field also satisfies $DBC$ and $NBC$, respectively, namely
\begin{eqnarray}\label{30}
&& (\phi_{0}^{(\tau=-1)}(x))^{2}|_{z=0,L}=0, \nonumber\\
&& \frac{\partial}{\partial z}(\phi_{0}^{(\tau=1)}(x))^{2}|_{z=0,L}=0.
\end{eqnarray}
\par We decompose the composite field in terms of its components in momentum space as 
$(\phi_{0}^{(\tau)}(x))^{2}= {\underset{j}{\sum}} \int d^{d-1}p exp(i\vec{p}.\vec{\rho}) 
\tilde{u}_{j}^{(\tau)}(z) (\phi_{0j}^{(\tau)}(\vec{p},j))^{2}$. Our goal is the determination of the new basis functions of the composite 
field $\tilde{u}_{j}^{(\tau)}(z)$ preferably in terms of the previous basis functions expressed in terms of sines and cosines. In order to 
achieve that, we first work out the representation of the composite field in connection with the product of two single fields computed at 
the same point, apply the differential operators over them and examine the consequences. Afterwards, we compare the results of the same 
operation performed on the above definition of the composite fields and the basis functions $\tilde{u}_{j}^{(\tau)}(z)$.
\par Using the representation of the field and taking the product of two fields 
at the same point, we find $\phi_{0}(x)\phi_{0}(x)= {\underset{j_{1},j_{2}}{\sum}} \int d^{d-1}k_{1} 
exp(i(\vec{k}_{1}+\vec{k}_{2}).\vec{\rho}) u_{j_{1}}(z) u_{j_{2}}(z)\phi_{0 j_{1}}(\vec{k}_{1})\phi_{0 j_{2}}(\vec{k}_{2})$. Recalling that 
$\bigtriangledown^{2}= \frac{\partial^{2}}{\partial \rho^{2}} + \frac{\partial^{2}}{\partial z^{2}}$, applying this operator to the product 
of two fields and inserting this in above equation, we find
\begin{eqnarray}\label{31}
&& \Bigl[\frac{\partial^{2}}{\partial z^{2}} - (p_{1}^{2}+p_{2}^{2}) - 2 \mu_{0}^{2} \Bigr]u_{j_{1}}^{(\tau)}(z)u_{j_{2}}^{(\tau)}(z)= 
2 \frac{du_{j_{1}}^{(\tau)}}{dz} 
\frac{du_{j_{2}}^{(\tau)}}{dz}.
\end{eqnarray} 
Restricting ourselves only to nonvanishing values for $j_{1},j_{2}$, we can show that 
$u_{j_{1}}^{(\tau)}(z) u_{j_{2}}^{(\tau)}(z)= \frac{1}{L}[cos(\tilde{\sigma}(j_{1}-j_{2})z) 
+ \tau cos(\tilde{\sigma}(j_{1}+j_{2})z)]$. This implies that $\frac{du_{j_{1}}^{(\tau)}}{dz} 
\frac{du_{j_{2}}^{(\tau)}}{dz}= \frac{\tilde{\sigma}^{2}j_{1}j_{2}}{L}[cos(\tilde{\sigma}(j_{1}-j_{2})z) 
- \tau cos(\tilde{\sigma}(j_{1}+j_{2})z)]$. Using the definition $\tilde{u}_{j_{1},j_{2}}\equiv u_{j_{1}}^{(\tau)}(z)u_{j_{2}}^{(\tau)}(z)$ into 
the above equation leads us to the identity $(p_{1}^{2}+p_{2}^{2} +\tilde{\sigma}^{2}j_{1}^{2} + \tilde{\sigma}^{2}j_{1}^{2} + 2\mu_{0}^{2})
\tilde{u}_{j_{1},j_{2}}=0$. These manipulations suggest the choice for the basis functions of the composite field:
\begin{equation}\label{32}  
\tilde{u}_{j}^{\tau}(z) = \frac{A}{L}cos(\tilde{\sigma}jz),
\end{equation}
which implies that $\tilde{u}_{j_{1},j_{2}}= \frac{1}{A}[\tilde{u}_{j_{1}-j_{2}} + \tau \tilde{u}_{j_{1}+j_{2}}]$. We can now determine the tensor 
associated to the vertex part $\Gamma^{(2,1)}$ . The typical object responsible for the appearance of these diagrams corresponds to 
\begin{equation}\label{33}
F^{(2,1)}= \int d^{d-1} \rho \int_{0}^{L} dz \phi_{0}^{(\tau)}(x)\phi_{0}^{(\tau)}(x)(\phi_{0}^{(\tau)}(x))^{2}.
\end{equation} 
Replacing the Fourier expansion of the fields and composite operator, we obtain
\begin{equation}\label{34}
F^{(2,1)}= {\underset{j_{1},j_{2},j}{\sum}}\hat{S}_{j_{1}j_{2}j}^{(\tau)}\int d^{d-1}p_{1}d^{d-1}p_{2}d^{d-1}p 
\delta^{d-1}(\vec{p}_{1} + \vec{p}_{2} + \vec{p})\phi_{0j_{1}}(\vec{p}_{1})\phi_{0j_{2}}(\vec{p}_{2})(\phi_{0j}(\vec{p}))^{2}, 
\end{equation} 
where 
\begin{equation}\label{35}
\hat{S}_{j_{1}j_{2}j}^{(\tau)}= \int_{0}^{L} u_{j_{1}}^{(\tau)}(z)u_{j_{2}}^{(\tau)}(z)\tilde{u}_{j}^{(\tau)}(z) dz.
\end{equation}
The constant $A$ can be determined by imposing the normalization condition $\int_{0}^{L} \tilde{u}_{j}(z)\tilde{u}_{j'}(z) dz 
= \frac{L}{2} [\delta(j-j') + \delta(j+j')]$ with $j,j' \neq 0$, which implies $A=L$. Therefore, it can be checked that 
\begin{equation}\label{36}
\hat{S}_{j_{1}j_{2}j}^{(\tau)}= \frac{1}{2}[\delta(j-j_{1}+j_{2}) + \delta(j+j_{1}-j_{2}) + \tau \delta(j+j_{1}+j_{2}) 
+ \tau \delta(j-j_{1}-j_{2})]. 
\end{equation}
\par We can invert the Fourier transform from the product of two fields computed 
at the same point, using the identity $\tilde{u}_{j_{1},j_{2}}= \frac{1}{L}[\tilde{u}_{j_{1}-j_{2}} + \tau \tilde{u}_{j_{1}+j_{2}}]$. First, we 
multiply it by $e^{-i\vec{p}.\vec{\rho}}$ and integrate over $d^{d-1} \rho$. After that, multiply the resulting expression by 
$\tilde{u}_{j'}(z)$ and integrate over $z$ using the orthonormality conditions of $\tilde{u}_{j}(z)$. We obtain:
\begin{equation}\label{37}
\phi_{0}^{(\tau)}(\vec{p},j)= \frac{1}{L}\int d^{d-1} \rho [\phi_{0}^{(\tau)}(x)]^{2} exp[-i\vec{p}.\vec{\rho}] \int_{0}^{L} \tilde{u}_{j}(z) dz.
\end{equation}
Replacing this result in the left-hand side of the expression which defines the Fourier transform of the composite field 
(with only one summation), we get to the closure relation $\frac{1}{L}\sum_{j} \tilde{u}_{j}(z)\tilde{u}_{j}(z') = \delta(z-z')$. This 
completes the basic properties of the basis functions of the composite field.  
\par Before analyzing the situation for $NBC$ let us make an important remark concerning the construction of arbitrary loop diagrams 
for the composite field. First, observe that even though this tensor is similar to the four-point vertex part tensor 
$\tilde{S}_{j_{1}j_{2}j_{3}j_{4}}^{(\tau)}$, the similarity is not complete. We would like to make a connection between the tensors 
$\hat{S}_{j_{1}j_{2}j}^{(\tau)}$ and $\tilde{S}_{j_{1}j_{2}j_{3}j_{4}}^{(\tau)}$, or in other words, the graphs of $\Gamma^{(2,1)}$ with those 
from $\Gamma^{(4)}$ by identifying, for instance two external legs of the latter with the composite field insertion, as is usual for 
infinite systems. 
\par The fact of the matter is that 
a simple way to link the two types of graphs which generalizes the situation of 
bulk systems is to perform identifications between the external quasi-momenta. If we set $j_{3}=j_{4}=j'$, we find 
$\tilde{S}_{j_{1}j_{2}j'j'}^{(\tau)}= \tilde{\sigma}[\delta(2j'+j_{1}+j_{2}) + 2\tau \delta(j_{1}+j_{2}) + \delta(2j'-j_{1}-j_{2}) 
+ \tau \delta(2j'-j_{1}+j_{2}) + \tau \delta(2j'+j_{1}-j_{2}) + 2 \delta(j_{1}-j_{2})]$  (no summation in the repeated index). 
\par Looking at Eq. (\ref{36}), it is tempting to perform the identification $2j'\equiv j$ in that formula, 
but this does not solve the problem completely, for the additional contributions showing up in $\tilde{S}_{j_{1}j_{2}j'j'}^{(\tau)}$ 
(proportional to $\delta(j_{1}\pm j_{2})$) are absent in $\hat{S}_{j_{1}j_{2}j}^{(\tau)}$. Since 
{\it the moding of the composite operator index can take zero values for both} $NBC$ and $DBC$, the construction of the $\Gamma^{(2,1)}$ 
graphs with the tensor $\hat{S}_{j_{1}j_{2}j}^{(\tau)}$ needs to be modified. In fact, the right combination to form an arbitrary loop graph 
should include the $j=0$ and $j\neq0$ pieces components from $\hat{S}_{j_{1}j_{2}j}^{(\tau)}$. To be precise, an arbitrary graph always 
contains the contributions proportional to $\hat{S}_{j_{1}j_{2}j}^{(\tau)} + \tau \hat{S}_{j_{1}j_{2}0}^{(\tau)}$ ($j=\pm (j_{1}+j_{2})$) 
contracted with $\tilde{S}_{j_{1}j_{2}j_{3}j_{4}}^{(\tau)}$ tensor stemming from the perturbative expansion in the coupling constant. In terms 
of the diagram the combination that should appear is 
\begin{eqnarray}\label{38}
&& \parbox{10mm}{\includegraphics[scale=0.05]{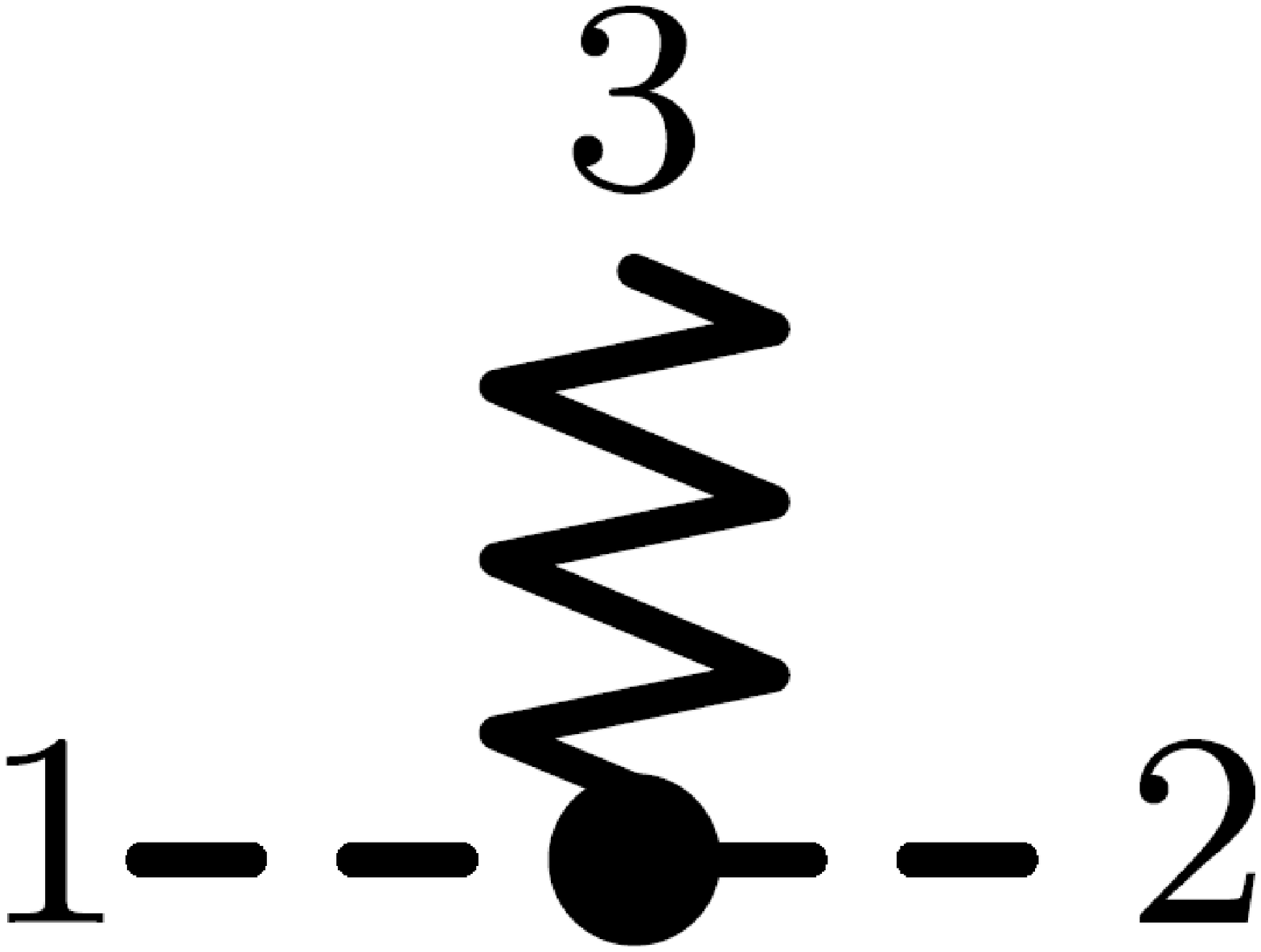}}\qquad\;= \parbox{10mm}{\includegraphics[scale=0.05]{fig27DN.eps}}|_
{(j=\pm(j_{1}+j_{2}))} + \tau \parbox{10mm}{\includegraphics[scale=0.05]{fig27DN.eps}}|_{(j=0)},
\end{eqnarray}
where the sign of $j$ is fixed. With this recipe utilized in the construction of each loop diagram, the renormalization of the composite 
field can be performed in a simple manner, as we are going to discuss later on.         
\par Let us focus now on the zero modes in Neumann boundary conditions whenever $j_{1}=0 (j_{2}\neq 0)$, $j_{2}=0 (j_{1} \neq 0)$ and 
$j_{1}=j_{2}=0$ present in the tensor $\hat{S}_{j_{1}j_{2}j}^{(\tau)}$. First, note that the zero mode composite basis functions are the same, 
namely, $\tilde{u}^{(1)}= cos(\tilde{\sigma}jz)$ and the product involving a zero mode becomes
$\tilde{u}_{0,j}^{(1)}(z) = u_{0}^{(1)} u_{j}^{(1)} = \frac{\sqrt{2}}{L}cos(\tilde{\sigma}jz)$. From the definition 
$\hat{S}_{0j_{2}j}^{(1)}= \int_{0}^{L} u_{0}^{(1)}(z)u_{j_{2}}^{(1)}(z)\tilde{u}_{j}^{(1)}(z) dz$ we read off the values 
$\hat{S}_{0j_{2}j}^{(1)}= \hat{S}_{j_{2}0j}^{(1)}= \frac{1}{\sqrt{2}}[\delta(j+j_{2}) + \delta(j-j_{2})]$. Analogously, it is not difficult to prove that $\hat{S}_{00j}^{(1)}=\delta(j)$.  
\par We can employ the framework just developed to compute arbitrary loop diagrams including 
multiplicatively renormalizable composite vertex operators. At the present moment we shall restrict ourselves in getting the expression 
corresponding to the one-loop graph for $\Gamma^{(2,1)}$. It is represented by
\begin{eqnarray}\label{39}
&& \parbox{13mm}{\includegraphics[scale=0.5]{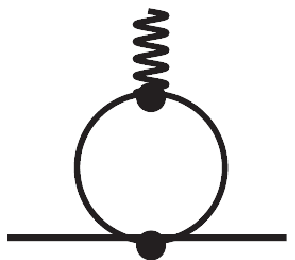}}\quad = \frac{(N+2)}{6} \overset{\infty}{\underset{j_{1},j_{2} \geq 0}{\sum}}\tilde{S}^{(\tau)}_{i_{1} i_{2} j_{1} j_2} \hat{S}^{(\tau)}_{j_{1} j_{2} j} \int d^{d-1}q G_{0}(q+k,j_{1}) G_{0}(q,j_{2}),
\end{eqnarray} 
Keep in mind that we wish to express this diagram into a unified form. Therefore, we choose $i_{1},{i_2}\neq 0$. Let us start 
explicitly with Neumann boundary conditions. The summations split in the form:
\begin{eqnarray}\label{40}
&& \overset{\infty}{\underset{j_{1},j_{2} \geq 0}{\sum}}\tilde{S}^{(1)}_{i_{1} i_{2} j_{1} j_2} \hat{S}^{(1)}_{j_{1} j_{2} j} 
\int d^{d-1}q G_{0}(q+k,j_{1}) G_{0}(q,j_{2})= \int  d^{d-1}q \Bigl[\tilde{S}^{(1)}_{i_{1} i_{2} 00} \hat{S}^{(1)}_{00 j} G_{0}(q+k,0) G_{0}(q,0) \nonumber\\
&& + \overset{\infty}{\underset{j_{1}=1}{\sum}}\tilde{S}^{(1)}_{i_{1} i_{2} j_{1} 0} \hat{S}^{(1)}_{j_{1} 0 j} G_{0}(q+k,j_{1}) 
G_{0}(q,0) + \overset{\infty}{\underset{j_{2}=1}{\sum}}\tilde{S}^{(1)}_{i_{1} i_{2} 0 j_2} \hat{S}^{(1)}_{0j_{2} j} G_{0}(q+k,0) 
G_{0}(q,j_{2}) + \overset{\infty}{\underset{j_{1},j_{2}=1}{\sum}}\tilde{S}^{(1)}_{i_{1} i_{2} j_{1} j_2} \nonumber\\
&& \;\; \times \hat{S}^{(1)}_{j_{1} j_{2} j} 
G_{0}(q+k,j_{1}) G_{0}(q,j_{2})\Bigr].
\end{eqnarray}
Using explicitly the tensor components we first obtain ($i_{\pm}\equiv i_{1} \pm i_{2}$):
\begin{eqnarray}\label{41}
&& \overset{\infty}{\underset{j_{1},j_{2} \geq 0}{\sum}}\tilde{S}^{(1)}_{i_{1} i_{2} j_{1} j_2} \hat{S}^{(1)}_{j_{1} j_{2} j} 
\int d^{d-1}q G_{0}(q+k,j_{1}) G_{0}(q,j_{2})= \frac{\tilde{\sigma}}{2}\int  d^{d-1}q \Bigl[2 \overset{\infty}{\underset{j_{1}=1}{\sum}}
[\delta(j-j_{1}) + \delta(j+j_{1})]\nonumber\\
&& [\delta(i_{+} + j_{1}) + \delta(i_{+} - j_{1}) + \delta(i_{-} + j_{1}) + \delta(i_{-} - j_{1})] 
G_{0}(q+k,j_{1}) G_{0}(q,0) 
+ 2 \overset{\infty}{\underset{j_{2}=1}{\sum}}
[\delta(j-j_{2}) \nonumber\\
&& + \delta(j+j_{2})][\delta(i_{+} + j_{2}) + \delta(i_{+} - j_{2}) + \delta(i_{-} + j_{2}) + \delta(i_{-} - j_{2})]
G_{0}(q+k,0) G_{0}(q,j_{2}) \nonumber\\
&& + \overset{\infty}{\underset{j_{1},j_{2}=1}{\sum}}[\delta(i_{+} + j_{1} + j_{2}) + \delta(i_{+} - j_{1}+ j_{2}) 
+ \delta(i_{+} + j_{1} - j_{2}) + \delta(i_{+} - j_{1} -j_{2}) + \delta(i_{-} + j_{1} \nonumber\\
&& + j_{2}) + \delta(i_{-} - j_{1}+ j_{2}) 
+ \delta(i_{-} + j_{1} - j_{2}) + \delta(i_{-} - j_{1} -j_{2})][\delta(j + j_{1} + j_{2}) + \delta(j - j_{1}+ j_{2}) \nonumber\\
&& + \delta(j + j_{1} - j_{2}) + \delta(j - j_{1} -j_{2})] 
G_{0}(q+k,j_{1}) G_{0}(q,j_{2}) + 4 \delta(j)[\delta(i_{+}) + \delta(i_{-})]G_{0}(q,0)\nonumber\\
&& \times \;\;G_{0}(q+k,0)]\Bigr].
\end{eqnarray}
In the first term of the first (second) single summation we perform the change $j_{1} \rightarrow -j_{1}$ ($j_{2} \rightarrow -j_{2}$). 
This extends the range of the summation to the interval ($-\infty, \infty$) but precluding the zero mode. We can include it in order to 
write the first summation as: 
\begin{eqnarray}\label{42}
&& \overset{\infty}{\underset{j_{1}=1}{\sum}}
[\delta(j-j_{1}) + \delta(j+j_{1})][\delta(i_{+} + j_{1}) + \delta(i_{+} - j_{1}) + \delta(i_{-} + j_{1}) + \delta(i_{-} - j_{1})] 
G_{0}(q+k,j_{1}) \nonumber\\
&& G_{0}(q,0) = \overset{\infty}{\underset{j_{1}=-\infty}{\sum}}
\delta(j+j_{1})[\delta(i_{+} + j_{1}) + \delta(i_{+} - j_{1}) + \delta(i_{-} + j_{1}) + \delta(i_{-} - j_{1})] 
G_{0}(q+k,j_{1}) \nonumber\\
&& \;\; \times G_{0}(q,0) - 2 \delta(j) [\delta(i_{+}) + \delta(i_{-})]G_{0}(q+k,0)G_{0}(q,0),
\end{eqnarray}
and similarly for the second single summation. The double summation can be manipulated using analogous steps and 
leads to
\begin{eqnarray}\label{43}
&& \overset{\infty}{\underset{j_{1},j_{2}=1}{\sum}}[\delta(i_{+}+j_{1}+j_{2}) + \delta(i_{+}-j_{1}+j_{2}) 
+ \delta(i_{+}+j_{1}-j_{2}) + \delta(i_{+}-j_{1}-j_{2}) + \delta(i_{-}+j_{1}+j_{2})\nonumber\\
&&  + \delta(i_{-}-j_{1}+ j_{2}) + \delta(i_{-}+j_{1}-j_{2}) + \delta(i_{-}-j_{1}-j_{2})][\delta(j+j_{1}+j_{2}) + \delta(j-j_{1}+j_{2}) + \delta(j \nonumber\\
&& + j_{1}-j_{2}) + \delta(j-j_{1}-j_{2})] 
G_{0}(q+k,j_{1}) G_{0}(q,j_{2}) = \overset{\infty}{\underset{j_{1},j_{2}=-\infty}{\sum}}[\delta(i_{+}+j_{1}+j_{2}) + \delta(i_{+}-j_{1}+j_{2})
\nonumber\\
&& + \delta(i_{+}+j_{1}-j_{2}) + \delta(i_{+}-j_{1}-j_{2}) + \delta(i_{-}+j_{1}+j_{2}) + \delta(i_{-}-j_{1}+j_{2}) 
+ \delta(i_{-}+j_{1}-j_{2})\nonumber\\
&& + \delta(i_{-}-j_{1}-j_{2})]\delta(j+j_{1}+j_{2}) 
G_{0}(q+k,j_{1}) G_{0}(q,j_{2}) - 2 \overset{\infty}{\underset{j_{1}=-\infty}{\sum}}
\delta(j+j_{1})[\delta(i_{+}+j_{1}) + \delta(i_{+}\nonumber\\
&& -j_{1}) + \delta(i_{-}+j_{1}) + \delta(i_{-}-j_{1})] 
G_{0}(q+k,j_{1}) G_{0}(q,0) - 2 \overset{\infty}{\underset{j_{2}=-\infty}{\sum}}
\delta(j+j_{2})[\delta(i_{+}+j_{2}) + \delta(i_{+}\nonumber\\
&& -j_{2}) + \delta(i_{-}+j_{2}) + \delta(i_{-}-j_{2})]G_{0}(q+k,0) G_{0}(q,j_{2}) + 4 \delta(j)[\delta(i_{+}) + \delta(i_{-})]G_{0}(q+k,0) 
\nonumber\\
&& \times G_{0}(q,0).
\end{eqnarray} 
Replacing the value of each expression inside the combination of the diagram, the single summations and the independent terms cancel among 
each other and we obtain:
\begin{eqnarray}\label{44}
&& \parbox{13mm}{\includegraphics[scale=0.5]{fig3DN.eps}}\quad = \frac{(N+2)}{6} \frac{\tilde{\sigma}}{2} 
\overset{\infty}{\underset{j_{1},j_{2}=-\infty}{\sum}} \delta(j+j_{1}+j_{2})[\delta(i_{+}+j_{1}+j_{2}) + \delta(i_{+}-j_{1}+j_{2})+ \delta(i_{+}+j_{1}\nonumber\\
&& -j_{2}) + \delta(i_{+}-j_{1}-j_{2}) + \delta(i_{-}+j_{1}+j_{2}) + \delta(i_{-}-j_{1}+j_{2}) + \delta(i_{-}+j_{1}-j_{2}) 
+ \delta(i_{-}-j_{1}\nonumber\\
&& -j_{2})] \int d^{d-1}q G_{0}(q+k,j_{1})G_{0}(q,j_{2}).
\end{eqnarray}
Recall that we wish to express the one-loop diagram involving at maximum a single summation. Performing the summations explictly, 
we find 
\begin{eqnarray}\label{45}
&& \parbox{13mm}{\includegraphics[scale=0.5]{fig3DN.eps}}\quad = \frac{(N+2)}{6} \frac{1}{2}\Bigl[(\delta(i_{+}+j) 
+ \delta(i_{+}-j) + \delta(i_{-}+j) + \delta(i_{-}-j))I_{2}(k,j,\tilde{\sigma},\mu_{0}) \nonumber\\
&& + \tilde{I}_{2}(k,\frac{i_{+}-j}{2},\frac{i_{+}+j}{2},\tilde{\sigma},\mu_{0}) 
+ \tilde{I}_{2}(k,\frac{i_{+}+j}{2},\frac{i_{+}-j}{2},\tilde{\sigma},\mu_{0}) 
+ \tilde{I}_{2}(k,\frac{i_{-}-j}{2},\frac{i_{-}+j}{2},\tilde{\sigma},\mu_{0})\nonumber\\
&&  + \tilde{I}_{2}(k,\frac{i_{-}+j}{2},\frac{i_{-}-j}{2},\tilde{\sigma},\mu_{0})\Bigr],
\end{eqnarray}  
where the integrals $I_{2}((k,i,\tilde{\sigma},\mu_{0})$ and $\tilde{I}_{2}(k,i,j,\tilde{\sigma},\mu_{0})$ are defined in Eqs. (\ref{26}).
\par For $DBC$ the situation is even simpler since we do not have to deal with zero modes. We can write
\begin{eqnarray}\label{46}
&& \parbox{13mm}{\includegraphics[scale=0.5]{fig3DN.eps}} = \frac{(N+2)}{6} \overset{\infty}{\underset{j_{1},j_{2}=1}{\sum}}\tilde{S}^{(-1)}_{i_{1} i_{2} j_{1} j_2} \hat{S}^{(-1)}_{j_{1} j_{2} j} \int d^{d-1}q G_{0}(q+k,j_{1}) G_{0}(q,j_{2}).
\end{eqnarray} 
Using the tensor components already derived, we can write the summation as
\begin{eqnarray}\label{47}
&& \overset{\infty}{\underset{j_{1},j_{2}=1}{\sum}}\tilde{S}^{(1)}_{i_{1} i_{2} j_{1} j_2} \hat{S}^{(1)}_{j_{1} j_{2} j} \int d^{d-1}q G_{0}(q+k,j_{1}) G_{0}(q,j_{2})= \overset{\infty}{\underset{j_{1},j_{2} =1}{\sum}}[\delta(i_{+}+j_{1}+j_{2}) + \delta(i_{+}-j_{1}
\nonumber\\
&& - j_{2}) - \delta(i_{-}+j_{1}+j_{2}) 
- \delta(i_{-}-j_{1}-j_{2}) - \delta(i_{+}-j_{1}+j_{2}) - \delta(i_{+}+j_{1}-j_{2}) + \delta(i_{-}-j_{1}\nonumber\\
&& +j_{2}) + \delta(i_{-}+j_{1}-j_{2})][\delta(j-j_{1}+j_{2}) + \delta(j+j_{1}-j_{2}) - \delta(j-j_{1}-j_{2}) - \delta(j+j_{1}+j_{2})]\nonumber\\
&& \;\;\times\;\; \int d^{d-1}q G_{0}(q+k,j_{1}) G_{0}(q,j_{2}).
\end{eqnarray} 
Now, in the second bracket perform the change $j_{2} \rightarrow -j_{2}$ in the first and fourth terms. This has the effect to produce a global 
factor of $(-1)$ in each coefficient of those two terms. The net result can be written in the form
\begin{eqnarray}\label{48}
&& \overset{\infty}{\underset{j_{1},j_{2}=1}{\sum}}\tilde{S}^{(1)}_{i_{1} i_{2} j_{1} j_2} \hat{S}^{(1)}_{j_{1} j_{2} j} \int d^{d-1}q G_{0}(q+k,j_{1}) G_{0}(q,j_{2})=-\overset{\infty}{\underset{j_{1}=1}{\sum}}\overset{\infty}{\underset{j_{2}=-\infty, j_{2}\neq 0}{\sum}}\delta(j+j_{1}+j_{2})[\delta(i_{+}+j_{1}\nonumber\\
&& +j_{2}) + \delta(i_{+}-j_{1} - j_{2}) - \delta(i_{-}+j_{1}+j_{2}) 
- \delta(i_{-}-j_{1}-j_{2}) - \delta(i_{+}-j_{1}+j_{2}) - \delta(i_{+}+j_{1}-j_{2}) \nonumber\\
&& + \delta(i_{-}-j_{1}+j_{2}) + \delta(i_{-}+j_{1}-j_{2})]\int d^{d-1}q G_{0}(q+k,j_{1}) G_{0}(q,j_{2}).
\end{eqnarray}   
By performing similar transformations in some terms involving $j_{1} (\rightarrow -j_{1})$, we can extend the summation over 
$j_{1}$ for negative values as well. The simplicity here is that no zero mode survives when we include the values $j_{1},j_{2}=0$ 
in the summations. The graph then reads:
\begin{eqnarray}\label{49}
&& \parbox{13mm}{\includegraphics[scale=0.5]{fig3DN.eps}} = - \frac{(N+2)}{6} \frac{\tilde{\sigma}}{2}
\overset{\infty}{\underset{j_{1},j_{2}=-\infty}{\sum}} \delta(j+j_{1}+j_{2})[\delta(i_{+}+j_{1}+j_{2}) - \delta(i_{+}-j_{1}+j_{2}) - \delta(i_{+}+j_{1}\nonumber\\
&& -j_{2}) + \delta(i_{+}-j_{1}-j_{2}) - \delta(i_{-}+j_{1}+j_{2}) + \delta(i_{-}-j_{1}+j_{2}) + \delta(i_{-}+j_{1}-j_{2}) 
- \delta(i_{-}-j_{1}\nonumber\\
&& -j_{2})] \int d^{d-1}q G_{0}(q+k,j_{1})G_{0}(q,j_{2}).
\end{eqnarray}  
Comparing this with Eq. (\ref{44}), $DBC$ ($\tau=-1$) and $NBC$ ($\tau=1$) can be unified very easily. When this diagram 
is written in terms of the integrals $I_{2}(k,i,\tilde{\sigma},\mu_{0})$ and $\tilde{I}_{2}(k,i,\tilde{\sigma},\mu_{0})$, the 
unified result is
\begin{eqnarray}\label{50}
&& \parbox{13mm}{\includegraphics[scale=0.5]{fig3DN.eps}} = \frac{(N+2)}{6} \frac{\tau}{2}\Bigl[(\delta(i_{+}+j) 
+ \delta(i_{+}-j) + \tau (\delta(i_{-}+j) + \delta(i_{-}-j)))I_{2}(k,j,\tilde{\sigma},\mu_{0}) \nonumber\\
&& + \tau[\tilde{I}_{2}(k,\frac{i_{+}-j}{2},\frac{i_{+}+j}{2},\tilde{\sigma},\mu_{0}) 
+ \tilde{I}_{2}(k,\frac{i_{+}+j}{2},\frac{i_{+}-j}{2},\tilde{\sigma},\mu_{0})] 
+ \tilde{I}_{2}(k,\frac{i_{-}-j}{2},\frac{i_{-}+j}{2},\tilde{\sigma},\mu_{0})\nonumber\\
&&  + \tilde{I}_{2}(k,\frac{i_{-}+j}{2},\frac{i_{-}-j}{2},\tilde{\sigma},\mu_{0})\Bigr].
\end{eqnarray} 
This can be rewritten in a more elegant, compact notation as:
\begin{eqnarray}\label{51}
&& \parbox{13mm}{\includegraphics[scale=0.5]{fig3DN.eps}} = \frac{(N+2)}{6} \frac{\tau}{2}
\overset{1}{\underset{\alpha,\beta=-1}{\sum}} \alpha^{\frac{1-\tau}{2}}\Bigl[\delta(i_{\alpha} - \beta j)I_{2}(k,i_{\alpha},\tilde{\sigma},\mu_{0}) 
+ \tau \tilde{I}_{2}(k,\frac{i_{\alpha}+\beta j}{2},\frac{i_{\alpha}-\beta j}{2},\nonumber\\
&& \;\;\tilde{\sigma},\mu_{0})\Bigr]. 
\end{eqnarray} 
\par This framework has the virtue of expressing all integrals in terms of infinite sums which makes it simple the comparison 
with the results from periodic and antiperiodic boundary conditions. Needless to say, all the massless integrals follow from the 
substitution $\mu_{0}=0$ in the integrals above and in the higher-loop contributions, which will be analyzed next.

\subsubsection{Higher loop diagrams for $\Gamma^{(2)}$, $\Gamma^{(4)}$ and $\Gamma^{(2,1)}$}    
\par Our goal now is just to write down the various 
graphs in terms of integrals which resemble those from periodic and antiperiodic boundary conditions, possibly with additional 
``nondiagonal'' terms. We are going to consider some two-loop graphs, although just a smaller subset of them will be necessary to our 
computation of the critical exponents within the present technique. We will analyze only one three-loop graph for the two-point function 
which is the only one needed for our purposes. We will save the construction of other nontrivial diagrams (external momentum dependent) 
including mass insertions for later when we will discuss the renormalization of this theory. 
\par Start with the two point function $\Gamma^{(2)}$, for instance. 
We can classify the diagrams in trivial and nontrivial contributions. The trivial contributions will be generically called 
``tadpole diagrams'' and are going to be discussed first. These contributions are displayed in the two-loop case for the sake of 
comparison with the one-loop case but we will not need them in the present computation.   
\par Of course, we can start from scratch with two four-index finite-size tensors and perform the apropriate contractions 
between them in pretty much the same way we did in the one-loop case. The result including both boundary conditions is:
\begin{eqnarray}\label{52}
&& \parbox{10mm}{\includegraphics[scale=0.32]{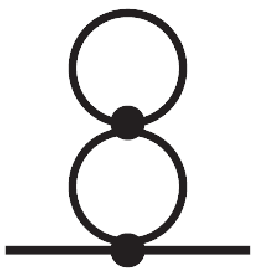}} = \left(\frac{N+2}{3}\right)^{2} \Bigl[\delta(i_{-})
D_{2}(\tilde{\sigma},\mu_{0})D_{1}(\tilde{\sigma},\mu_{0}) + \tau \tilde{I}_{2}(0,\frac{i_{-}}{2},\frac{i_{-}}{2},\tilde{\sigma},\mu_{0})
D_{1}(\tilde{\sigma},\mu_{0}) + \tau I_{2}(0,i_{-},\nonumber\\
&& \tilde{\sigma},\mu_{0})\tilde{D}_{1}(\frac{i_{-}}{2},\tilde{\sigma},\mu_{0})
+ \tau \delta(i_{+})D_{2}(\tilde{\sigma},\mu_{0})D_{1}(\tilde{\sigma},\mu_{0}) 
+ \tilde{I}_{2}(0,\frac{i_{+}}{2},\frac{i_{+}}{2},\tilde{\sigma},\mu_{0})D_{1}(\tilde{\sigma},\mu_{0})
+ I_{2}(0,i_{+},\nonumber\\
&& \tilde{\sigma},\mu_{0}) \tilde{D}_{1}(\frac{i_{+}}{2},\tilde{\sigma},\mu_{0})
+ \overset{\infty}{\underset{l=-\infty}{\sum}}\Bigr[\tilde{I}_{2}(0,\frac{i_{-}}{2} + l, \frac{i_{-}}{2} - l,\tilde{\sigma},\mu_{0})
\tilde{D}_{1}(l,\tilde{\sigma},\mu_{0}) + \tau \tilde{I}_{2}(0,\frac{i_{+}}{2} + l, \frac{i_{+}}{2} - l,\tilde{\sigma},\nonumber\\
&& \mu_{0})\tilde{D}_{1}(l,\tilde{\sigma},\mu_{0})\Bigr]\Bigr],
\end{eqnarray}
where
\begin{eqnarray}\label{53}
&& D_{2}(\tilde{\sigma},\mu_{0})= \tilde{\sigma}\overset{\infty}{\underset{l=-\infty}{\sum}} \int \frac{d^{d-1}q}
{[q^{2}+\tilde{\sigma}^{2}l^{2}+\mu_{0}^{2}]^{2}}.
\end{eqnarray}
\par The nontrivial two-loop graph of the two-point function can be 
obtained most easily by taking the four-point graph, making the identification $i_{1}=i_{3}=j$, introducing a new propagator with momentum 
$q_{2}$, mode $j$, performing a summation over $j$ and an integrating over $d^{d-1}q_{2}$. We find:
\begin{eqnarray}\label{54}
&& \parbox{10mm}{\includegraphics[scale=0.32]{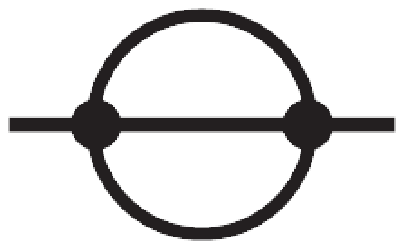}}  \quad = \left(\frac{N+2}{3}\right)\Bigl[(\delta(i_{-})
+\tau \delta(i_{+}))I_{3}(k,i_{1},\tilde{\sigma},\mu_{0}) + 3 \tilde{I}_{3}(k,\frac{i_{+}}{2},\frac{i_{-}}{2},\tilde{\sigma},\mu_{0})
+ 3 \tau  \tilde{I}_{3}(k,\frac{i_{-}}{2},\frac{i_{+}}{2},\nonumber\\
&& \tilde{\sigma},\mu_{0}) \Bigr],
\end{eqnarray}
where,
\begin{subequations}\label{55}
\begin{eqnarray}
&& I_{3}(k,j,\tilde{\sigma},\mu_{0}) = \tilde{\sigma}^{2} \overset{\infty}{\underset{j_{1},j_{2}=-\infty}{\sum}} 
\int \frac{d^{d-1}q_{1} d^{d-1}q_{2}}
{[q_{2}^{2}+\tilde{\sigma}^{2}j_{2}^{2}+\mu_{0}^{2}][(q_{1}+q_{2}+k)^{2}
+\tilde{\sigma}^{2}(j_{1}+j_{2}+j)^{2}+\mu_{0}^{2}]}\nonumber\\
&& \;\;\;\quad \times \frac{1}{[q_{1}^{2}+\tilde{\sigma}^{2}j_{1}^{2}+\mu_{0}^{2}]},\\
&& \tilde{I}_{3}(k,i,j,\tilde{\sigma},\mu_{0})= \tilde{\sigma}^{2} \overset{\infty}{\underset{l=-\infty}{\sum}} 
\int \frac{d^{d-1}q_{1} d^{d-1}q_{2}}
{[q_{1}^{2}+\tilde{\sigma}^{2}l^{2}+\mu_{0}^{2}][q_{2}^{2}+\tilde{\sigma}^{2}i^{2}+\mu_{0}^{2}]}\nonumber\\
&& \qquad\qquad\qquad \times \qquad \frac{1}{[(q_{1}+q_{2}+k)^{2}
+\tilde{\sigma}^{2}(l+j)^{2}+\mu_{0}^{2}]}.
\end{eqnarray}
\end{subequations}
\par The nontrivial three-loop graph contributing to the two-point function can be determined similarly. The complete result is:
\begin{eqnarray}\label{56}
&& \parbox{10mm}{\includegraphics[scale=0.32]{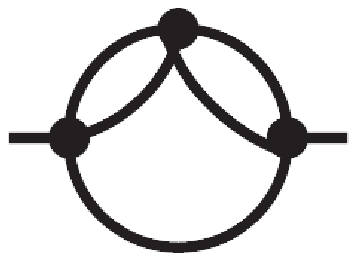}}  \quad =  \frac{(N+2)(N+8)}{27}
\Bigl[(\delta(i_{-})+ \tau \delta(i_{+}))I_{5}(k,i_{1},\tilde{\sigma},\mu_{0}) 
+ \tilde{I}_{5}(k,\frac{i_{+}}{2},\frac{i_{-}}{2},\tilde{\sigma},\mu_{0}) + \tau \tilde{I}_{5}(k,\frac{i_{-}}{2},\nonumber\\
&& \frac{i_{+}}{2},\tilde{\sigma},\mu_{0})
+  \hat{I}_{5}(k,\frac{i_{+}}{2},\frac{i_{-}}{2},i_{2},\tilde{\sigma},\mu_{0})
+  \tau \hat{I}_{5}(k,\frac{i_{-}}{2},\frac{i_{+}}{2},i_{2},\tilde{\sigma},\mu_{0}) 
+ 2 \hat{I}_{5}(k,\frac{i_{+}}{2},\frac{i_{-}}{2},i_{1},\tilde{\sigma},\mu_{0}) + 2 \tau \hat{I}_{5}\nonumber\\
&& (k,\frac{i_{-}}{2},\frac{i_{+}}{2},i_{1},\tilde{\sigma},\mu_{0})
+ \check{I}_{5}(k,i_{1},\frac{i_{-}}{2},\frac{i_{+}}{2},\tilde{\sigma},\mu_{0}) 
+  \tau \check{I}_{5}(k,i_{1},\frac{i_{+}}{2},\frac{i_{-}}{2},\tilde{\sigma},\mu_{0}) 
+ \bar{I}_{5}(k,i_{1},\frac{i_{+}}{2},\frac{i_{-}}{2},\tilde{\sigma},\mu_{0})\nonumber\\ 
&& + \tau \bar{I}_{5}(k,i_{1},\frac{i_{-}}{2},\frac{i_{+}}{2},\tilde{\sigma},\mu_{0}) 
+ \dot{I}_{5}(k,\frac{i_{+}}{2},\frac{i_{-}}{2},i_{1},\tilde{\sigma},\mu_{0}) 
+ \tau \dot{I}_{5}(k,\frac{i_{-}}{2},\frac{i_{+}}{2},-i_{2},\tilde{\sigma},\mu_{0}) \Bigr]
\end{eqnarray}
where the integrals above are defined by (the arguments ($k,l,m,n,\tilde{\sigma},\mu_{0}$) were suppressed in the integrals with 
three external quasi-momenta indices for sake of simplicity)
\begin{subequations}\label{57}
\begin{eqnarray}
&& I_{5}(k,i,\tilde{\sigma},\mu_{0})= \tilde{\sigma}^{3} \overset{\infty}{\underset{j_{1},j_{2},j_{3}=-\infty}{\sum}} 
\int \frac{d^{d-1}q_{1} d^{d-1}q_{2}d^{d-1}q_{3}}
{[q_{1}^{2}+\tilde{\sigma}^{2}j_{1}^{2}+\mu_{0}^{2}][q_{2}^{2}+\tilde{\sigma}^{2}j_{2}^{2}+\mu_{0}^{2}][q_{3}^{2}+\tilde{\sigma}^{2}j_{3}^{2}+\mu_{0}^{2}]}\nonumber\\
&& \times \frac{1}{[(q_{1}+q_{2}+k)^{2}
+\tilde{\sigma}^{2}(j_{1}+j_{2}+i)^{2}+\mu_{0}^{2}][(q_{1}+q_{3}+k)^{2}
+\tilde{\sigma}^{2}(j_{1}+j_{3}+i)^{2}+\mu_{0}^{2}]},\\
&& \tilde{I}_{5}(k,i,j,\tilde{\sigma},\mu_{0})= \tilde{\sigma}^{3} \overset{\infty}{\underset{j_{1},j_{2}=-\infty}{\sum}} 
\int \frac{d^{d-1}q_{1} d^{d-1}q_{2}d^{d-1}q_{3}}
{[q_{1}^{2}+\tilde{\sigma}^{2}i^{2}+\mu_{0}^{2}][q_{2}^{2}+\tilde{\sigma}^{2}j_{1}^{2}+\mu_{0}^{2}][q_{3}^{2}+\tilde{\sigma}^{2}j_{2}^{2}+\mu_{0}^{2}]}\nonumber\\
&& \times \frac{1}{[(q_{1}+q_{2}+k)^{2}
+\tilde{\sigma}^{2}(j_{1}+j)^{2}+\mu_{0}^{2}][(q_{1}+q_{3}+k)^{2}
+\tilde{\sigma}^{2}(j_{2}+j)^{2}+\mu_{0}^{2}]},\\
&& \hat{I}_{5}=\tilde{\sigma} \overset{\infty}{\underset{j=-\infty}{\sum}} 
\int \frac{d^{d-1}q \tilde{I}_{2}(q+k,l,j+m) I_{2}(q+k,j+n)}{q^{2} + \sigma^{2}j^{2} + \mu_{0}^{2}},\\
&& \check{I}_{5}=\tilde{\sigma} \overset{\infty}{\underset{j_{1},j_{2}=-\infty}{\sum}} 
\int \frac{d^{d-1}q \tilde{I}_{2}(q+k,j_{2},j_{2}+j_{1}+l)\tilde{I}_{2}(q+k,j_{2}+m,j_{2}+j_{1}+n)}{q^{2} + \sigma^{2}j_{1}^{2} + \mu_{0}^{2}},\\
&& \bar{I}_{5}= \tilde{\sigma} \overset{\infty}{\underset{j_{1},j_{2}=-\infty}{\sum}} 
\int \frac{d^{d-1}q \tilde{I}_{2}(q+k,j_{2},j_{2}+j_{1}+l)\tilde{I}_{2}(q+k,j_{1}+j_{2}+m,j_{2}+n)}{q^{2} + \sigma^{2}j_{1}^{2} + \mu_{0}^{2}},\\
&& \dot{I}_{5}= \tilde{\sigma} \overset{\infty}{\underset{j_{1},j_{2}=-\infty}{\sum}} 
\int \frac{d^{d-1}q \tilde{I}_{2}(q+k,j_{1}+l,m)\tilde{I}_{2}(q+k,j_{2},2j_{2}+j_{1}+n)}{q^{2} + \sigma^{2}j_{1}^{2} + \mu_{0}^{2}}.
\end{eqnarray}
\end{subequations}
\par Let us turn our attention to the four-point vertex part. The simplest contribution at two-loops is the double bubble. Using the 
fusion of two one-loop diagrams of the four-point function results in the following unified expression 
$(I_{2}(k,i,\tilde{\sigma},\mu_{0})\equiv I_{2}(k,i), \tilde{I}_{2}(k,i,j,\tilde{\sigma},\mu_{0}) \equiv \tilde{I}_{2}(k,i,j)$)
\begin{eqnarray}\label{58}
&& \parbox{10mm}{\includegraphics[scale=0.3]{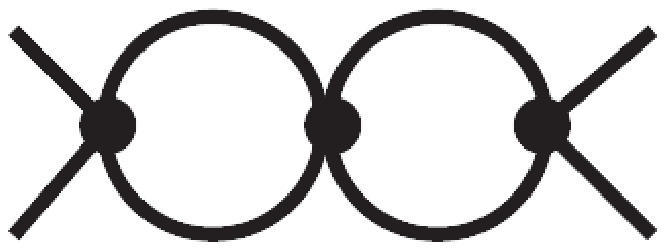}}\qquad =  \frac{(N^{2}+6N+20)}{27} \tilde{\sigma}
\Bigl\{[\delta(i_{+}+\tilde{i}_{+}) + \tau \delta(i_{-}+\tilde{i}_{+}) + \delta(i_{+}-\tilde{i}_{+}) + \tau \delta(i_{-}-\tilde{i}_{+})]
I_{2}^{2}(k,\tilde{i}_{+})\nonumber\\
&& +[\tau \delta(i_{+}+\tilde{i}_{-}) + \delta(i_{-}+\tilde{i}_{-}) + \tau \delta(i_{+}-\tilde{i}_{-}) 
+ \delta(i_{-}-\tilde{i}_{-})]I_{2}^{2}(k,\tilde{i}_{-})+2\tau(I_{2}(k,\tilde{i}_{+}) + I_{2}(k,i_{+}))\nonumber\\
&& \;\;\times\;\; \tilde{I}_{2}(k,\frac{i_{+}-\tilde{i}_{+}}{2},\frac{i_{+}+\tilde{i}_{+}}{2})+2\tau(I_{2}(k,\tilde{i}_{-}) + I_{2}(k,i_{-}))
\tilde{I}_{2}(k,\frac{i_{-}-\tilde{i}_{-}}{2},\frac{i_{-}+\tilde{i}_{-}}{2})+2(I_{2}(k,\tilde{i}_{-}) \nonumber\\
&& \;\;+\;\; I_{2}(k,i_{+}))\tilde{I}_{2}(k,\frac{i_{+}-\tilde{i}_{-}}{2},\frac{i_{+}+\tilde{i}_{-}}{2})+2(I_{2}(k,\tilde{i}_{+}) + I_{2}(k,i_{-}))
\tilde{I}_{2}(k,\frac{i_{-}-\tilde{i}_{+}}{2},\frac{i_{-}+\tilde{i}_{+}}{2}) \nonumber\\
&& +2\overset{\infty}{\underset{j=-\infty}{\sum}}
\Bigl(\tilde{I}_{2}(k,j,j+i_{+})\Bigr[\tilde{I}_{2}(k,j+\frac{i_{+}-\tilde{i}_{+}}{2},j+\frac{i_{+}+\tilde{i}_{+}}{2}) 
+ \tau \tilde{I}_{2}(k,j+\frac{i_{+}-\tilde{i}_{-}}{2},j+\frac{i_{+}+\tilde{i}_{-}}{2})\Bigr]\nonumber\\
&&  + \tilde{I}_{2}(k,j,j+i_{-})\Bigl[\tilde{I}_{2}(k,j+\frac{i_{-}-\tilde{i}_{-}}{2},j+\frac{i_{-}+\tilde{i}_{-}}{2}) 
+ \tau \tilde{I}_{2}(k,j+\frac{i_{-}-\tilde{i}_{+}}{2},j+\frac{i_{-}+\tilde{i}_{+}}{2})\Bigr]\Bigr) \Bigr\}.
\end{eqnarray}
The important point which simplifies our task is to realize that the last terms like, for instance, 
$\overset{\infty}{\underset{j=-\infty}{\sum}}\tilde{I}_{2}(k,j,j+i_{+})\tilde{I}_{2}(k,j+\frac{i_{+}-\tilde{i}_{+}}{2},j+\frac{i_{+}+\tilde{i}_{+}}{2})$ are $O(\epsilon^{0})$ and do not contribute to the singular part of this diagram. These terms can 
be ignored. 
\par Consider the evaluation of the nontrivial two-loop contribution of the four-point 
vertex function. Following the same line of thought we obtain ($P=k_{1} + k_{2}$)
\begin{eqnarray}\label{59}
&& \parbox{10mm}{\includegraphics[scale=0.32]{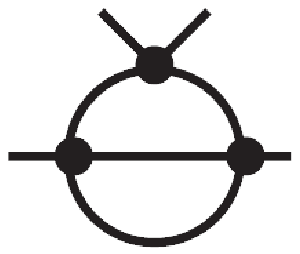}}\quad =\frac{(5N+22)}{27} \tilde{\sigma} 
\Bigl\{[\delta(i_{+}-\tilde{i}_{+}) + \tau \delta(i_{+}-\tilde{i}_{-})]I_{4}(P,k_{3},i_{+},i_{3}) + [\delta(i_{-}-\tilde{i}_{-}) 
+ \tau \delta(i_{-}-\tilde{i}_{+})]\nonumber\\
&& \times I_{4}(P,k_{3},i_{-},i_{3})  
+ [\delta(i_{+}+\tilde{i}_{+}) + \tau \delta(i_{+}+\tilde{i}_{-})]I_{4}(P,k_{3},i_{+},-i_{3}) + [\delta(i_{-}+\tilde{i}_{-}) 
+ \tau \delta(i_{-}+\tilde{i}_{+})]\times \nonumber\\
&& I_{4}(P,k_{3},i_{-},-i_{3}) + \tilde{I}_{4}(P,k_{3},\frac{i_{+}+\tilde{i}_{-}}{2},
\frac{i_{+}-\tilde{i}_{-}}{2},\frac{i_{+}-\tilde{i}_{+}}{2}) + \tau \tilde{I}_{4}(P,k_{3},\frac{i_{-}+\tilde{i}_{-}}{2},
\frac{i_{-}-\tilde{i}_{-}}{2},\frac{i_{-}-\tilde{i}_{+}}{2}) \nonumber\\
&& + \tau \tilde{I}_{4}(P,k_{3},\frac{i_{+}-\tilde{i}_{+}}{2},
\frac{i_{+}+\tilde{i}_{+}}{2},\frac{i_{+}-\tilde{i}_{-}}{2}) + \tilde{I}_{4}(P,k_{3},\frac{i_{-}-\tilde{i}_{+}}{2},
\frac{i_{-}+\tilde{i}_{+}}{2},\frac{i_{-}-\tilde{i}_{-}}{2}) + \tau \tilde{I}_{4}(P,k_{3},\frac{i_{+}}{2}\nonumber\\
&& +\frac{\tilde{i}_{+}}{2},
\frac{i_{+}-\tilde{i}_{+}}{2},\frac{i_{+}-\tilde{i}_{-}}{2}) + \tilde{I}_{4}(P,k_{3},\frac{i_{-}+\tilde{i}_{+}}{2},
\frac{i_{-}-\tilde{i}_{+}}{2},\frac{i_{-}-\tilde{i}_{-}}{2}) + \tilde{I}_{4}(P,k_{3},\frac{i_{+}-\tilde{i}_{-}}{2},
\frac{i_{+}+\tilde{i}_{-}}{2},\frac{i_{+}}{2}\nonumber\\
&& +\frac{\tilde{i}_{+}}{2}) + \tau \tilde{I}_{4}(P,k_{3},\frac{i_{-}-\tilde{i}_{-}}{2},
\frac{i_{-}+\tilde{i}_{-}}{2},\frac{i_{-}+\tilde{i}_{+}}{2})+ 2\Bigl[\hat{I}_{4}(P,k_{3},i_{+},\frac{i_{+}+\tilde{i}_{-}}{2},
\frac{i_{+}-\tilde{i}_{+}}{2}) + \tau \hat{I}_{4}(P,k_{3},\nonumber\\
&& i_{-},\frac{i_{-}+\tilde{i}_{-}}{2},
\frac{i_{-}-\tilde{i}_{+}}{2}) + \tau \hat{I}_{4}(P,k_{3},i_{+},\frac{i_{+}-\tilde{i}_{+}}{2},
\frac{i_{+}+\tilde{i}_{-}}{2}) + \hat{I}_{4}(P,k_{3},i_{-},\frac{i_{-}-\tilde{i}_{+}}{2},
\frac{i_{-}+\tilde{i}_{-}}{2}) \;\;+ \nonumber\\
&& \hat{I}_{4}(P,k_{3},i_{+},\frac{i_{+}-\tilde{i}_{-}}{2},
\frac{i_{+}+\tilde{i}_{+}}{2}) + \tau \hat{I}_{4}(P,k_{3},i_{-},\frac{i_{-}-\tilde{i}_{-}}{2},
\frac{i_{-}+\tilde{i}_{+}}{2}) + \tau \hat{I}_{4}(P,k_{3},i_{+},\frac{i_{+}+\tilde{i}_{+}}{2},
\frac{i_{+}}{2}\nonumber\\
&& -\frac{\tilde{i}_{-}}{2}) + \hat{I}_{4}(P,k_{3},i_{-},\frac{i_{-}+\tilde{i}_{+}}{2},
\frac{i_{-}-\tilde{i}_{-}}{2})\Bigr]\Bigr\}, 
\end{eqnarray} 
where the objects $I_{4}(k,k',i,j)(\equiv I_{4}(k,k',i,j;\tilde{\sigma},\mu_{0})), 
\tilde{I}_{4}(k,k',i,j,l)(\equiv I_{4}(k,k',i,j,l;\tilde{\sigma},\mu_{0}))$ and 
$\hat{I}_{4}(k,k',i,j,l)(\equiv I_{4}(k,k',i,j,l;\tilde{\sigma},\mu_{0}))$ are defined, respectively, 
by
\begin{subequations}\label{60}
\begin{eqnarray}
&& I_{4}(k,k',i,j)= \tilde{\sigma}^{2}\overset{\infty}{\underset{l,m=-\infty}{\sum}}\int \frac{d^{d-1}q_{1}d^{d-1}q_{2}}
{[q_{1}^{2}+\tilde{\sigma}^{2}l^{2}+\mu_{0}^{2}][(q_{1}-k)^{2}+\tilde{\sigma}^{2}(l-i)^{2}+\mu_{0}^{2}]
[q_{2}^{2}+\tilde{\sigma}^{2}m^{2}+\mu_{0}^{2}]}\nonumber\\
&& \;\;\times\;\; \frac{1}{[(q_{1}-q_{2}+k')^{2}+\tilde{\sigma}^{2}(l-m-j)^{2}+\mu_{0}^{2}]},\\
&& \tilde{I}_{4}(k,k',i,j,l)=\tilde{\sigma}^{2}\overset{\infty}{\underset{m=-\infty}{\sum}}\int \frac{d^{d-1}q_{1}d^{d-1}q_{2}}
{[q_{1}^{2}+\tilde{\sigma}^{2}i^{2}+\mu_{0}^{2}][(q_{1}-k)^{2}+\tilde{\sigma}^{2}j^{2}+\mu_{0}^{2}]
[q_{2}^{2}+\tilde{\sigma}^{2}m^{2}+\mu_{0}^{2}]}\nonumber\\
&& \;\;\times\;\; \frac{1}{[(q_{1}-q_{2}+k')^{2}+\tilde{\sigma}^{2}(l-m)^{2}+\mu_{0}^{2}]},\\
&& \hat{I}_{4}(k,k',i,j,l)=\tilde{\sigma}^{2}\overset{\infty}{\underset{m=-\infty}{\sum}}\int \frac{d^{d-1}q_{1}d^{d-1}q_{2}}
{[q_{1}^{2}+\tilde{\sigma}^{2}m^{2}+\mu_{0}^{2}][(q_{1}-k)^{2}+\tilde{\sigma}^{2}(m-i)^{2}+\mu_{0}^{2}]
[q_{2}^{2}+\tilde{\sigma}^{2}j^{2}+\mu_{0}^{2}]}\nonumber\\
&& \;\;\times\;\; \frac{1}{[(q_{1}-q_{2}+k')^{2}+\tilde{\sigma}^{2}(l-m)^{2}+\mu_{0}^{2}]}.
\end{eqnarray}
\end{subequations}
Note that the pole in $\epsilon$ coming from the computation of the summation is absent in the integral $\hat{I}_{4}$. Since we are only 
interested in the singular part of the above diagram, we can simply neglect the contribution coming from this integral. 
\par We are now left with the two-loop diagrams of the vertex part $\Gamma^{(2,1)}(k_{1},k_{2}; Q,i_{1},i_{2},j, \mu_{0}, \tilde{\sigma})$. Sticking to the method above described using a similar simplifying notation for 
the integrals by omitting the arguments $(\mu_{0}, \tilde{\sigma})$, whereas performing the identifications $(k_{i})\equiv(k_{1},k_{2};Q), 
k=k_{1}+k_{2}$, it is not difficult to find the following expressions ($i_{\pm}=i_{1} \pm i_{2}$):
\begin{subequations}\label{61}
\begin{eqnarray}
&& \parbox{10mm}{\includegraphics[scale=0.4]{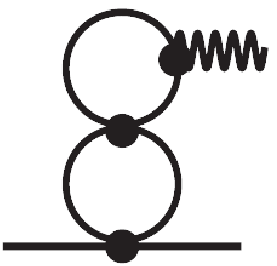}} \quad(k_{i}) \quad = \frac{\tau}{2}\;\frac{(N+2)^{2}}{108} 
\Bigl[(\delta(i_{+}+j) + \delta(i_{+}-j) + \tau \delta(i_{-}+j) + \tau \delta(i_{-}+j))I_{2}^{2}(k,j) \nonumber\\
&& + 2 \tau \tilde{I}_{2}(k,\frac{j+i_{+}}{2},\frac{j-i_{+}}{2})[I_{2}(k,j) + I_{2}(k,i_{+})] + + 2 \tilde{I}_{2}(k,\frac{j+i_{-}}{2},
\frac{j-i_{-}}{2})[I_{2}(k,j) + I_{2}(k,i_{-})]\nonumber\\
&& + 2\overset{\infty}{\underset{l=-\infty}{\sum}} \Bigl(\tilde{I}_{2}(k,\frac{j-i_{+}-2l}{2},\frac{j+i_{+}+2l}{2}) 
\tilde{I}_{2}(k,l,l+i_{+}) + \tilde{I}_{2}(k,\frac{j-i_{-}-2l}{2},\frac{j+i_{-}+2l}{2})\nonumber\\
&& \times \;\; \tilde{I}_{2}(k,l,l+i_{-})\Bigr)\Bigr],\\
&& \parbox{10mm}{\includegraphics[scale=0.4]{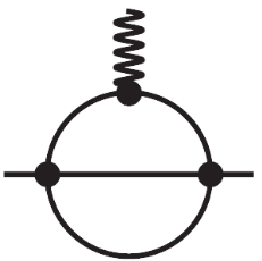}}\quad(k_{i}) \quad = \frac{\tau}{2}\;\frac{(N+2)}{36} \Bigl\{(\delta(i_{+}+j) 
+ \delta(i_{+}-j))I_{4}(k,Q,i_{+},i_{2})) + \tau (\delta(i_{-}+j) + \delta(i_{-}-j))\nonumber\\
&& \times I_{4}(k,Q,i_{-},i_{2}) + \tilde{I}_{4}(k,Q,\frac{j-i_{-}}{2},\frac{j+i_{-}}{2},\frac{j-i_{+}}{2}) + \tau \tilde{I}_{4}(k,Q,\frac{j+i_{+}}{2},\frac{j-i_{+}}{2},
\frac{j+i_{-}}{2}) + \tilde{I}_{4}(k,Q,\nonumber\\
&& \frac{j+i_{-}}{2},\frac{j-i_{-}}{2},\frac{j+i_{+}}{2}) + \tau \tilde{I}_{4}(k,Q,\frac{j-i_{+}}{2},\frac{j+i_{+}}{2},
\frac{j-i_{-}}{2}) + 2 \tau \hat{I}_{4}(k,Q,j,\frac{j-i_{+}}{2},\frac{j-i_{-}}{2}) \nonumber\\
&& + 2 \hat{I}_{4}(k,Q,j,\frac{j+i_{-}}{2},\frac{j+i_{+}}{2}) + 2 \tau \hat{I}_{4}(k,Q,j,\frac{j+i_{+}}{2},\frac{j+i_{-}}{2}) + 2 \hat{I}_{4}(k,Q,j,\frac{j-i_{-}}{2},\frac{j-i_{+}}{2})\Bigr\}
\end{eqnarray}
\end{subequations}
\par We have now a set of integrals which are very similar to those coming from periodic boundary conditions 
(see Ref. \cite{BL}). From our discussion, getting the expression of the diagrams is the laborious part of the 
method. Since we have already discussed the periodic integrals, only minor modifications are necessary to perform the 
computations aimed as we shall see next.

\section{Massive fields for $DBC$ and $NBC$ in the exponential representation} 
\par The vertex parts depend on the boundary condition, but we will suppress that dependence in what follows. By now it is rather 
clear that our unified description in the last section leaves no ambiguity to define the renormalization algorithm simultaneously for 
both boundary conditions. 
\par Before discussing explicitly the renormalization algorithm, it is interesting to restrict ourselves to the minimal number of 
diagrams to perform our computation of the critical exponents. The argument for the massive theory regarding infinite systems appeared 
recently \cite{CLJMP}. We will summarize the steps here and use the results directly. Start by defining the three-loop bare mass 
$\tilde{\mu}_{0}$ in terms of the the tree-level bare mass as $\tilde {\Gamma}^{(2)}(k=0,j,\tilde{\sigma},\mu_{0}, \lambda_{0}) \equiv \tilde{\mu}_{0}^{2} = \Gamma^{(2)}(k=0,j,\tilde{\sigma},\mu_{0}, \lambda_{0}) 
-\tilde{\sigma}^{2}j^{2}$. Then, inverting this equation to get the tree-level $\mu_{0}$ in terms of the three-loop bare mass 
$\tilde{\mu_{0}}$ and all the diagrams computed at zero external momenta has the virtue of eliminating all tadpole diagrams 
for they do not depend on the external momentum. Next, expanding $\mu_{0}=\mu_{0}(\tilde{\mu}_{0},\lambda_{0})$ in each 
primitively divergent vertex part at first order in $\lambda_{0}$ eliminates all the remaining graphs containing one-loop mass 
insertions. Here is one of the most nontrivial results and we are going to discuss this topic carefully in what follows.
\par Recall that the two-point function always includes a factor $S_{i_{1}i_{2}}=\delta(i_{1}-i_{2})$ in arbitrary loop diagrams. The set of 
steps just described permits one to write the tree-level bare mass in terms of the three-loop bare mass up to $O(\lambda_{0}$) through
\begin{equation}\label{62}
\mu_{0}^{2} = \tilde{\mu}_{0}^{2} - \frac{\lambda_{0}}{2} \parbox{10mm}{\includegraphics[scale=0.2]{fig1DN.eps}}.
\end{equation}
Notice that when we perform this expansion in any integral, a complication occurs: the tadpole graph has finite size 
(kind of ``internal symmetry'') indices, say $l,m$, which will be attached to a propagator of a certain internal line of an 
arbitrary graph. So these indices should be contracted with the set of indices taking place naturally in the diagram. Consider 
the two-loop diagram of the four-point function with a ``mass insertion'':
\begin{eqnarray}\label{63}
&& \parbox{10mm}{\includegraphics[scale=0.32]{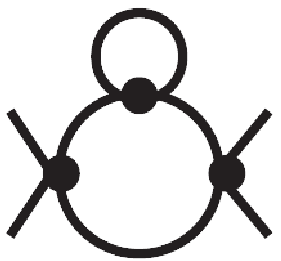}}\quad = \frac{(N+8)(N+2)}{27} 
\overset{\infty}{\underset{j_{1},j_{2},j_{3},j_{4} \geq 0}{\sum}}\tilde{S}^{(\tau)}_{i_1 i_2 j_1 j_2} 
\tilde{S}^{(\tau)}_{j_1 j_3 j_4 j_4} \tilde{S}^{(\tau)}_{j_2 j_3 i_3 i_4} \int d^{d-1}q_{1}  d^{d-1}q_{2} G_{0}(q_{1}+k,j_{1})\nonumber\\
&& \;\;\;G_{0}(q_{1}+k,j_{3}) G_{0}(q_{1},j_{2}) G_{0}(q_{2},j_{4}).
\end{eqnarray}
(The condensed nomenclature $\tilde{S}^{(\tau)}_{i_1 i_2 j_1 j_2}= \tilde{\sigma}[i_1 i_2 j_1 j_2]_{\tau}$ is sometimes useful.) In this 
expression we can set $\mu_{0}=\tilde{\mu}_{0}$ since the correction will produce three-loop terms which are not going to be useful 
to our analysis of the four-point function up to two-loop order and is going to be neglected.  
\par Let us take a look in Eq. (\ref{19}). There the propagator is defined as $G_{0}(q,j) \equiv G_{0}(q,j, \mu_{0})$. When we replace the 
mass $\mu_{0}$ by $\tilde{\mu}_{0}$ into the propagator, the left of the graph carries index $j$ from the original propagator before the 
insertion. Due to the internal character of the propagator which carries finite-size symmetry index, the coupling of the left hand side with 
the inserted mass tadpole is implemented as follows: the index $j$ must be free in the resulting expression and all the the other indices 
appearing there should be contracted with summations. This is equivalent to the following simple rule 
\begin{eqnarray}\label{64}
&& G_{0}(q,j,\mu_{0}) = G_{0}(q,j,\tilde{\mu}_{0})\Bigl[1-\sum_{l \geq 0}^{\infty}\frac{\lambda_{0}}{2[q^{2} + \tilde{\sigma}^{2}l^{2} 
+ \tilde{\mu}_{0}^{2}]} 
\Bigl[\parbox{10mm}{\includegraphics[scale=0.2]{fig1DN.eps}}\Bigr]_{j l}\Bigr]^{-1} \nonumber\\
&& = G_{0}(q,j,\tilde{\mu}_{0}) + \frac{\lambda_{0}}{2}  \sum_{l \geq 0}^{\infty}G_{0}(q,j,\tilde{\mu}_{0})  
G_{0}(q,l,\tilde{\mu}_{0}) \Bigl[\parbox{10mm}{\includegraphics[scale=0.2]{fig1DN.eps}}\Bigr]_{j l}.
\end{eqnarray}  
Now, replace this result in Eq. (\ref{19}). The one-loop diagram has two propagators which contribute the same amount to the 
correction due to the mass insertion (after some redefinitions). Upon substitution of Eq. (\ref{8}) for the tadpole, we find
\begin{eqnarray}\label{65}
&& \parbox{10mm}{\includegraphics[scale=0.32]{fig2DN.eps}}\quad = \frac{(N+8)}{9} 
\overset{\infty}{\underset{l_{1},l_{2} \geq 0}{\sum}}\tilde{S}^{(\tau)}_{i_1 i_2 l_1 l_2} \tilde{S}^{(\tau)}_{l_1 l_2 i_3 i_4} \int d^{d-1}q G_{0}(q+k,l_{1}, \tilde{\mu}_{0}) G_{0}(q,l_{2}, \tilde{\mu_{0}})\nonumber\\ 
&& + \lambda_{0} \frac{(N+8)(N+2)}{27} 
\overset{\infty}{\underset{j_{1},j_{2},j_{3},j_{4} \geq 0}{\sum}}\tilde{S}^{(\tau)}_{i_1 i_2 j_1 j_2} 
\tilde{S}^{(\tau)}_{j_1 j_3 j_4 j_4} \tilde{S}^{(\tau)}_{j_2 j_3 i_3 i_4} \int d^{d-1}q_{1}  d^{d-1}q_{2} G_{0}(q_{1}+k,j_{1},\tilde{\mu}_{0})\nonumber\\
&& \;\;\;G_{0}(q_{1}+k,j_{3},\tilde{\mu}_{0}) G_{0}(q_{1},j_{2},\tilde{\mu}_{0}) G_{0}(q_{2},j_{4},\tilde{\mu}_{0}).
\end{eqnarray}
\par Then, we conclude that the two-loop correction term due to the expansion of the internal propagators in the one-loop diagram 
of the four-point function produces the mass insertion whose value is equal exactly to that from Eq. (\ref{63}). This statement is 
equivalent to the important graphical identity:
\begin{eqnarray}\label{66}
&& \parbox{10mm}{\includegraphics[scale=0.32]{fig2DN.eps}}\quad |_{\mu_{0}} = \parbox{10mm}{\includegraphics[scale=0.32]{fig2DN.eps}}\quad |_{\tilde{\mu}_{0}} + \lambda_{0} \;\; \parbox{10mm}{\includegraphics[scale=0.32]{fig11DN.eps}}\;|_{\tilde{\mu}_{0}}.
\end{eqnarray}  
By following the same prescription, all the primitively divergent vertex parts with mass (tadpole) insertion are eliminated. For instance, 
the reader can check from what we have been discussing that the following diagrammatic identity also holds
\begin{eqnarray}\label{67}
&& \parbox{10mm}{\includegraphics[scale=0.32]{fig5DN.eps}}\quad |_{\mu_{0}} = \parbox{10mm}{\includegraphics[scale=0.32]{fig5DN.eps}}\quad |_{\tilde{\mu}_{0}} + \frac{3 \lambda_{0}}{2} \parbox{10mm}{\includegraphics[scale=0.4]{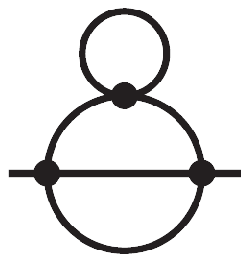}}|_{\tilde{\mu}_{0}}\quad.
\end{eqnarray}
It is easy to conclude that similar diagrammatic identities are valid for the composite operator $\Gamma^{(2,1)}$ due to its 
similarity with $\Gamma^{(4)}$. Despite the apparently complicated tensorial structure of this finite-size field-theoretic 
formulation, the mass insertions are cancelled for arbitrary external quasi-momenta modes as shown above in a simple manner. This 
cancellation takes place since different orders in $\lambda_{0}$ in the perturbative expression have different signs, and the 
combinatorial factors coming from the mass insertion corrections in each diagram exactly matches those from the perturbation 
expansion of the original diagram (one power of $\lambda_{0}$ higher) in $\mu_{0}$. This suffices to 
prove the claim explicitly at the perturbative order considered in the present paper. See \cite{CLJMP} for more details.
\par The resulting vertex parts now depend exclusively on $\tilde{\mu}_{0},\lambda_{0}$. Their diagrammatic expansion in 
terms of the reduced number of diagrams are:  
\begin{subequations}\label{68}
\begin{eqnarray}
&& \tilde{\Gamma}^{(2)}(k,j,\tilde{\sigma},\tilde{\mu}_{0},\lambda_{0}) =  k^{2} + \tilde{\mu}_{0}^{2} - \frac{\lambda_{0}^{2}}{6}\Bigl(
\parbox{10mm}{\includegraphics[scale=0.32]{fig5DN.eps}}\quad\bigg|_{\tilde{\mu}_{0}} - \parbox{10mm}{\includegraphics[scale=0.32]{fig5DN.eps}}\quad\bigg|_{k=0, \tilde{\mu}_{0}}\Bigr)\;+ \; \frac{\lambda_{0}^{3}}{4}\Bigl(
\parbox{10mm}{\includegraphics[scale=0.32]{fig6DN.eps}}\quad\bigg|_{\tilde{\mu}_{0}} \nonumber\\
&& \quad  - \; \parbox{10mm}{\includegraphics[scale=0.32]{fig6DN.eps}}\quad\bigg|_{k=0,\tilde{\mu}_{0}}\Bigr), \label{68a}\\
&&\Gamma^{(4)}_{i_{1}i_{2}i_{3}i_{4}}(k_{i},\tilde{\sigma},\mu_{0},\lambda_{0}) =  \lambda_{0}\tilde{S}_{i_{1}i_{2}i_{3}i_{4}} 
- \frac{\lambda_{0}^{2}}{2}
 \Bigl(\Bigl[\parbox{10mm}{\includegraphics[scale=0.32]{fig2DN.eps}}\;\;\Bigr]_{\tilde{\mu}_{0}}(k_{1}+k_{2})+2perms.\Bigr) 
+ \frac{\lambda_{0}^{3}}{2}\Bigl\{\Bigl(\Bigl[\parbox{10mm}{\includegraphics[scale=0.32]{fig8DN.eps}}\;\Bigr]_{\tilde{\mu}_{0}}\nonumber\\
&& (k_{i}) + 5perms. \Bigr) + \frac{1}{2}\Bigl(\Bigl[\parbox{10mm}{\includegraphics[scale=0.2]{fig7DN.eps}}\quad \Bigr]_{\tilde{\mu}_{0}} (k_{1}+k_{2})+2perms.\Bigr) \Bigl\}, \label{68b} \\
&& \Gamma^{(2,1)}_{i_{1},i_{2},j}(k_{1},k_{2};Q_{3},\tilde{\sigma},\mu_{0},\lambda_{0}) = \tau \Gamma^{(2,1)}_{i_{1},i_{2},j=0} + \Gamma^{(2,1)}_{i_{1},i_{2},j=i_{1}+i_{2}}, \label{68c}\\
&& \Gamma^{(2,1)}_{i_{1},i_{2},j}(k_{1},k_{2};Q_{3},\tilde{\sigma},\mu_{0},\lambda_{0}) = \parbox{10mm}{\includegraphics[scale=0.05]{fig27DN.eps}} 
- 3 \lambda_{0} \Bigl[\parbox{10mm}{\includegraphics[scale=0.32]{fig3DN.eps}}\Bigr]_{\tilde{\mu}_{0}}(k_{1}+k_{2};Q_{3}) 
+ 3 \lambda_{0}^{2}\Bigl[\parbox{10mm}{\includegraphics[scale=0.32]{fig9DN.eps}}\Bigr]_{\tilde{\mu}_{0}}(k_{1}+k_{2};Q_{3}) \nonumber\\
&& + 6\lambda_{0}^{2}\Bigl[\parbox{10mm}{\includegraphics[scale=0.32]{fig10DN.eps}}\Bigr]_{\tilde{\mu}_{0}}(k_{1},k_{2};Q_{3}). \label{68d}
\end{eqnarray} 
\end{subequations}
We choose not to display explicitly the tensor $S_{i_{1}i_{2}}$ for the vertex $\Gamma^{(2)}$ since 
after using the value of this tensor, the two-point function depends upon only one external quasi-momentum (mode). 
Note that Eq. (\ref{68c}) takes into account that each diagram appears in the combination mentioned before in Eq. (\ref{38}). We appplied a 
similar simplifying notation for the vertex part $\Gamma^{(2,1)}_{i_{1}i_{2}j}$ and 
choose not to show the dependence on the tensor $\hat{S}_{i_{1},i_{2},j}$ in the zeroth order diagram because of that implicit 
combination. (We could have done the same to the vertex part $\Gamma^{(4)}_{i_{1},i_{2},i_{3},i_{4}}$ provided we include the zero 
order graph in the diagrammatic expansion.) We omit hereafter the lower indices (``internal'' finite-size modes) whenever 
referring to an arbitrary vertex part but keep then on the argument of that vertex.
\par The vertex part $\tilde{\Gamma}^{(2)}(k,j,\tilde{\sigma},\tilde{\mu}_{0},\lambda_{0})\equiv \Gamma^{(2)}(k,j,\tilde{\sigma},\tilde{\mu}_{0},\lambda_{0}) -\tilde{\sigma}^{2}j^{2}$  has the advantage of not possessing the 
tree-level term $\tilde{\sigma}^{2}j^{2}$ and looks like the two-point vertex part from the bulk case. It has a logarithmic divergence 
as well as $\Gamma^{(4)}(k_{i},j_{i},\tilde{\sigma},\tilde{\mu}_{0},\lambda_{0})$. In our specific case of a given order in 
the coupling constant expansion, we want to go up to three-loop order in the 
expansion of $\tilde{\Gamma}^{(2)}$. Define the renormalized quantity $\tilde{\Gamma}_{R}^{(2)}(k,j,\tilde{\sigma},\tilde{\mu}_{0},\lambda_{0})= Z_{\phi}\tilde{\Gamma}^{(2)}(k,j,\tilde{\sigma},\tilde{\mu}_{0},\lambda_{0})$. The introduction of the 
normalization function  $Z_{\phi}$ produces the cancellation of the logarithmic divergence aforementioned. After that, we define a 
renormalized finite mass at three-loop order as $\mu^{2}=Z_{\phi}\tilde{\mu}_{0}^{2}$. On the other hand, the four-point vertex part starts 
proportional to the bare coupling constant, so we can define a renormalized coupling constant up to two-loop level by writing 
$g=Z_{\phi}^{2}\lambda_{0}$. First, write $Z_{\phi}= 1 + g z_{1} + g^{2} z_{2} + g^{3} z_{3}+...$ (where $z_{n}$ are divergent quantities). 
From the absence of linear terms in the coupling constant perturbative expansion of $\tilde{\Gamma}_{R}^{(2)}$, we immediately obtain 
$z_{1}=0$. At third order, we can express it entirely in terms 
of renormalized quantities, $\tilde{\Gamma}_{R}^{(2)}(k,j,\tilde{\sigma},\mu,g) \equiv \tilde{\Gamma}_{R}^{(2)}(k,j,\tilde{\sigma},\tilde{\mu}_{0},\lambda_{0})$. The same happens to the other vertex part, since up to two-loop order the renormalized object defined by 
$\Gamma_{R}^{(4)}(k_{i},j_{i},\tilde{\sigma},\mu,g)= Z_{\phi}^{2}\Gamma^{(4)}(k,j_{i},\tilde{\sigma},\tilde{\mu}_{0},\lambda_{0})$ is automatically given in terms only of renormalized mass and coupling constant, as the reader can easily verify. The same argument carries through to the 
renormalized composed vertex part obtained from the logarithmic divergent bare one via 
$\Gamma_{R}^{(2,1)}(k,j,\tilde{\sigma},\mu,g)= Z_{\phi^{2}}Z_{\phi}\Gamma^{(2)}(k,j,\tilde{\sigma},\tilde{\mu}_{0},\lambda_{0})$. It requires 
another normalization function $Z_{\phi^{2}}$ and no longer depends on bare quantities at this order.
\par In general, we can go to arbitrary loop order and get rid of all the bare quantities in the definition of generic vertex parts which 
are renormalized multiplicatively.  Multiplicative renormalizability amounts to say that an arbitrary vertex part 
$\Gamma^{(L,M)} ((L,M)\neq (0,2))$ including composite operators can be renormalized through the functions $Z_{\phi}^{(\tau)}, 
Z_{\phi^{2}}^{(\tau)}$ such that the vertex parts defined by 
\begin{equation}\label{69}
\Gamma_{R}^{(L,M)}(p_{n}, i_{n}, Q_{n'}, j_{n'}, g, \mu)= (Z_{\phi}^{(\tau)})^{\frac{M}{2}}(Z_{\phi^{2}}^{(\tau)})^{L} \Gamma^{(L,M)}(p_{n}, i_{n}, Q_{n'}, j_{n'}, \lambda_{0}, \tilde{\mu}_{0}, \Lambda),
\end{equation} 
are automatically finite. (In the argument above $\Gamma_{R}^{(L,M)} \equiv \tilde{\Gamma}_{R}^{(L,M)}$ for $(L,M)=(2,0)$.) 
Here, $g$, $\mu$ and $\Lambda$ are the renormalized coupling constant, mass and cutoff, respectively. The 
argument works well for all kinds of regulators, which are used to express infinite (divergent) quantities in terms of functions of 
these parameters (regulators). We employ the cutoff when deriving differential equations satisfied by renormalized vertex parts. From 
the explicit operational viewpoint, however, we shall focus here in dimensional regularization, so that 
divergences are written in terms of poles in $\epsilon=4-d$ ($\propto \epsilon^{-l}$, where $l$ is an integer positive number, 
usually the number of loops), where 
$d$ is the space dimension of the system. The symbol $n$ labels the external momenta $p_{n}$ as well as the mode 
of the external quasi-momenta $i_{n}$ associated to the external legs ($n=1,...,L$) of 
the fields. The label $n'$ is connected to the external momenta $Q_{n'}$ and the mode 
$j_{n'}$ of the quasi-momentum corresponding to the external legs ($n'=1,...,M$) 
of the composite fields ($\phi^{2}$ insertions) in an arbitrary $1PI$ diagram. 
\par The normalization conditions on the primitively divergent vertex parts for the massive theory are chosen at zero external 
momenta and nonzero quasi-momenta. From the structure of the diagrams, it is obvious that the condition on the external 
momenta are not sufficient to simplify our task. In fact, there are other renormalization schemes, for instance minimal 
subtraction, which do not require any fixation of the external momenta. In order to give a simpler prescription useful for 
all other renormalization methods required from an {\it ab initio} formulation, we choose the external quasi-momenta to be 
set at especial values. The symmetry point here is defined at external quasi-momenta from the external legs chosen to be 
equal and set to the value $i$ (arbitrary nonvanishing positive integer). The finiteness of the theory can be achieved 
through the conditions   
\begin{subequations}\label{70}
\begin{eqnarray}
&& \tilde{\Gamma}_{R}^{(2)}(k=0 , i, g, \mu) = \mu^{2},
\label{70a} \\
&& \frac{\partial\tilde{\Gamma}_{R}^{(2)}(k, i, g, \mu )}{\partial k^{2}}\Bigl|_{k^{2}=0} \
= 1,\label{70b} \\
&& \Gamma_{R}^{(4)}(k_{l}=0, i_{l}=i, g, \mu ) \equiv \Gamma_{R}^{(4)}\Bigl|_{SP}  = 3\tilde{\sigma}g  , \label{70c}\\
&& \Gamma_{R}^{(2,1)}(k=0, i_{1}=i_{2}=i, Q=0, j, g, \mu) \equiv \Gamma_{R}^{(2,1)}\Bigl|_{\overline{SP}} = \frac{3 \tau}{2} .\label{70d}
\end{eqnarray}
\end{subequations}  
\par We can now discuss the renormalization group invariance of arbitrary renormalized vertex functions. At the critical dimension $d=4$ utilize the cutoff as the regulator. The infinite cutoff limit in the integrals appearing in 
each vertex part multiplicatively renormalized is taken solely after the 
renormalization prescription is established. Apply the derivative $\frac{\partial}{\partial \tilde{\mu}_{0}^{2}}$ over 
the bare vertex part $\Gamma^{(N,M)}$ ($(N,M)\neq(0,2)$) at fixed 
$\lambda_{0}, \Lambda$ in order to obtain the 
vertex function $\Gamma^{(N,M+1)} (p_{n},i_{n}, Q_{n'}, i'_{n'}; 0; \lambda_{0}, 
\tilde{\mu}_{0}, \Lambda)$ at zero inserted momentum. Next, rewrite the 
remaining bare vertex parts in terms of the renormalized ones. After similar manipulations performed in the case of periodic and 
antiperiodic boundary conditions in \cite{BL} but utilizing the above normalization conditions, one learns that the renormalization 
group invariance of the renormalized vertex parts in different renormalized mass scales is substantiated in the following form of 
the Callan-Symanzik equation:
\begin{eqnarray}\label{71}
&&\left(\mu \frac{\partial}{\partial \mu} + 
\beta^{(\tau)} \frac{\partial}{\partial g}
- \frac{N}{2} \gamma_{\phi}^{(\tau)} + M \gamma_{\phi^{2}}^{(\tau)}\right)
\Gamma_{R}^{(N,M)} (p_{l}, i_{l}, Q_{l}, i'_{l}, g,
\mu)= \\ \nonumber
&& \mu^{2} (2 - \gamma_{\phi}^{(\tau)})\Gamma_{R}^{(N,M+1)} (p_{l}, i_{l}, 
Q_{l}, i'_{l};0, g, \mu)  \;\;, 
\end{eqnarray}
where $\beta^{(\tau)}(\mu, g) = \mu \frac{\partial g}{\partial \mu}$, 
$\gamma_{\phi}^{(\tau)} = \mu \frac{\partial ln Z_{\phi}^{(\tau)}}{\partial \mu}$ and 
$\gamma_{\phi^{2}}^{(\tau)}= - \mu \frac{\partial ln Z_{\phi^{2}}^{(\tau)}}{\partial \mu}$. Even for  $i \neq 0$ in the present unified 
framework, this form resembles very much the situation for $PBC$. 
\par When the field is not at the critical dimension, the annoying dimensionful aspect of the coupling constant can be 
circumvented by writing the bare (renormalized) coupling constant in terms of a genuine dimensionless bare (renormalized) 
coupling constant $u_{0}$ ($u$) as $\lambda_{0}=\mu^{\epsilon}u_{0}$ ($g=\mu^{\epsilon}u$). The flow function 
$\beta^{(\tau)}(\mu, g)$ can be written in terms Gell-Mann-Low function defined by 
$[\beta(g,\mu)]_{GL}= -\epsilon g + \beta(g, \mu)$. Using the Gell-Mann-Low function into the $CS$ equation, along with the 
dimensionful quantities defined in terms of dimensionless amounts, we find that 
$[\beta(g,\mu)]_{GL} \frac{\partial}{\partial g}= \beta(u) \frac{\partial}
{\partial u}$. It turns out that resulting renormalization-group picture involves only the dimensionless renormalized coupling 
constant and possesses a well defined scaling limit \cite{Vladimirov,Naud}. The Callan-Symanzik equation now reads
\begin{eqnarray}\label{72}
&&\left(\mu \frac{\partial}{\partial \mu} + 
\beta(u) \frac{\partial}{\partial u}
- \frac{L}{2} \gamma_{\phi}^{(\tau)} + M \gamma_{\phi^{2}}^{(\tau)}\right)
\Gamma_{R}^{(L,M)} (p_{n}, i_{n}, Q_{n}, i'_{n'}, u, \mu)= \\ \nonumber
&& \mu^{2}(2 - \gamma_{\phi}^{(\tau)}) \Gamma_{R}^{(L,M+1)} (p_{n}, i_{n}, 
Q_{n'}, i'_{n'};0, u, \mu)  \;\;,
\end{eqnarray} 
where 
\begin{subequations}\label{73}
\begin{eqnarray}
\beta^{(\tau)}(u)&=& -\epsilon \left(\frac{\partial lnu_{0}^{(\tau)}}{\partial u}\right), \label{73a}\\
\gamma_{\phi}^{(\tau)}(u) &=& \beta^{(\tau)}(u)
\left(\frac{\partial ln Z_{\phi}^{(\tau)}}{\partial u}\right), \label{73b}\\
\gamma_{\phi^{2}}^{(\tau)}(u) &=& \beta^{(\tau)}(u) 
\left(\frac{\partial ln Z_{\phi^{2}}^{(\tau)}}{\partial u}\right). \label{73c}
\end{eqnarray}
\end{subequations}
Another function that will be useful to our purposes and utilizes the definition 
$\bar{Z}_{\phi^{2}}^{(\tau)}= Z_{\phi^{2}}^{(\tau)}Z_{\phi}^{(\tau)}$ is written as
\begin{equation} \label{74}
\bar{\gamma}_{\phi^{2}}^{(\tau)} = \beta^{(\tau)} 
\left(\frac{\partial ln \bar{Z}_{\phi^{2}}^{(\tau)}}{\partial u}\right).
\end{equation}   
The solution of the Callan-Symanzik equation has already been described in the ultraviolet scaling regime. The right-hand side 
becomes negligible at the scaling limit, $\frac{p}{\mu} \rightarrow \infty$ \cite{BLZ1,BLZ2}, showing that the solution admits 
a ultraviolet fixed point. We will just make use of these results here and prove that the definitions above are sufficient to our 
computation of the fixed point and its aftermath in the evaluation of critical indices.     
\par Let us start the discussion of the asymptotic limits for both boundary conditions. The basic objects to be computed 
are the one-loop integrals $I_{2}(k,i,\tilde{\sigma},\mu)$ and $\tilde{I}_{2}(k,i,j,\tilde{\sigma},\mu)$ 
which belong to the one-loop graph of the four-point function. The resemblence of these integrals for $DBC$, $NBC$, $ABC$ and $PBC$ 
will permit a more economical approach to this topic and the reader is advised to consult Ref. \cite{BL} to grasp many details. We 
prefer to omit them herein.
\par The integral $I_{2}(k,i,\tilde{\sigma},\mu)$ is identical in the form 
to $I_{2}^{(\tau)}(k,i,\sigma,\mu)$ for periodic boundary condition ($\tau=0$ in Eq. (10) from \cite{BL}). First, we factor out the 
mass $\mu$ from the integral, rescale all momenta and 
define $\tilde{r}=\frac{\tilde{\sigma}}{\mu}$. Second, 
we introduce a Feynman parameter and solve the integral over the momenta. Third, we perform the summation using the representation 
from Ref. \cite{BF}. Fourth, transform the argument of the resulting (factorial) $\Gamma$ function from $(d-1)$ to $d$. Then, divide 
the integral by the area of the unit sphere at $d$ dimensions. These sets of 
steps lead to the following result (see also Appendix A) 
\begin{eqnarray}\label{75}
&& I_{2}(k,i;\tilde{r}) \equiv \frac{I_{2}(k,i;\tilde{\sigma}, \mu)}{S_{d}}= \frac{\mu^{-\epsilon}}{\epsilon}\Bigl(\Bigl(1-\frac{\epsilon}{2}\Bigr)\int_{0}^{1} dx [x(1-x)(k^{2} + \tilde{r}^{2} i^{2}) + 1]^{ - \frac{\epsilon}{2}} \nonumber\\
&& \;\;+ \;\;\frac{\epsilon}{2} \Gamma\Bigl(2-\frac{\epsilon}{2}\Bigr) F_{\frac{\epsilon}{2}}(k,i;\tilde{r})\Bigr),
\end{eqnarray} 
where 
\begin{subequations}\label{76}
\begin{eqnarray}
&&\tilde{F}_{\alpha}(k,i;\tilde{r})= \tilde{r}^{-2 \alpha} \int_{0}^{1} dx 
f_{\frac{1}{2} + \alpha}\Bigl(xi, h(k, i, \tilde{r}) \Bigr),\label{76a} \\
&& f_{\alpha}(a,b)= 4\overset{\infty}{\underset{m=1}{\sum}} cos(2\pi ma) \bigl(\frac{\pi m}{b}\bigr)^{\alpha -\frac{1}{2}}K_{\alpha -\frac{1}{2}}(2 \pi mb),
\label{76b} \\
&& h(k, i, \tilde{r})= \tilde{r}^{-1} \sqrt{x(1-x)(k^{2} + \tilde{r}^{2} i^{2}) +1}, \label{76c}
\end{eqnarray}
\end{subequations}
where $K_{\nu}(x)$ is the modified Bessel function of the second kind.
\par Using a similar chain of reasoning, it is not difficult to find that the other integral required reads:
\begin{eqnarray}\label{77}
& \tilde{I}_{2}(k,i,j;\tilde{r}) \equiv \frac{\tilde{I}_{2}(k,i,j;\tilde{\sigma}, \mu)}{S_{d}}= \frac{\tilde{r}\mu^{-\epsilon}}{2}\int_{0}^{1} dx \bigl(\tilde{r}^{2}[xi^{2} + (1-x)j^{2}] + k^{2}x(1-x)+1\bigr)^{-\frac{1}{2}}.
\end{eqnarray} 
\par Recall that in order to compute the one-loop diagram, owing to our use of normalization conditions 
in this massive framework, we need the previous integrals evaluated at vanishing external momenta $(k=0)$. Not only this: 
specifically, we need $I_{2} (0,2i,\tilde{r}), I_{2} (0,0,\tilde{r}), \tilde{I}_{2} (0,0,0;\tilde{r}), \tilde{I}_{2} (0,i,i;\tilde{r})$ 
and $\tilde{I}_{2}(k,2i,0;\tilde{r}) $. In that case, setting $k=0$ and $\epsilon=0$ in the $O(\epsilon^{0})$ terms from 
$I_{2}(k,2i;\tilde{r})$  we find:
\begin{eqnarray}\label{78}
I_{2} (0,2i,\tilde{r})= \mu^{-\epsilon} \frac{1}{\epsilon}
\Bigl(1-\frac{\epsilon}{2} - \frac{\epsilon}{2} \int_{0}^{1} ln[4x(1-x)\tilde{r}^{2}i^{2} +1] dx  
+ \frac{\epsilon}{2} \tilde{F}_{0}(0,i,\tilde{r})\Bigr).
\end{eqnarray} 
It is also simple to demonstrate that $\tilde{I}_{2}(0,0,0;\tilde{r})=\frac{\tilde{r}\mu^{-\epsilon}}{2}$, 
$\tilde{I}_{2}(k,2i,0;\tilde{r})= \frac{\tilde{r}\mu^{-\epsilon}}{1+\sqrt{1+4\tilde{r}^{2}i^{2}}}$ and  
$\tilde{I}_{2}(k,i,i;\tilde{r})= \frac{\tilde{r}\mu^{-\epsilon}}{2\sqrt{1+\tilde{r}^{2}i^{2}}}$. Replacing these 
results in the expression  of the one-loop four-point diagram for the particular choices of external momenta and quasi-momenta, we can write
\begin{subequations}\label{79}
\begin{eqnarray}
& \parbox{12mm}{\includegraphics[scale=0.4]{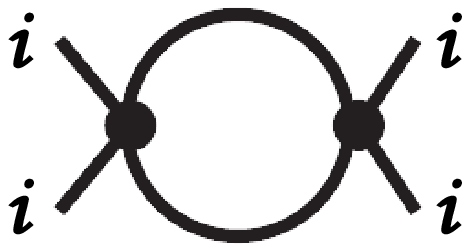}}\quad\;= 3\tilde{\sigma}\frac{(N+8)}{9\epsilon}\mu^{-\epsilon}\Bigl[1-\frac{\epsilon}{2} + \epsilon \zeta^{(\tau)}(i;\tilde{r})\Bigr],\label{79a} \\
& \zeta^{(\tau)}(i;\tilde{r})= \frac{1}{6}\tilde{F}_{0}(0,2i,\tilde{r}) + \frac{1}{3}\tilde{F}_{0}(0,0,\tilde{r}) 
+ \frac{\tilde{r}\tau}{3} +\frac{2\tilde{r}\tau}{3[1+\sqrt{1 + 4\tilde{r}^{2}i^{2}}]} + \frac{2\tilde{r}}{3\sqrt{1 + \tilde{r}^{2}i^{2}}}\nonumber\\ 
& -\frac{1}{3}\Bigl[\frac{\sqrt{1+\tilde{r}^{2}i^{2}}}{\tilde{r}i}arcsinh(\tilde{r}i) -1\Bigr].\label{79b}
\end{eqnarray}
\end{subequations}
In the above equation $\tilde{r} \propto (\frac{L}{\xi})^{-1}$ where $\xi (\sim \mu^{-1})$ is the bulk (infinite 
system) correlation length. We will analyze this correction term in order to study its asymptotical behavior for 
large as well as small values of $L$.  
\par Let us start by the first scaling regime, namely  $\frac{L}{\xi} \rightarrow \infty$ ($\tilde{r} \rightarrow 0$) limit. It is easy to demonstrate that {\it i)} the linear terms in $\tilde{r}$ go to zero, 
{\it ii)} $\tilde{F}_{0}(0,0,\tilde{r}\rightarrow 0) \rightarrow 0$  \cite{BL} and {\it iii)} the last term also vanishes. 
It is not difficult to demonstrate that $\tilde{F}_{0}(0,2i,\tilde{r} \rightarrow 0)$ tends to zero as follows. From Eq. (\ref{76c}), 
$\underset{\tilde{r} \rightarrow 0}{lim}  h(0,2i,\tilde{r}) \rightarrow \tilde{r}^{-1}$, and we can solve 
the integral by writing explicitly the integrand in terms of a summation involving Bessel functions. The coefficient of each term 
in the summation  is equal to $sin(4 \pi mi)$ which is zero for integer $i$, therefore proving the assertion. Then, 
the finite-size correction reduces to the bulk result ($\zeta^{(\tau)}(i;\tilde{r} \rightarrow 0)=0$) whenever 
$\frac{L}{\xi} \rightarrow \infty$.
\par Next we consider the limit $\frac{L}{\xi} \rightarrow 0$ ($\tilde{r} \rightarrow \infty$). The several terms inside the 
correction function are going to be examined separately. Note that $\underset{\tilde{r} \rightarrow \infty}{lim} \frac{2\tilde{r}\tau}{3[1+\sqrt{1 + 4\tilde{r}^{2}i^{2}}]} \rightarrow \frac{\tau}{3i}$, 
$\underset{\tilde{r} \rightarrow \infty}{lim}\frac{2\tilde{r}}{3\sqrt{1 + \tilde{r}^{2}i^{2}}} \rightarrow \frac{2}{3i}$, which are 
convergent and $\tilde{r}$-independent. In addition, 
 $\underset{\tilde{r} \rightarrow \infty}{lim} -\frac{1}{3}\Bigl[\frac{\sqrt{1 + \tilde{r}^{2}i^{2}}}{\tilde{r}i} arcsinh(\tilde{r}i) -1\Bigr] \rightarrow -\frac{1}{3}[ln(2\tilde{r}i) -1]$, and we have a logarithmic divergence in the desired limit. 
\par We now focus on the contributions from $\tilde{F}_{0}(0,2i,\tilde{r})$ and $\tilde{F}_{0}(0,0,\tilde{r})$ in this limit. 
The function
\begin{equation}\label{80}
f_{\frac{1}{2}}(2ix,h(k=0,2i,\tilde{r})= 4\overset{\infty}{\underset{m=1}{\sum}} cos(4 \pi mix) K_{0}(2 \pi mh(k=0,2i,\tilde{r}),
\end{equation}
can be better understood through the identity \cite{GR}
\begin{eqnarray}\label{81}
&&\sum_{n=1}^{\infty} K_{0}(n \hat{x}) cos(n \hat{x} t) = \;\; \frac{1}{2} \left[\gamma
+ ln(\frac{\hat{x}}{4 \pi})\right] \;\; 
+ \;\;\frac{\pi}{2\hat{x} \sqrt{1+t^{2}}} \nonumber\\
&& + \;\; \frac{\pi}{2} 
\sum_{n=1}^{\infty} \left[\frac{1}{\sqrt{\hat{x}^{2} + (2n\pi + t\hat{x})^{2}}}
- \frac{1}{2 n \pi}\right] + \frac{\pi}{2} \sum_{n=1}^{\infty} \left[\frac{1}{\sqrt{\hat{x}^{2} + (2n\pi - t\hat{x})^{2}}} - \frac{1}{2 n \pi}\right], 
\end{eqnarray}
where $\gamma = 0.57721566...$ is the Euler-Mascheroni constant. It is valid for positive values of the variable $\hat{x}$. Performing the  
the identifications $\hat{x}= 2 \pi \tilde{r}^{-1}\sqrt{1+4\tilde{r}^{2} i^{2} x(1-x)}$, 
$t= \frac{2\tilde{r} ix}{\sqrt{1+4\tilde{r}^{2} i^{2} x(1-x)}}$ we get to
\begin{eqnarray}\label{82}
&& f_{\frac{1}{2}}(2ix,h(k=0,2i,\tilde{r})= 2\gamma + 2 ln\Bigl[\frac{\sqrt{1+4\tilde{r}^{2} i^{2} x(1-x)}}{2\tilde{r}}\Bigr] 
+ \frac{\tilde{r}}{\sqrt{1+4\tilde{r}^{2} i^{2}x}}\nonumber\\
&&   +  \sum_{n=1}^{\infty}\Bigl[\frac{1}{\sqrt{\tilde{r}^{-2} + 4i(i-n)x + n^{2}}}-\frac{1}{n}\Bigr]
+ \sum_{n=1}^{\infty}\Bigl[\frac{1}{\sqrt{\tilde{r}^{-2} + 4i(i+n)x + n^{2}}}-\frac{1}{n}\Bigr].
\end{eqnarray}
In particular, for $i=0$
\begin{eqnarray}\label{83}
&& f_{\frac{1}{2}}(0,h(k=0,0,\tilde{r})= 2\gamma - 2 ln(2\tilde{r}) + \tilde{r} 
+ 2 \sum_{n=1}^{\infty}\Bigl[\frac{1}{\sqrt{\tilde{r}^{-2} + n^{2}}} - \frac{1}{n}\Bigr].
\end{eqnarray}
This coincides exactly with $\tilde{F}_{0}(0,0,\tilde{r})$. Taking the limit (setting $\tilde{r}= \infty$ into  the last term)
we find $\underset{\tilde{r} \rightarrow \infty}{lim}\tilde{F}_{0}(0,0,\tilde{r}) \rightarrow  \tilde{r} -2ln(2\tilde{r})$. 
Finally, we have to analyze $\tilde{F}_{0}(0,2i,\tilde{r})$ which is given by
\begin{eqnarray}\label{84}
&& \tilde{F}_{0}(k=0,2i,\tilde{r})= 2\gamma + 2 \int_{0}^{1} dx ln\Biggl[\frac{\sqrt{1+4\tilde{r}^{2} i^{2} x(1-x)}}{2\tilde{r}}\Biggr] 
+ \int_{0}^{1} dx \frac{\tilde{r}}{\sqrt{1+4\tilde{r}^{2} i^{2}x}} +  \sum_{n=1}^{\infty} \int_{0}^{1} dx \nonumber\\
&& \times \;\;\Bigl[\frac{1}{\sqrt{\tilde{r}^{-2} + 4i(i-n)x + n^{2}}}-\frac{1}{n}\Bigr]
+ \sum_{n=1}^{\infty} \int_{0}^{1} dx\Bigl[\frac{1}{\sqrt{\tilde{r}^{-2} + 4i(i+n)x + n^{2}}}-\frac{1}{n}\Bigr].
\end{eqnarray} 
One can show that $\underset{\tilde{r} \rightarrow \infty}{lim} \int_{0}^{1} dx ln\Bigl[\frac{\sqrt{1+4\tilde{r}^{2} i^{2} x(1-x)}}{2\tilde{r}}\Bigr] \rightarrow lni -1$, which is finite in this limit. The third term can also be shown to be finite, 
namely, $\underset{\tilde{r} \rightarrow \infty}{lim} \int_{0}^{1} dx \frac{\tilde{r}}{\sqrt{1+4\tilde{r}^{2} i^{2}x}} \rightarrow \bigl(\frac{1}{i})\bigr)$. 
\par The summations conceal an underlying subtlety for $i \neq 0$. Looking at Eq. (\ref{83}) there was no problem whatsoever in replacing 
directly $\tilde{r}= \infty$ when $i=0$ at the summations. That was the way we renormalized $PBC$ and $ABC$ with $i=0$. Indeed,  
setting directly ($r=2\tilde{r}$ in those boundary conditions) $\tilde{r}^{-2}=0$, these terms contribute 
zero to $PBC$ (just as here) and $(2ln2-1)$ for $ABC$. In other words, they contribute a finite constant $\tilde{r}$-independent, which do not 
have a pathological behavior in the limit $\tilde{r} \rightarrow \infty$. Looking at $PBC$, if we do not choose to renormalize the theory 
at $i=0$, but choose instead all external quasi-momenta equal to $i>0$, then the form of these summations is identical to those 
which appear in $DBC$ and $NBC$ here. In $ABC$ the occurrence of this feature is a little bit different since one has to add the finite term 
above mentioned.
\par Now consider $i \neq 0$ and see what happens when we set $\tilde{r}^{-2}=0$ in the summations. Using the power series 
expansion from  $(1+y)^{-\frac{1}{2}}$, keeping all terms in $y$ and performing the integrals over the Feynman parameter $x$, we obtain
\begin{eqnarray}\label{85}
&\underset{\tilde{r} \rightarrow \infty}{lim} \Biggl(\overset{\infty}{\underset{n=1}{\sum}} \int_{0}^{1} dx \Bigl[\frac{1}{\sqrt{\tilde{r}^{-2} + 4i(i-n)x + n^{2}}}
-\frac{1}{n}\Bigr]+ \overset{\infty}{\underset{n=1}{\sum}} \int_{0}^{1} dx\Bigl[\frac{1}{\sqrt{\tilde{r}^{-2} + 4i(i+n)x + n^{2}}}-\frac{1}{n}\Bigr]\Biggr)= \nonumber\\
& 2 \overset{\infty}{\underset{n=1}{\sum}} i^{2n}\zeta(2n+1), 
\end{eqnarray}
where $\zeta(2n+1)= \overset{\infty}{\underset{p=1}{\sum}}\frac{1}{p^{2n+1}}$.
\par It is clear that the series diverges for $i \geq 1$. This divergent number is harmless in the limit 
$\tilde{r}\rightarrow \infty$, since {\it i)} it does not depend on 
$\tilde{r}$, {\it ii)} it can be eliminated by suitable choices of external quasi-momenta in $PBC$, $NBC$ and $ABC$ ({\it e.g.}, $i=0$). As far 
as the series does not depend explicitly on $\tilde{r}$, it can be safely considered regular in this limit in comparison with poles in 
$\epsilon$. In order to eliminate the divergent term for $PBC$, $ABC$, $DBC$ and $NBC$ when we choose $i \neq 0$, which is 
clearly an artifact of the method, we can define a ``normal ordered'' (regularized) correction 
$:\zeta^{(\tau)}(i,\tilde{r}): = \zeta^{(\tau)}(i,\tilde{r}) - \frac{1}{3} \sum_{n=1}^{\infty}i^{2n}\zeta(2n+1)$ that is still divergent in the 
limit $L \rightarrow 0$ but without the nonphysical divergence associated to this summation. This normal ordering operation ammounts to say that 
if one replaces directly  $\tilde{r} = \infty$ in the infinite sumations Eq. (\ref{81}) (or alternatively into Eqs. (\ref{82}), (\ref{84})) 
whenever $i \neq 0$ in analogy with which is done in the case $i=0$, one shoud be careful to subtract this term aformentioned. That is the 
actual quantity that should be used in the limit of small values of $L$. 
\par Putting together all the terms examined in this limit, we find that the regularized finite-size correction function reduces to 
\begin{eqnarray}\label{86}
&&\underset{\tilde{r} \rightarrow \infty}{lim} :\zeta^{(\tau)}(i,\tilde{r}): \rightarrow \frac{1}{6i}(2\tau + 5) 
- ln(2\tilde{r}) + \frac{\tilde{r}(1+\tau)}{3}.
\end{eqnarray}
\par The finite-size correction is clearly boundary condition dependent. From this expression we can obtain two distinct 
behaviors in this limit whenever the system is not located at the bulk critical temperature $(t\neq 0)$. For $DBC (\tau=-1)$, the 
divergence is purely logarithimic like in $ABC$, whereas the dominant term for $NBC (\tau=1)$ diverges linearly with 
$\tilde{r}$ just like in $PBC$. When these terms become comparable to the dimensional pole in $\epsilon$ through decreasing values of $L$ 
the dimensional crossover takes place.  
\par Let us turn now our attention to the two- and three-loop diagrams. They are required to getting the 
critical exponents $\eta$ and $\nu$ perturbatively in the safe scaling region discussed above where $L$ is not so small and 
$\epsilon$-expansion results are valid. We will just write down the solution of the integrals. The interested reader should consult 
the details in Appendix A. 
\par Consider the two-loop diagrams of the four-point vertex function. They consist of ``diagonal terms'' which are responsible for the 
leading singularities in $\epsilon$ as well as ``nondiagonal terms''. The double bubble can be easily obtained from the one-loop diagram 
just discussed by noting that the last piece comprising the products of two $\tilde{I}_{2}$'s and including an infinite summation are 
nonsingular in $\epsilon$ and can be neglected away from the crossover region. We then find at zero external momenta and special external 
quasi-momenta $i$ the following result
\begin{eqnarray}\label{87}
\parbox{10mm}{\includegraphics[scale=0.3]{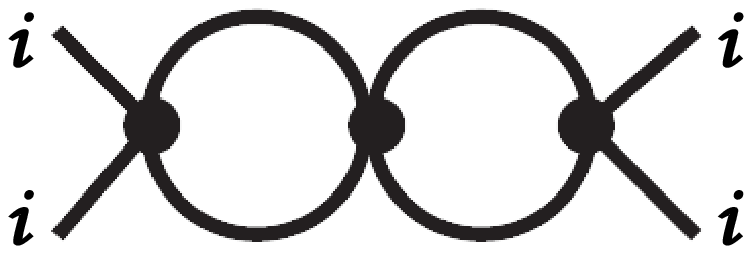}}\qquad\quad = 3\tilde{\sigma} \frac{(N^{2}+6N+20)}{27} \mu^{-2\epsilon}\Bigl\{\frac{1}{\epsilon^{2}}\Bigl(1- \epsilon 
+ 2\epsilon \zeta^{(\tau)}(i,\tilde{r})\Bigr)\Bigr\}.
\end{eqnarray} 
\par A similar observation permits our computation of the nontrivial two-loop 
diagram since the integrals $\hat{I}_{4}$ do not contribute to the singular terms of this graph. At zero external momenta and nonvanishing 
external quasi-momenta set to the value $i$ in all external legs, it turns out to be given by:
\begin{eqnarray}\label{88}
\parbox{10mm}{\includegraphics[scale=0.4]{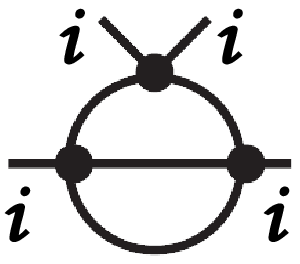}}\quad = 3\tilde{\sigma} \frac{(5N+22)}{27} \mu^{-2\epsilon}\Bigl\{\frac{1}{2\epsilon^{2}}\Bigl(1-\frac{\epsilon}{2} + 2\epsilon 
\zeta^{(\tau)}(i,\tilde{r})\Bigr)\Bigr\}.
\end{eqnarray}  
\par In normalization conditions, we shall need the derivative of the two-point vertex part diagrams computed at zero external 
momenta in this massive framework. The ``sunsetting'' two-loop diagram involves the integral 
$\frac{I_{3}(k=0,i,\tilde{\sigma},\mu)}{S_{d}^{2}} \equiv I_{3}(k=0,i,\tilde{r})$ and we are interested in the object 
$I_{3}^{\prime}(i,\tilde{r}) \equiv \frac{\partial I_{3}(k,i,\tilde{r})}{\partial k^{2}}|_{k=0}$ together with the other tilded integrals . 
The three-loop graph has a similar structure in terms of the integrals 
$\frac{I_{5}(k=0,i,\tilde{\sigma},\mu)}{S_{d}^{3}} \equiv I_{5}(k=0,i,\tilde{r})$. 
The utilization of the combinations previously prescribed as well as the result of the integrals in the Appendix A imply that those 
diagrams can be cast in the form
\begin{subequations}\label{89}
\begin{eqnarray}
&& \frac{\partial}{\partial k^{2}}\left(\parbox{10mm}{\includegraphics[scale=0.3]{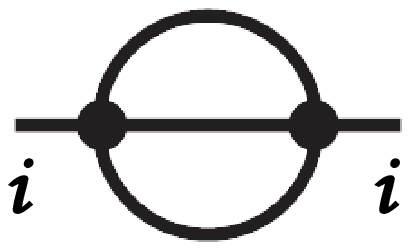}}\quad\right)_{k^{2}=0}\quad =-\frac{\mu^{-2\epsilon} (N+2)}{24\epsilon}\Bigl[1-\frac{\epsilon}{4} + \epsilon \tilde{W}^{(\tau)}(i,\tilde{r})\Bigr],\\
&& \frac{\partial}{\partial k^{2}}\left(\parbox{10mm}{\includegraphics[scale=0.3]{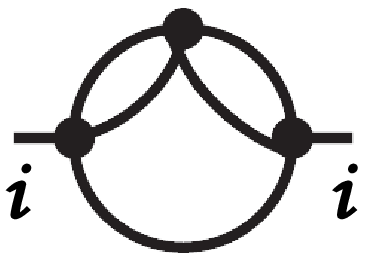}}\quad\right)_{k^{2}=0}\quad =-\frac{\mu^{-3\epsilon}(N+2)(N+8)}{162 \epsilon^{2}}\Bigl[1-\frac{\epsilon}{4} + \frac{3\epsilon}{2} \tilde{W}^{(\tau)}(i;\tilde{r})\Bigr].
\end{eqnarray}
\end{subequations} 
Some useful definitions similar to those occurring in $PBC$ and $ABC$ boundary conditions are:
\begin{subequations}\label{90}
\begin{eqnarray}
&& \tilde{F}_{\alpha,\beta}^{(\tau)}(k,i;\tilde{r}) \equiv \frac{1}{S_{d}}\tilde{r} \overset{\infty}{\underset{j=-\infty}{\sum}} 
\int d^{d-1}q \frac{\tilde{F}_{\alpha}^{(\tau)}(q+k,j+i;\tilde{r})}{[(q)^{2} + \tilde{r}^{2}j^{2}+ 1]^{\beta}},\label{90a}\\ 
&& \tilde{F}_{\alpha}^{\prime (\tau)}(i;\tilde{r}) \equiv \frac{\partial \tilde{F}_{\alpha,1}^{(\tau)}(k,i;\tilde{r})}
{\partial k^{2}}\big|_{k=0},\label{90b}\\
&& \tilde{\mathcal{F}}_{\alpha,\beta}^{(\tau)}(k,j,i;\tilde{r}) \equiv \frac{1}{S_{d}}\tilde{r}  
\int d^{d-1}q \frac{\tilde{F}_{\alpha}^{(\tau)}(q+k,i;\tilde{r})}{[(q)^{2} + \tilde{r}^{2}j^{2}+ 1]^{\beta}},\label{90c}\\
&& \tilde{\mathcal{F}}_{\alpha}^{\prime (\tau)}(i,j;\tilde{r}) \equiv \frac{\partial \tilde{\mathcal{F}}_{\alpha,1}^{(\tau)}(k,i,j;\tilde{r})}
{\partial k^{2}}\big|_{k=0}.\label{90d}
\end{eqnarray}
\end{subequations}
When expressed in terms of these definitions, the above finite size correction to the higher loop two-point graphs represented by 
the amount $\tilde{W}^{(\tau)}(i,\tilde{r})$ can be written as 
\begin{eqnarray}\label{91}
&& \tilde{W}^{(\tau)}(i;\tilde{r}) = -\frac{1}{2} - 2\int_{0}^{1} \int_{0}^{1}dx dy (1-y)ln\Biggl[y(1-y)i^{2} + (1-y)\tilde{r}^{-2} 
+ \frac{y\tilde{r}^{-2}}{x(1-x)}\Biggr]\nonumber\\ 
&& + 2 \int_{0}^{1} \int_{0}^{1}dx dy (1-y)f_{\frac{1}{2}}\Biggl(iy,\sqrt{y(1-y)i^{2} + (1-y)\tilde{r}^{-2} 
+ \frac{y\tilde{r}^{-2}}{x(1-x)}}\Biggr) - 4\tilde{F}'_{0}(i;\tilde{r}) \nonumber\\
&& + 6\tilde{r}\Biggl[\int_{0}^{1} \int_{0}^{1}dx dy (1-y)\Bigl(\tilde{r}^{2}i^{2}(1-y)
+ y\Bigl(\frac{1}{x(1-x)}-1\Bigr) +1\Bigr)^{-\frac{1}{2}}\nonumber\\ 
&& + \tau \int_{0}^{1} \int_{0}^{1}dx dy (1-y)\Bigl(\tilde{r}^{2}i^{2} y + y\Bigl(\frac{1}{x(1-x)}-1\Bigr) +1\Bigr)^{-\frac{1}{2}}\Biggr]  \nonumber\\
&& -12[\mathcal{{F}}_{0}^{\prime (\tau)}(i,0;\tilde{r}) + \tau \mathcal{{F}}_{0}^{\prime (\tau)}(0,i;\tilde{r})  \Bigr] 
\end{eqnarray}
\par Finally let us present the solution of the vertex $\Gamma^{(2,1)}$ graphs at a suitable symmetry point (zero external momenta). 
These diagrams can be written in terms of those integrals of the four-point function. For instance, the one-loop diagram, according to 
the rule outlined in the diagrammatic Eq. (\ref{38}) can be written as
\begin{eqnarray}\label{92}
&& \left(\parbox{13mm}{\includegraphics[scale=0.4]{fig3DN.eps}} \right)_{SP}\quad = \tau \parbox{13mm}{\includegraphics[scale=0.4]{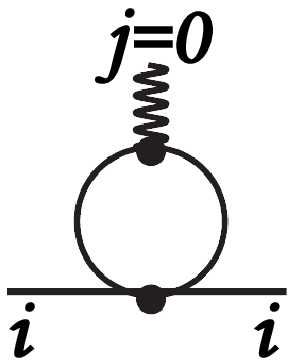}} 
+ \parbox{13mm}{\includegraphics[scale=0.4]{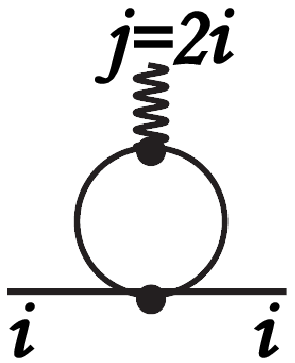}}= \frac{3 \mu^{-\epsilon} \tau}{2} \frac{(N+2)}{18 \epsilon} 
\Biggl[1-\frac{\epsilon}{2} + \epsilon \zeta^{(\tau)}(i;\tilde{r})\Biggr].
\end{eqnarray}
We are also interested in the solution for the two-loop diagrams. Using the same rule, we first obtain for the trivial two-loop diagram 
\begin{eqnarray}\label{93}
&& \left(\parbox{13mm}{\includegraphics[scale=0.3]{fig9DN.eps}}\right)_{SP}\quad = \tau \parbox{13mm}{\includegraphics[scale=0.4]{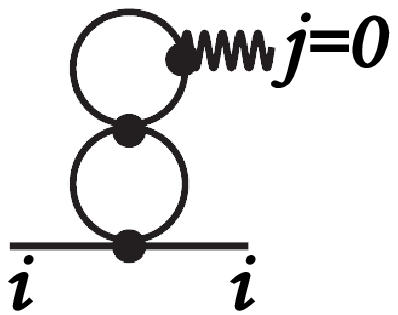}}\quad 
+ \quad\parbox{13mm}{\includegraphics[scale=0.3]{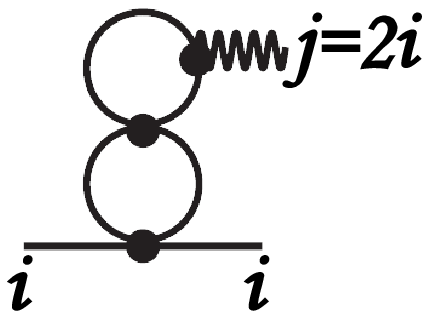}}= \frac{3 \mu^{-2\epsilon} \tau}{2} \frac{(N+2)^{2}}{108 \epsilon^{2}} 
[1- \epsilon + 2\epsilon \zeta^{(\tau)}(i;\tilde{r})].
\end{eqnarray}
The nontrivial two-loop, on the other hand is given by:
\begin{eqnarray}\label{94}
&& \left(\parbox{13mm}{\includegraphics[scale=0.4]{fig10DN.eps}} \right)_{SP}\quad = \tau \parbox{13mm}{\includegraphics[scale=0.4]{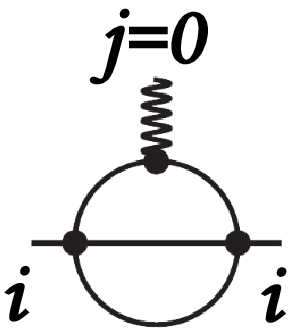}} 
+ \parbox{13mm}{\includegraphics[scale=0.4]{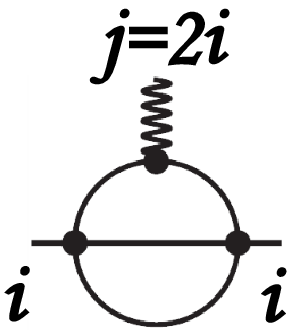}}= \frac{3 \mu^{-2\epsilon} \tau}{2} \frac{(N+2)}{72 \epsilon^{2}} \Biggl[1-\frac{\epsilon}{2} + 2 \epsilon \zeta^{(\tau)}(i;\tilde{r})\Biggr].
\end{eqnarray}
\par A glance in the results obtained from the graphs analyzed so far demonstrates the nontrivial character of the finite-size correction 
terms for $DBC$ and $NBC$ in comparison to the $PBC$ and $ABC$ counterparts. The nondiagonal terms produce singular contributions at 
two-loop order in all primitively divergent vertex functions (and three-loop order in the two-point function) which modify the correction 
terms from the periodic and antiperiodic cases.
\par With these data we can proceed to compute the critical exponents at least up to two-loop level. We carry out this task in the 
following section.
\section{$DBC$ and $NBC$ critical exponents in the massive finite-size regime}
\par In order to compute the fixed point and finally the critical exponents in a parallel plate geometries with one direction of finite 
extent in the apropriate regime, we write the dimensionless 
bare coupling constant in terms of the renormalized one as $u_{0}^{(\tau)}= u(1 + a_{1}^{(\tau)} u + a_{2}^{(\tau)} u^{2})$. Although we 
are going to show that the fixed point is boundary condition dependent, we shall omit henceforth the explicit dependence on $\tau$ in 
both bare and renormalized dimensionless coupling constants. The renormalization 
functions are written as  $Z_{\phi}^{(\tau)} = 1 + b_{2}^{(\tau)} u^{2} + b_{3}^{(\tau)} u^{3}$ and 
$\bar{Z}_{\phi^{2}}^{(\tau)} = 1 + c_{1}^{(\tau)} u + c_{2}^{(\tau)} u^{2}$.     
\par The bare vertex parts which can be renormalized multiplicatively at zero external momenta and nonvanishing external quasi-momenta $i$ 
can be written in the form 
\begin{subequations}\label{95}
\begin{eqnarray}
&& \frac{\partial \tilde{\Gamma}^{(2)}}{\partial k^{2}}\Bigl|_{k^{2}=0}= 
1 - \tilde{B}_{2}^{(\tau)}u_{0}^{2} + \tilde{B}_{3}^{(\tau)}u_{0}^{3},\label{95a} \\
&& \Gamma^{(4)}\Bigl|_{SP}= 3\tilde{\sigma}u_{0}\mu^{\epsilon}[1 - \tilde{A}_{1}^{(\tau)}u_{0} 
+ (\tilde{A}_{2}^{(1\; \tau)} + \tilde{A}_{2}^{(2\; \tau)})u_{0}^{2}], \label{95b} \\
&& \Gamma^{(2,1)}\Bigl|_{\overline{SP}}= \frac{3\tau}{2} - \tilde{C}_{1}^{(\tau)}u_{0}
+ (\tilde{C}_{2}^{(1\; \tau)} + \tilde{C}_{2}^{(2\; \tau)})u_{0}^{2}.\label{95c}
\end{eqnarray}
\end{subequations}
In the previous equations, the following identifications can be made:
\begin{subequations}\label{96}
\begin{eqnarray} 
&& \tilde{B}_{2}^{(\tau)}= \frac{\mu^{2\epsilon}}{6}\frac{\partial}{\partial k^{2}}\left(\parbox{10mm}{\includegraphics[scale=0.3]{fig16DN.eps}}\quad\right)_{k^{2}=0}\;=-\frac{(N+2)}{144\epsilon}\Bigl[1-\frac{\epsilon}{4} + \epsilon \tilde{W}^{(\tau)}(i,\tilde{r})\Bigr],\\
&& \tilde{B}_{3}^{(\tau)}=\frac{\mu^{3\epsilon}}{4}\frac{\partial}{\partial k^{2}}\left(\parbox{10mm}{\includegraphics[scale=0.3]{fig17DN.eps}}\quad\right)_{k^{2}=0};=-\frac{(N+2)(N+8)}{648 \epsilon^{2}}\Bigl[1-\frac{\epsilon}{4} + \frac{3\epsilon}{2} \tilde{W}^{(\tau)}(i,\tilde{r})\Bigr],\\
&& \tilde{A}_{1}^{(\tau)}=\frac{\mu^{\epsilon}}{2\tilde{\sigma}}\parbox{12mm}{\includegraphics[scale=0.4]{fig13DN.eps}}\quad\;= \frac{(N+8)}{6\epsilon}\Bigl[1-\frac{\epsilon}{2} + \epsilon \zeta^{(\tau)}(i;\tilde{r})\Bigr],\\
&& \tilde{A}_{2}^{(1\; \tau)}= \frac{\mu^{2\epsilon}}{4\tilde{\sigma}}\parbox{10mm}{\includegraphics[scale=0.3]{fig14DN.eps}}\qquad = 
\frac{(N^{2}+6N+20)}{36} \Bigl\{\frac{1}{\epsilon^{2}}(1- \epsilon 
+ 2\epsilon \zeta^{(\tau)}(i,\tilde{r}))\Bigr\},\\
&& \tilde{A}_{2}^{(2\; \tau)})=\frac{\mu^{2\epsilon}}{\tilde{\sigma}}\parbox{10mm}{\includegraphics[scale=0.4]{fig15DN.eps}}\quad = 
\frac{(5N+22)}{18\epsilon^{2}}\Bigl(1-\frac{\epsilon}{2} + 2\epsilon 
\zeta^{(\tau)}(i,\tilde{r})\Bigr),\\
&& \tilde{C}_{1}^{(\tau)} = 3 \mu^{\epsilon} \left(\parbox{13mm}{\includegraphics[scale=0.4]{fig3DN.eps}} \right)_{SP}\quad = \frac{3 \tau}{2} \frac{(N+2)}{6 \epsilon} \Biggl[1-\frac{\epsilon}{2} + \epsilon \zeta^{(\tau)}(i;\tilde{r})\Biggr],\\
&& \tilde{C}_{2}^{(1\; \tau)} = 3 \mu^{2\epsilon}\left(\parbox{13mm}{\includegraphics[scale=0.3]{fig9DN.eps}}\right)_{SP}\quad = 
\frac{3 \tau}{2} \frac{(N+2)^{2}}{36 \epsilon^{2}} 
[1- \epsilon + 2\epsilon \zeta^{(\tau)}(i;\tilde{r})].,\\
&& \tilde{C}_{2}^{(2\; \tau)} = 6 \mu^{2\epsilon}\left(\parbox{13mm}{\includegraphics[scale=0.4]{fig10DN.eps}} \right)_{SP}\quad = 
\frac{3 \tau}{2} \frac{(N+2)}{12 \epsilon^{2}} \Biggl[1-\frac{\epsilon}{2} + 2 \epsilon \zeta^{(\tau)}(i;\tilde{r})\Biggr].
\end{eqnarray}
\end{subequations}
Next, we apply the normalization conditions Eqs. (\ref{70}). Using the definition of the renormalized vertex 
parts discussed before, we can determine the normalization functions by imposing finiteness of the renormalized quantities. We start by 
replacing the dimensionless bare coupling constant in the diagrammatic expansion of the two- and four-point function. Use the definition 
from $Z_{\phi}^{(\tau)}$ into Eq. (\ref{70b}) in order to determine $b_{2}^{(\tau)}$. Replace the dimensionless 
bare coupling constant by the renormalized one at $O(u_{0}^{2})$ since the correction will be $O(u^{3})$ and can be neglected. We then find 
\begin{equation}\label{97}
b_{2}^{(\tau)} = -\frac{(N+2)}{144 \epsilon}\Bigl[1 -\frac{\epsilon}{4} + \epsilon \tilde{W}^{(\tau)}(i,\tilde{r})\Bigr].
\end{equation}
Using the above expression in the definition of the renormalized four-point function is sufficient to obtain
\begin{subequations}\label{98}
\begin{eqnarray}
&& a_{1}^{(\tau)} = \frac{(N+8)}{6 \epsilon} \Bigl[1 - \frac{\epsilon}{2} + \epsilon \zeta^{(\tau)}(i;\tilde{r})\Bigr], \label{101a} \\
&& a_{2}^{(\tau)} =  \Biggl[\frac{(N+8)}{6 \epsilon}\Biggr]^{2} \Bigl[1 + 2\epsilon \zeta^{(\tau)}(i;\tilde{r})\Bigr] -\frac{(2N^{2}+41N+170)}{72\epsilon}.\label{101b}
\end{eqnarray}
\end{subequations}
Using these data, we can compute $b_{3}^{(\tau)}$ which can be shown to be given by
\begin{equation}\label{99}
b_{3}^{(\tau)}= -\frac{(N+2)(N+8)}{1296\epsilon^{2}} \Bigl[1 -\frac{7 \epsilon}{4} + 3 \epsilon \zeta^{(\tau)}(i;\tilde{r})\Bigr].
\end{equation}
The renormalization function which defines the renormalized composite field can be determined in a very similar manner using the above 
expressions for $a_{1}^{(\tau)}, a_{2}^{(\tau)}$ and the diagrammatic expansion for that vertex part. We then find
\begin{subequations}\label{100}
\begin{eqnarray}
&& c_{1}^{(\tau)} = \frac{(N+2)}{6 \epsilon} \Bigl[1 - \frac{\epsilon}{2} + \epsilon \zeta^{(\tau)}(i;\tilde{r})\Bigr],\label{100a}\\
&& c_{2}^{(\tau)} = \frac{(N+2)(N+5)}{36 \epsilon^{2}} - \frac{(N+2)(2N+13)}{72\epsilon} + \frac{(N^{2}+7N+10)}{18 \epsilon}\zeta^{(\tau)}(i;\tilde{r}).
\label{100b}
\end{eqnarray}
\end{subequations}
The Wilson functions defined in Eqs. (\ref{73}) can be written in terms of these coefficients as
\begin{subequations}\label{101}
\begin{eqnarray}
&& \beta^{(\tau)}(u)= -\epsilon u[1-a_{1}^{(\tau)}u + 2((a_{1}^{(\tau)})^{2} - a_{2}^{(\tau)})u^{2}],\label{101a} \\
&& \gamma_{\phi}^{(\tau)}(u)=-\epsilon u[2b_{2}^{(\tau)} + (3b_{3}^{(\tau)}-2b_{2}^{(\tau)}a_{1}^{(\tau)}) u^{2}],\label{101b} \\
&& \bar{\gamma}_{\phi^{2}}^{(\tau)}(u)=\epsilon u[c_{1}^{(\tau)} + (2c_{2}^{(\tau)} -(c_{1}^{(\tau)})^{2} - a_{1}^{(\tau)}c_{1}^{(\tau)})u].\label{101c}
\end{eqnarray}
\end{subequations}
In order to evaluate the critical exponents, we need the fixed point of the dimensionless coupling constant, which is determined by 
the condition $\beta^{(\tau)}(u_{\infty})=0$. It is a simple task to demonstrate that it is given by the following expression:
\begin{equation}\label{102}
u_{\infty}=\Bigl[\frac{6 \epsilon}{N+8}\Bigr] \Bigl\{1+\Bigl[\frac{(9N+42)}{(N+8)^{2}}+\frac{1}{2}
-\zeta^{(\tau)}(i;\tilde{r})\Bigr]\epsilon \Bigr\}.
\end{equation} 
The exponent $\eta$ is identified with $\gamma_{\phi}^{(\tau)}(u_{\infty})$ which implies the following three-loop result
\begin{equation}\label{103}
\eta \equiv \gamma_{\phi}^{(\tau)}(u_{\infty})= \frac{(N+2)}{2(N+8)^{2}}\epsilon^{2} \Bigl\{1 + \epsilon \Biggl[\frac{6(3N+14)}{(N+8)^{2}} -\frac{1}{4}\Biggr]\Bigr\}.
\end{equation} 
The other Wilson function at the fixed point is given by $\bar{\gamma}_{\phi^{2}}^{(\tau)}(u_{\infty})=\frac{(N+2)}{(N+8)}\epsilon 
\Bigl[1+\frac{(6N+18)}{(N+8)^{2}}\epsilon\Bigr]$. The exponent $\nu$ is related to the exponent $\eta$ and the last expression through 
$\nu^{-1}=2 -\bar{\gamma}_{\phi^{2}}^{(\tau)}(u_{\infty}) - \eta$, which yields
\begin{equation}\label{104}
\nu= \frac{1}{2} + \frac{(N+2)}{4(N+8)}\epsilon + \frac{(N+2)(N^{2}+23N+60)}{8(N+8)^{3}}\epsilon^{2}.
\end{equation} 
\par These universal results are independent of the boundary conditions and reproduce the bulk critical exponents. Despite the fixed point 
and the Wilson functions carry residual nonuniversal information from the normalization conditions employed encoded in the nontrivial 
correction function $\zeta^{(\tau)}(i;\tilde{r})$, the cancellations take place exactly to eliminate all the finite-size information 
resulting in the bulk critical exponents. A similar feature was shown to be valid for $PBC$ and $ABC$. In the present case, even 
though the boundary conditions have not the same smooth character from $PBC$ and $ABC$, we have just proven that the previous conjecture 
made by Nemirovsky and Freed that the critical exponents are the same from the bulk theory also for $DBC$ and $NBC$. 
\par In order to complete the analysis of the critical behavior from finite-size systems confined between parallel plates, let us discuss 
the situation from the viewpoint of masless fields, generalizing the picture just developed involving only massive fields. We shall use 
other set of normalization conditions and will show that the critical exponents agree with those obtained in the present section. 

\section{Massless fields formulation: DBC and NBC in the exponential representation}
\par We have determined in previous sections the exact form of the several diagrams in perturbation theory. Now we perform the 
substitution of the massive integrals by their zero mass analogues. Then, it is possible to define a simple set of normalization conditions 
which allow us to consider the same type of diagrams already considered in the massive framework.    
\par Let us get started by describing the normalization conditions. The restriction to the minimal number of diagrams is 
analogous to what was discussed in the massive theory. We define the quantity 
$\tilde{\Gamma}^{(2)}(k,j,\mu=0,\tilde{\sigma}) \equiv \Gamma^{(2)}(k,j,\mu=0,\tilde{\sigma}) - \tilde{\sigma}^{2}j^{2}$. 
This means that the term $\tilde{\sigma}^{2}j^{2}$ does not need to be renormalized by the normalization functions, just 
like we did for the massive fields. (We careless kept the original vertex $\Gamma^{(2)}(k,j=0,\mu=0,\sigma)$ 
in our treatment for $ABC$ and $PBC$ such that the similar term $\sigma^{2} \tau^{2}$ appeared therein. It has no effect 
for $PBC$, but produces a small deviation in the normalization as discussed in Ref. \cite{BL} for $ABC$. The correct form 
to get rid of this inconvenience is to define the object $\tilde{\Gamma}^{(2)}(k,j=0,\mu=0,\sigma) \equiv \Gamma^{(2)}(k,j=0,\mu=0,\sigma) - \sigma^{2}\tau^{2}$ for $ABC$ as we did above for $DBC$ and $NBC$. This maneuver 
do not alter the normalization constants obtained in  Ref. \cite{BL}, since the argument there implicitly took into 
account this feature). We choose a symmetry point with the following properties: {\it i)} since the theory now possesses 
infrared divergences, we have to renormalize the primitively divergent vertices at nonzero external momenta; {\it ii)} we also choose nonvanishing external quasi-momenta owing to the boundary conditions. If $k_{i}$ are the external momenta 
associated to the infinite $(d-1)$-dimensional subspace and $j$ is the moding attached to the external quasi-momentum 
characterizing the distance between the plates, the theory is renormalized at fixed external momentum scale $\kappa$ and 
arbitrary nonvanishing moding $i$. The symmetric point is defined on the infinite subspace with the condition 
$k_{i}.k_{j}=\frac{\kappa^{2}}{4}(4\delta_{ij}-1)$, whereas all the external quasi-momenta mode of any primitively 
divergent one-particle irreducible ($1PI$) vertex part is chosen in the value $i$ (except for $\Gamma^{(2,1)}$; see below). 
\par The multiplicative renormalization can be successfully implemented through the following normalization conditions on the 
primitively divergent vertex functions:
\begin{subequations}\label{105}
\begin{eqnarray}
&& \tilde{\Gamma}_{R}^{(2)}(k=0, i, g, 0) = 0, \label{105a}\\
&& \frac{\partial\tilde{\Gamma}_{R}^{(2)}(k=\kappa, j, g, 0)}{\partial k^{2}}\Bigl|_{k^{2}=\kappa^{2}} \
= 1,  \label{105b}\\
&& \Gamma_{R}^{(4)}(k_{l}, i_{l}=i, g, 0)\Bigl|_{SP} = 3\tilde{\sigma}g  ,  \label{105c}\\
&& \Gamma_{R}^{(2,1)}(k_{1}, i_{1}=i, k_{2}, i_{2}=i, Q, j, g, 0)\Bigl|_{\overline{SP}} = 3 \frac{\tau}{2} . \label{105d}
\end{eqnarray}
\end{subequations}   
Note that the symmetry point is such that the insertion momentum is related to the other momenta in last equation 
through $Q^{2}=(k_{1}+k_{2})^{2}$. Second, recall from our discussion from the massive fields that the same rule applies 
for the vertex $\Gamma_{R}^{(2,1)}(k_{1}, i_{1}, k_{2}, i_{2}, Q, j, g, 0)|_{\overline{SP}}$: the tree-level vertex diagram 
corresponds to the combination 
\begin{eqnarray}\label{106}
&& \parbox{10mm}{\includegraphics[scale=0.075]{fig27DN.eps}}\qquad\;= \parbox{10mm}{\includegraphics[scale=0.075]{fig27DN.eps}}
\quad|_{(j=\pm(j_{1}+j_{2})} + \tau \parbox{10mm}{\includegraphics[scale=0.075]{fig27DN.eps}}\quad|_{(j=0)},
\end{eqnarray}
where the signal of $j$ is fixed. We attach to the external legs quasi-momentum $i$, such that the insertion quasi-momentum has two 
contributions: either $j=2i$ or $j=0$. 
\par In order to get rid of the mass insertions, we can follow two different trends. The first one resembles the massive theory 
and can be formulated as follows. Start with a tree-level bare mass $\mu_{0}$. Impose the condition 
$\tilde{\Gamma}^{(2)}(k=0,\mu_{0}, j,\lambda_{0})=\tilde{\mu}_{0}^{2}$. Invert this equation to obtain $\mu_{0}(\tilde{\mu}_{0})$ 
just as we done in the massive case. This $\mu_{0}$ could be interpreted as the shift in the bulk critical temperature.
Replace this back into the diagrammatic expression of the vertex part 
$\tilde{\Gamma}^{(2)}(k=0,\mu_{0}, j,\lambda)=\tilde{\mu}_{0}^{2}$. This eliminates all tadpole diagrams. Next, express 
$\mu_{0}(\tilde{\mu}_{0})$ up to $O(\lambda_{0})$ and Taylor expand the remaining vertex part around $k^{2}=\kappa^{2}$. The net 
effect is to eliminate all mass insertions in all primitively divergent vertex parts. Finally set $\tilde{\mu}_{0}=0$. 
\par Although this argument makes perfect sense from the point of view of statistical mechanics where the mass is identified with the 
reduced temperature (distance to the critical temperature), the reader working on quantum field theory might feel uneasy by 
starting with a nonvanishing bare mass, defining a three-loop bare mass and set it to zero afterwards, since it resembles a 
dynamical mass generation induced perturbatively. It is then worthwhile to develop the most traditional argument in which the 
masss is zero in all orders in perturbation theory as we will describe now.
\par Let us start directly with $\mu_{0}=0$, which implies $\tilde{\Gamma}^{(2)}(k=0,\mu_{0}=0, j,\lambda_{0})=0$. Consequently, one 
finds the diagrammatic expression up to three-loop order:
\begin{eqnarray} \label{107}
&&  \frac{\lambda_{0}^{2}}{6} \parbox{10mm}{\includegraphics[scale=0.32]{fig5DN.eps}}\;\;\bigg|_{k=0} - 
\frac{\lambda_{0}^{3}}{4} \Bigl(\parbox{10mm}{\includegraphics[scale=0.32]{fig6DN.eps}}\;\;\bigg|_{k=0} 
+ \parbox{10mm}{\includegraphics[scale=0.32]{fig12DN.eps}}\;\;\bigg|_{k=0}\Bigr) = 
\frac{\lambda_{0}}{2} \parbox{10mm}{\includegraphics[scale=0.3]{fig1DN.eps}}\nonumber\\
&&  - \frac{\lambda_{0}^{2}}{4} 
\parbox{10mm}{\includegraphics[scale=0.3]{fig4DN.eps}} \quad 
+ \lambda_{0}^{3}\Bigl[\frac{1}{8} \parbox{10mm}{\includegraphics[scale=0.2]{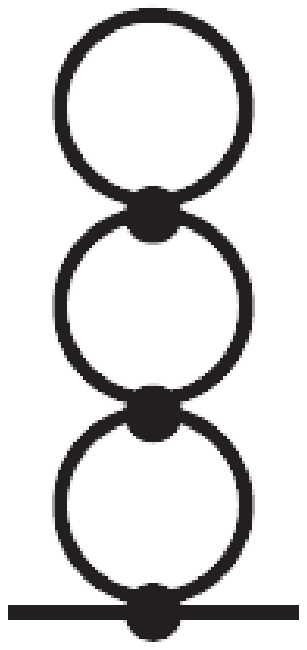}} \quad 
+ \frac{1}{8} \parbox{10mm}{\includegraphics[scale=0.25]{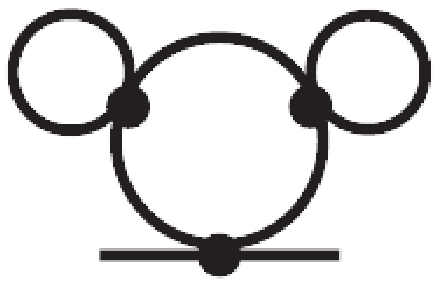}} \quad 
+ \frac{1}{12} \parbox{10mm}{\includegraphics[scale=0.25]{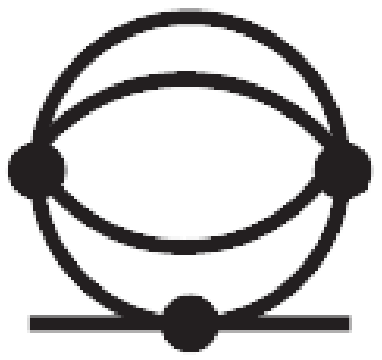}}\;\;\Bigr].
\end{eqnarray}
In the right hand side are all tadpole diagrams up to this order. Since the tadpoles do not depend on the external momenta, 
we can replace this expressios directly in the diagrammatic expression in order to get rid of them. The right hand side 
diagrams computed at zero external momenta survive. The mass insertions cancel most easily: at one-loop order the condition 
$\tilde{\Gamma}^{(2)}(k=0,\mu_{0}=0,\lambda_{0})=0= \frac{\lambda_{0}}{2} \parbox{10mm}{\includegraphics[scale=0.3]{fig1DN.eps}}$ $\;\;$ implies that the contributions with tadpole insertions appearing in {\it all} primitively divergent vertex parts are zero 
identically up to the perturbative order considered.
\par Plugging together all these facts into the bare vertex part $\tilde{\Gamma}^{(2)}(k,j,\mu_{0}=0,\tilde{\sigma})$ 
(for further details, the reader is advised to consult Ref. \cite{CLJMP}), we end up with
\begin{eqnarray}\label{108}
&& \tilde{\Gamma}^{(2)}(k,\mu_{0}=0,\lambda_{0}) =  k^{2} - 
\frac{\lambda^2}{6}\Biggl[\parbox{10mm}{\includegraphics[scale=0.32]{fig5DN.eps}}\;\; -\;  
\parbox{10mm}{\includegraphics[scale=0.32]{fig5DN.eps}}\bigg|_{k=0}\Biggr] 
+ \frac{\lambda^{3}}{4}\Biggl[\parbox{10mm}{\includegraphics[scale=0.32]{fig6DN.eps}}\; -\; \parbox{10mm}{\includegraphics[scale=0.32]{fig6DN.eps}}\bigg|_{k=0}\Biggr]. 
\end{eqnarray}
The situation here is completely similar to the massive case, since the number of diagrams are the same, only the 
renormalization symmetry point is chosen differently due to the infrared divergences occurring in the massless theory. We will 
keep the same normalization point for the external quasi-momentum modes $i$ and a nonvanishing external momenta.    
\par An arbitrary vertex part including composed operators which is multiplicatively renormalizable in a massless theory is defined by the 
following expression
\begin{equation}\label{109}
\Gamma_{R}^{(L,M)}(p_{n}, i_{n}, Q_{n'}, j_{n'}, g,0)= (Z_{\phi}^{(\tau)})^{\frac{M}{2}}(Z_{\phi^{2}}^{(\tau)})^{L} \Gamma^{(L,M)}(p_{n}, i_{n}, Q_{n'}, j_{n'}, \lambda_{0},0, \Lambda),
\end{equation} 
where the quantities $Z_{\phi}^{(\tau)}$ ,$Z_{\phi^{2}}^{(\tau)}$ are the renormalization functions which make the above defined 
renormalized vertex automatically finite (even when the cutoff $\Lambda$ goes to infinity). As discussed previously our perturbative 
analysis will utilize only dimensional regularization and forget about the cuttof from now on, since formally the divergences of the 
integrals manifest themselves as dimensional poles represented by negative powers of the parameter $\epsilon=4-d$. In order to prevent 
confusion, in the above expresssion the identification $\Gamma_{R}^{(2,0} \equiv \tilde{\Gamma}_{R}^{(2)}$ should be taken into account.
\par The external momentum scale where the theory is renormalized induces a flux in space parameter such that the renormalized vertex 
parts satisfy the renormalization group equation
\begin{eqnarray}\label{110}
&&\left(\kappa \frac{\partial}{\partial \kappa} + 
\beta(u) \frac{\partial}{\partial u}
- \frac{L}{2} \gamma_{\phi}^{(\tau)} + M \gamma_{\phi^{2}}^{(\tau)}\right)
\Gamma_{R}^{(L,M)} (p_{n}, i_{n}, Q_{m}, i'_{m}, u, 0)=  0,
\end{eqnarray} 
where $\beta^{(\tau)}(u)= -\epsilon \left(\frac{\partial lnu_{0}^{(\tau)}}{\partial u}\right)$, $\gamma_{\phi}^{(\tau)}(u) = \beta^{(\tau)} 
\left(\frac{\partial ln Z_{\phi}^{(\tau)}}{\partial u}\right)$ 
and $\gamma_{\phi^{2}}^{(\tau)} = \beta^{(\tau)} 
\left(\frac{\partial ln Z_{\phi^{2}}^{(\tau)}}{\partial u}\right)$. The 
combinations 
$\bar{Z}_{\phi^{2}}^{(\tau)}= Z_{\phi^{2}}^{(\tau)}Z_{\phi}^{(\tau)}$ and 
$\bar{\gamma}_{\phi^{2}}^{(\tau)} = \beta^{(\tau)} 
(\frac{\partial ln \bar{Z}_{\phi^{2}}^{(\tau)}}{\partial u})$ will be important in what follows. The renormalized theory is defined at fixed 
$L$. The reader is advised to consult Refs. \cite{amit,BL}.
\par We turn our attention to the dimensional crossover regime in the massless theory. We will analyze the one-loop four 
point contribution and investigate its properties in the limits $L \rightarrow \infty$ and $L \rightarrow 0$. The latter 
should indicate how small $L$ can be without affecting the validity of the $\epsilon$-expansion. 
\par The graph required involves the combination of two integrals which follow directly from our discussion in the massive case, namely
\begin{subequations}
\begin{eqnarray}\label{111}
&& I_{2}(k,i,\tilde{\sigma})= \tilde{\sigma} \overset{\infty}{\underset{l= -\infty}{\sum}} \int 
\frac{d^{d-1}q}{[q^{2} + \tilde{\sigma}^{2}l^{2}][(q+k)^{2}+\tilde{\sigma}^{2}(l+i)^{2}]}, \label{114a}\\
&& \tilde{I}_{2}(k,i,j,\tilde{\sigma})= \tilde{\sigma} \int \frac{d^{d-1}q}{[q^{2} + \tilde{\sigma}^{2}i^{2}][(q+k)^{2}+\tilde{\sigma}^{2}j^{2}]}.
\label{114b}
\end{eqnarray}
\end{subequations}
We begin with the calculation of the the integral $I_{2}(k,i,\tilde{\sigma})$. Utilize a Feynman parameter $x$. Next integrate over the 
momentum using Eq. (\ref{A2}) from Appendix A. We are left with 
\begin{eqnarray} \label{112}
&& I_{2}(k,i,\tilde{\sigma}) = \frac{\tilde{\sigma} S_{d-1} \Gamma(\frac{d-1}{2}) \Gamma(2 -\frac{(d-1)}{2})}{2} 
\overset{\infty}{\underset{l= -\infty}{\sum}} \int_{0}^{1} dx  \Bigl[x(1-x)(k^{2}+ \tilde{\sigma}^{2} i^{2}) \nonumber\\
&& \qquad + \quad \tilde{\sigma}^{2} (l+ix)^{2} \Bigr]^{\frac{(d-1)}{2}-2} .
\end{eqnarray}
Next perform the summation using the generalized thermal function as practiced before. Transforming $S_{d-1}$ into $S_{d}$ as previously 
prescribed, dividing the result by $S_{d}$ and performing the 
$\epsilon$-expansion of the Gamma functions, we find
\begin{eqnarray} \label{113}
I_{2}(k,i,\tilde{\sigma}) &=& \frac{1}{\epsilon} \Bigl(1 + \frac{\epsilon}{2} \Bigr) 
[\tilde{\sigma}^{2} i^{2} + k^{2}]^{-\frac{\epsilon}{2}} 
+ \frac{\tilde{\sigma}^{- \epsilon}}{2}  \int_{0}^{1}dx 
f_{\frac{1}{2}+\frac{\epsilon}{2}}\Bigl(ix,\sqrt{x(1-x)\Bigl[i^{2} + \frac{k^{2}}{\tilde{\sigma}^{2}}\Bigr]}\Bigr).
\end{eqnarray}
In the massless theory, we factor out the external momentum scale. We define 
$\hat{F}_{\alpha}(k,i;\tilde{\sigma})= \tilde{\sigma}^{-2 \alpha}\int_{0}^{1}dx 
f_{\frac{1}{2}+\frac{\epsilon}{2}}\Bigl(ix,\sqrt{x(1-x)\Bigl[i^{2} + \frac{k^{2}}{\tilde{\sigma}^{2}}\Bigr]}\Bigr)$, factorize 
the external momenta by defining the quantity $\hat{r}=\frac{\tilde{\sigma}}{\kappa}$ and identify 
$I_{2}(\kappa,i,\tilde{\sigma}) \equiv I_{2}(\kappa,i,\hat{r})$ (with similar identification for $\hat{F}_{\alpha}$) after this, which is the external momenta scale we need at the symmetry point. By expanding 
in $\epsilon$ gives the following result:
\begin{equation} \label{114}
I_{2}(\kappa,i,\hat{r}) = \kappa^{-\epsilon}\frac{1}{\epsilon}\Bigl[1 + \frac{\epsilon}{2} - \frac{\epsilon}{2}ln(1+\hat{r}^{2}i^{2}) 
+ \frac{\epsilon}{2}\hat{F}_{0}(\kappa,i;\hat{r})\Bigr].
\end{equation}
\par Following a similar trend it is a simple task to compute the nondiagonal integral whose solution can be written as
\begin{equation} \label{115}
\tilde{I}_{2}(\kappa,i,j,\hat{r})= \frac{\hat{r}\kappa^{-\epsilon}}{2}\int_{0}^{1}dx[x(1-x) + \hat{r}^{2}(i^{2}x+j^{2}(1-x))]^{-\frac{1}{2}}.
\end{equation}
In order to get the diagram computed at the symmetry point, we choose the external momentum and quasi-momentum at fixed nonvanishing 
values. We have already discuss the combinations of last integrals entering in the process of computing the required diagram. After using 
standard manipulations (see for instance Ref. \cite{GR}), implies the outcome 
\begin{subequations} \label{116}
\begin{eqnarray}
& \parbox{12mm}{\includegraphics[scale=0.4]{fig13DN.eps}}\quad\;= 3\tilde{\sigma}\frac{(N+8)}{9\epsilon}\kappa^{-\epsilon}\Bigl[1+\frac{\epsilon}{2} + \epsilon \hat{\zeta}^{(\tau)}(i;\hat{r})\Bigr], \label{116a}\\
& \hat{\zeta}^{(\tau)}(i;\hat{r})= -\frac{1}{6} ln(1+4\hat{r}^{2}i^{2}) + \frac{1}{6}\hat{F}_{0}(\kappa,2i,\hat{r}) 
+ \frac{1}{3}\hat{F}_{0}(\kappa,0,\hat{r}) 
+ \frac{\pi \hat{r}\tau}{3} \nonumber\\
& +\frac{2\hat{r}\tau}{3} (1+2\tau)\arcsin\Biggl[\frac{1}{\sqrt{1 + 4\hat{r}^{2} i^{2}}}\Biggr]. \label{116b}
\end{eqnarray}
\end{subequations}
The finite-size contribution function above has the same general structure as that from the massive theory, as it is going to 
be shown in a moment. Recall that the above coefficient $\tilde{\sigma}$ appears in the definition of the four-point renormalized vertex 
function. Just like in the massive approach, this factor will multiply the diagrams of this vertex function in arbitrary loop order and 
can be factored out from our asymptotic analysis.  
\par  Before focusing on the asymptotical values for the correction function, it is important to mention that the would be scaling 
variable in the massless theory $\frac{L}{\xi}$ goes to zero for finite values of $L$ as $\xi \equiv \infty$ in that case. So the cases 
that could be assessed here are given respectively by the regions $\frac{L}{\xi} \leq 1$. The limit $L \rightarrow \infty$ corresponds 
to $\frac{L}{\xi} \approx 1$, whereas small values for the distance between the boundary plates are characterized by the limit 
$\frac{L}{\xi} \rightarrow 0$. In the latter, the question again is how small $L$ can be such that the $\epsilon$-expansion is still valid.
\par We commence with the large $L$ limit, which is the same as $\hat{r} \rightarrow 0$. Since the linear terms in $\hat{r}$ goes to zero 
trivially, we restrict ourselves to the evaluation of the terms $\hat{F}_{0}(\kappa, 2i, \hat{r})$ and $\hat{F}_{0}(\kappa, 0, \hat{r})$. 
Using the definition along with the Boschi-Farina representation in terms of a summation of modified Bessel functions and taking that 
limit we find
\begin{eqnarray}\label{117}
&& \hat{F}_{0}(\kappa, 2i, \hat{r}) = 4 \sum_{n=1}^{\infty} \int_{0}^{1} dx cos(4 \pi n i x)K_{0}(2 \pi n \sqrt{x(1-x) 
[\hat{r}^{-2} + 4i^{2}]}) \rightarrow 4 \sum_{n=1}^{\infty} \int_{0}^{1} dx cos(4 \pi n i x)\nonumber\\
&& \;\;\times \;\; K_{0}(2 \pi n \sqrt{x(1-x)} \hat{r}^{-1}) < 4 \sum_{n=1}^{\infty} \int_{0}^{1} dx K_{0}(2 \pi n \sqrt{x(1-x)} \hat{r}^{-1}),  
\end{eqnarray}
where the latter is precisely  $\hat{F}_{0}(\kappa, 0, \hat{r})$, and we will show henceforth its convergence in that limit. Using the 
asymptotic values of the functions for large values of $z$, namely, $K_{0}(z)= \sqrt{\frac{\pi}{2z}} e^{-z}(1+ O(\frac{1}{z}))$, we have
\begin{eqnarray} \label{118}
&&  \hat{F}_{0}(\kappa, 0, \hat{r}) = 2\hat{r}^{\frac{1}{2}} \sum_{n=1}^{\infty} n^{-\frac{1}{2}} \int_{0}^{1} dx [x(1-x)]^{-\frac{1}{4}} 
exp(-n \hat{B}),
\end{eqnarray}
where $B=2 \pi \sqrt{x(1-x)} \hat{r}^{-1}$. If we interchange the summation and the integral, we have to compute the summation
$\sum_{n=1}^{\infty} n^{-\frac{1}{2}} exp(-n \hat{B}) < \sum_{n=1}^{\infty} exp(-n \hat{B}) = \frac{1}{exp(\hat{B})-1}= 
\frac{1}{\hat{B} + \sum_{n=1}^{\infty} \frac{\hat{B}^{n}}{n!}} < \hat{B}^{-1}$. Therefore,
\begin{eqnarray} \label{119}
&&  \hat{F}_{0}(\kappa, 0, \hat{r})< \hat{r}^{\frac{3}{2}} \frac{\Gamma(\frac{1}{4})^{2}}{\Gamma(\frac{1}{2})} \rightarrow 0.
\end{eqnarray}
 Last equation is the rigorous proof that the correction goes to zero and one recovers the bulk result.
\par Let us take the opposite limit $\hat{r} \rightarrow \infty$. We split the several pieces from 
which $\hat{\zeta}^{(\tau)}(i;\hat{r})$ is made of and take this limit on each of them. We can easily obtain in that 
limit the asymptotic behaviors: $-\frac{1}{6} ln(1+4\hat{r}^{2}i^{2}) \rightarrow - \frac{1}{3}ln(\hat{r}i)$; $\frac{2\hat{r}\tau}{3} 
(1+2 \tau) \arcsin{\Biggl[\frac{1}{\sqrt{1 + 4\hat{r}^{2} i^{2}}}\Biggr]} \rightarrow \frac{(\tau+2)}{3i}$. We are left with the nontrivial 
task of determining the terms proportional to $\hat{F}_{0}(\kappa, 2i, \hat{r})$ and  $\hat{F}_{0}(\kappa, 0, \hat{r})$. Using Eq. (\ref{84}), 
it is actually simple to prove that 
$\frac{1}{3}\hat{F}_{0}(\kappa, 0, \hat{r}) \rightarrow -\frac{2}{3} ln\hat{r} + \frac{\pi}{3} \hat{r}$. 
\par Performing the identifications $\hat{x}= 2 \pi \sqrt{x(1-x)[\hat{r}^{-2} + 4i^{2}]}$ and $t=\frac{2ix}{\sqrt{x(1-x)[\hat{r}^{-2} + 4i^{2}]}}$ into Eq. (\ref{81}) leads to the following expression 
\begin{eqnarray} \label{120}
&& \hat{F}_{0}(\kappa, 2i, \hat{r}) = 2 \gamma + 2 \int_{0}^{1} dx ln\Biggl[\frac{\sqrt{x(1-x)[\hat{r}^{-2} + 4i^{2}]}}{2}\Biggr] 
+ \int_{0}^{1} dx \frac{1}{\sqrt{x(1-x) \hat{r}^{-2} + 4i^{2} x}} \nonumber\\ 
&& +   \sum_{n=1}^{\infty} \int_{0}^{1} dx \Biggl(\frac{1}{\sqrt{\hat{r}^{-2} x(1-x) + 4i(i-n)x + n^{2}}} - \frac{1}{n} \Biggr) \nonumber\\
&&  +   \sum_{n=1}^{\infty} \int_{0}^{1} dx \Biggl(\frac{1}{\sqrt{\hat{r}^{-2} x(1-x) + 4i(i+n)x + n^{2}}} - \frac{1}{n} \Biggr).
\end{eqnarray} 
In this expression we can set $\hat{r}^{-1} \rightarrow 0$. Although the dimensional crossover regime has to do with the behavior 
of large $\hat{r}$ (small values of $L$) we choose to keep the explicit dependence on $i$ for the asymptotic behavior and neglect the other 
constants. The second and third terms are trivial in this limit, and yield the contributions $2ln2i$ and $\frac{1}{i}$, respectively. The 
two last terms involving the infinite summation and the parametric integral are identical to those already worked out in the massive theory. In this limit they produce the quantity $2 \overset{\infty}{\underset{n=1}{\sum}} i^{2n}\zeta(2n+1)$. Defining again the regularized correction for 
the massless theory 
$:\hat{\zeta}(i;\hat{r}): =  \hat{\zeta}(i;\hat{r}) - \frac{1}{3} \overset{\infty}{\underset{n=1}{\sum}} i^{2n}\zeta(2n+1)$ which is identical to that 
for small values of $L$ in the massive theory. Altogether, the regularized finite-size correction 
for small values of $L$ ($\hat{r} \rightarrow \infty$) in the massless case reads
\begin{eqnarray}\label{121}
&& :\hat{\zeta}^{(\tau)}(i;\hat{r}): \rightarrow -ln \hat{r} + \frac{\pi \hat{r}(1+\tau)}{3} + \frac{(2\tau+5)}{6i} .
\end{eqnarray}
It is simple to see that the finiteness correction is compatible in massive and massless cases. Notice the absence of terms proportional to $lni$ 
which got cancelled along the way in {\it both massless and massive theories}. Since the presence of this sort of term in conventional field 
theory defined for infinite systems implies that all lenght scales are coupled, the absence of them in the mode of external quasi-momentum 
$i$ shows that there is nothing fundamental on the sector of the finite-size correction which depends on $i$. In other words the layered geometry 
do not themselves interact strongly up to all length scales, from one plate to the total number of them, since this information is encoded in 
the logarithm which is absent in the correction. The apperance of the relevant lenght scale $L$ marks a profound difference with the bulk 
critical behavior even though the expoents are the same. It is a good explanation why surface effects are not important in this regime.  
\par If we choose the Neumann boundary condition $\tau=1$, we find the asymptotic behavior
\begin{equation}\label{122}
:\hat{\zeta}^{(\tau)}(i;\hat{r}): \rightarrow -ln\hat{r} + \frac{2 \pi \hat{r}}{3},   
\end{equation}
and the leading divergence on $\hat{r}$ is linear as happens with $PBC$. This is in agreement with the massive case.
\par On the other hand, if we take Dirichlet boundary condition we have instead
\begin{equation} \label{123}
:\hat{\zeta}^{(\tau)}(i;\hat{r}): \rightarrow -ln\hat{r},   
\end{equation}
and the divergence is purely logarithmic as that for $ABC$, which is just compatible with the massive case as well. If $L$ is too small, 
those singularities take over the poles in $\epsilon$ of the one-loop four-point function therefore invalidating the $\epsilon$-expansion 
results. This is the meaning of the dimensional crossover regime, where finite-size correction effects get bigger than the leading 
singularities in $\epsilon$ as already discussed for $PBC$ and $ABC$ in Ref. \cite{BL}.
\par The reader might have noticed that, except for the $i$-dependent terms appearing in the finite-size correction function in the 
$L \rightarrow 0$ limit, the rigorous arguments used to determine the correction function for $PBC$ and $ABC$ carries over for $DBC$ 
and $NBC$. Since we are neglecting the dimensional crossover regime, that is the main 
reason we refrain from writing another appendix with the derivation of the results for the massless integrals. 
\par With the resources furnished in the present paper in conjumination with those details worked out in Ref. \cite{BL} for $ABC$ and $PBC$, 
the other two-loop  diagrams of the four-point function at the symmetry point can be shown to be given by
\begin{subequations} \label{124}
\begin{eqnarray}
&& \parbox{10mm}{\includegraphics[scale=0.3]{fig14DN.eps}}\qquad\quad = 3\tilde{\sigma} \frac{(N^{2}+6N+20)}{27} \kappa^{-2\epsilon}\Bigl\{\frac{1}{\epsilon^{2}}\Bigl(1+ \epsilon 
+ 2\epsilon \hat{\zeta}^{(\tau)}(i,\hat{r})\Bigr)\Bigr\}, \label{124a}\\
&& \parbox{10mm}{\includegraphics[scale=0.4]{fig15DN.eps}}\quad = 3\tilde{\sigma} \frac{(5N+22)}{27} \kappa^{-2\epsilon}\Bigl\{\frac{1}{2\epsilon^{2}}\Bigl(1+\frac{3 \epsilon}{2} + 2\epsilon 
\hat{\zeta}^{(\tau)}(i,\hat{r})\Bigr)\Bigr\}. \label{124b}
\end{eqnarray}  
\end{subequations}
Similarly, the derivative of the two- and three-loop diagrams (with respect to $k^{2}$ at the symmetry point) of the two-point vertex part can 
be computed similarly. From our discussion so far, 
it is not difficult to find out the following (divergent) expressions 
\begin{subequations} \label{125}
\begin{eqnarray}
&& \frac{\partial}{\partial k^{2}}\left(\parbox{10mm}{\includegraphics[scale=0.3]{fig16DN.eps}}\quad\right)_{k^{2}=0}\quad =-\frac{\kappa^{-2\epsilon} (N+2)}{24\epsilon}\Bigl[1+\frac{5\epsilon}{4} - 2 \epsilon \hat{W}^{(\tau)}(i,\hat{r})\Bigr], \label{125a}\\
&& \frac{\partial}{\partial k^{2}}\left(\parbox{10mm}{\includegraphics[scale=0.3]{fig17DN.eps}}\quad\right)_{k^{2}=0}\quad =-\frac{\kappa^{-3\epsilon}(N+2)(N+8)}{162 \epsilon^{2}}\Bigl[1+ 2\epsilon - 3 \epsilon \hat{W}^{(\tau)}(i;\hat{r})\Bigr]. \label{125b}
\end{eqnarray}
\end{subequations} 
Here the finite-size correction for the two-point function as well as the quantities which define it can be written in the form
\begin{subequations} \label{126}
\begin{eqnarray}
&& \hat{W}^{(\tau)}(i;\hat{r})= \frac{1}{2}ln[1+\hat{r}^{2}i^{2}] + 2\hat{F'}_{0}(\kappa,i;\hat{r}) - \hat{\bar{F}}_{0}(\kappa,i;\hat{r}) 
- 3 \hat{r}[\hat{H}_{0}(\kappa,i,0;\hat{r}) \nonumber\\
&& + \tau \hat{H}_{0}(\kappa,0,i;\hat{r})] + 6\hat{r}[\hat{\mathcal{H}}_{0}(\kappa,i,0;\hat{r}) 
+ \tau \hat{\mathcal{H}}_{0}(\kappa,0,i;\hat{r})], \label{126a}\\
&&\hat{F}_{\alpha}(\kappa,i,\hat{r})=\frac{\tilde{\sigma}^{-2 \alpha}}{S_{d}} \int_{0}^{1} f_{\frac{1}{2}+\alpha}(ix, 
\sqrt{x(1-x)[\hat{r}^{-2} + i^{2}]}), \label{126b}\\ 
&&\hat{F}_{\alpha,\beta}(k,i;\tilde{\sigma})= \frac{\tilde{\sigma}}{S_{d}} \sum_{j=-\infty}^{\infty} \int d^{d-1}q 
\frac{\hat{F}_{\alpha}(q+k,j+i,\tilde{\sigma})}
{[q^{2} + \tilde{\sigma}^{2}j^{2}]^{\beta}}, \label{126c}\\
&& \hat{\bar{F}}_{\alpha}(\kappa,i;\hat{r})= \tilde{\sigma}^{-2 \alpha} \int_{0}^{1} dx x^{\frac{\alpha}{2}} (1-x) f_{\frac{1}{2}+\alpha}(ix, 
\sqrt{x(1-x)[\hat{r}^{-2} + i^{2}]}), \label{126d}\\
&& \hat{H}_{0}(\kappa,i,j;\hat{r})= \int_{0}^{1} dx (1-x)[x(1-x) + \hat{r}^{2}(i^{2}(1-x) + j^{2} x)^{-\frac{1}{2}}], \label{126e}
\end{eqnarray}
\end{subequations}
where $\hat{F'}_{\alpha}(\kappa,i;\hat{r})= \frac{\partial \hat{F}_{\alpha,1}(k,i;\hat{r})}{\partial k^{2}}\Bigl|_{k^{2}=\kappa^{2}}$.The other amounts appearing in the equation for the correction are defined by
\begin{subequations} \label{127}
\begin{eqnarray}
&& \hat{\mathcal{H}}_{0}(\kappa,i,j;\hat{r})= \kappa^{2 \epsilon} \hat{\mathcal{F}}_{0}^{\prime (\tau)}(\kappa,i,j;\hat{r}), \label{127a}\\
&& \hat{\mathcal{F}}_{\alpha,\beta}^{(\tau)}(k,i,j;\tilde{\sigma}) \equiv \frac{1}{S_{d}}\tilde{\sigma}  
\int d^{d-1}q \frac{\hat{F}_{\alpha}^{(\tau)}(q+k,j;\tilde{\sigma})}{[q^{2} + \tilde{\sigma}^{2}i^{2}]^{\beta}}, \label{127b}\\
&& \hat{\mathcal{F}}_{\alpha}^{\prime (\tau)}(\kappa,i,j;\hat{r}) \equiv \frac{\partial \hat{\mathcal{F}}_{\alpha,1}^{(\tau)}(k,i,j;\hat{r})}
{\partial k^{2}}\big|_{k^{2}=\kappa^{2}}. \label{127c}
\end{eqnarray}
\end{subequations}
\par We have all elements to compute the critical exponents using the massless fields framework. As we are going to show in the next section, 
all dependence on the finite-size corrections for the several vertex parts disappears in the Wilson functions at the nontrivial attractive 
infrared fixed point. The main consequence of this is that in the finite-size regime the critical exponents are boundary condition 
independent.
\section{Critical exponents in the massless sector}      
\par We are going to be rather brief in the present section. We shall focus solely on the computation of critical exponents using 
normalization conditions, albeit the results in the present work should be worthwhile in the evaluation of critical exponents using 
minimal subtraction. The algorithm was already set in the massive case. All we have to do is to replace the 
massive diagrams by its massless counterpart. It is important to realize that in the perturbative expansion of the primitively bare vertex 
parts, the bare coupling constant is defined in terms of the bare dimensionless coupling constant through 
$\lambda_{0}=u_{0} \kappa^{\epsilon}$. This implies that for arbitrary fixed $\kappa$, the perturbative expansion consisting of the 
multiplication of suitable powers of $\lambda_{0}$ by the diagram eliminate all the $\kappa$ prefactors appearing in each diagram as just 
discussed in the previous section. 
\par Firstly, use the expansion of $u_{0}^{(\tau)}$, $Z_{\phi}^{(\tau)}$ and $\bar{Z}_{\phi^{2}}$ in powers of $u$ in exactly the same form as we did 
before in the massive case. Secondly, employ the analogue of Eq. (\ref{95}) recalling that now the symmetry point occurs for nonzero momentum and 
the diagrams correspond to massless fields. We refrain from giving the massless version of Eq. (\ref{96}), rather we implicitly plug them in the 
definition of the primitively divergent massless vertex parts. We then obtain the following solution for the renormalization functions 
(i.e., $u_{0}^{(\tau)}$, $Z_{\phi}^{(\tau)}$ and $\bar{Z}_{\phi^{2}}$) coefficients 
\begin{subequations} \label{128}
\begin{eqnarray}
&& a_{1}^{(\tau)}= \frac{(N+8)}{6\epsilon}\Biggl[1 + \frac{\epsilon}{2} + \epsilon \hat{\zeta}^{(\tau)}(i;\hat{r})\Biggr], \label{128a}\\
&& a_{2}^{(\tau)}= \Biggl[\frac{(N+8)}{6\epsilon}\Biggr]^2\Bigl[1+ 2\epsilon\hat{\zeta}^{(\tau)}(i;\hat{r})\Bigr] 
+ \frac{2N^{2}+23N +86}{72\epsilon}, \label{128b}\\
&& b_{2}^{(\tau)}= -\frac{(N+2)}{144 \epsilon} \Biggl[1+\frac{5 \epsilon}{4} - 2 \epsilon \hat{W}^{(\tau)}(i;\hat{r})\Biggr], \label{128c}\\
&& b_{3}^{(\tau)} =  -\frac{(N+2)(N+8)}{1296 \epsilon^{2}} \Biggl[1 + \frac{5 \epsilon}{4} + 3 \hat{\zeta}^{(\tau)}(i;\hat{r})\Biggr], \label{128d}\\
&& c_{1}^{(\tau)} = \frac{(N+2)}{6\epsilon}\Biggl[1 + \frac{\epsilon}{2} + \epsilon \hat{\zeta}^{(\tau)}(i;\hat{r})\Biggr], \label{128e}\\
&& c_{2}^{(\tau)}= \frac{(N+2)(N+5)}{36 \epsilon^{2}} + \frac{2N^{2}+11N+14}{72 \epsilon} + \frac{N^{2}+7N+10}{18 \epsilon} 
\hat{\zeta}^{(\tau)}(i;\hat{r}). \label{128f}
\end{eqnarray}
\end{subequations}
\par We are going to use this information in the calculation of the Wilson functions. We just have to plug these coefficients into 
Eqs. (\ref{101}). The nontrivial infrared fixed point value of the coupling constant $u^{*}$ is obtained from the condition 
$\beta^{(\tau)}(u^{*})=0$. It is given by
\begin{equation}\label{129}
u^{*} = \Bigl[\frac{6\epsilon}{N+8}\Bigr] \left\{1 + \Biggl[\frac{(9N+42)}{(N+8)^{2}} -\frac{1}{2} 
- \hat{\zeta}^{(\tau)}(i;\hat{r})\Biggr]\epsilon \right\}.
\end{equation}
\par Upon substitution of the fixed point into $\gamma_{\phi}(u)$ one obtains the exponent $\eta$ (the anomalous dimension of the field) 
up to three-loop order, namely
\begin{equation}\label{130}
\eta \equiv  \gamma_{\phi}(u^{*})= \frac{(N+2)}{2(N+8)^{2}}\epsilon^{2} \Bigl\{1 + \epsilon \Biggl[\frac{6(3N+14)}{(N+8)^{2}} -\frac{1}{4}\Biggr]\Bigr\}.
\end{equation}    
\par Finally, using the relation $\nu^{-1}=2 -\bar{\gamma}_{\phi^{2}}^{(\tau)}(u^{*}) - \eta$, leads us to the result
\begin{equation}\label{131}
\nu= \frac{1}{2} + \frac{(N+2)}{4(N+8)}\epsilon + \frac{(N+2)(N^{2}+23N+60)}{8(N+8)^{3}}\epsilon^{2}.
\end{equation} 
\par In spite of being quite different as far as the $\epsilon$-expansion of the diagrams are concerned, the massless and massive settings 
yield the same critical exponents as expected from the universality hypothesis. Ultraviolet and infrared analysis are completely equivalent 
in the evaluation of critical exponents, which reproduce those from the bulk as in $PBC$ and $ABC$ conditions. 
\par What is really appealing in this new formulation inspired in the Nemirovsky and Freed construction for massive fields is that no 
surface fields are required to implement successfully the finite-size renormalization program for $DBC$ and $PBC$.  
\section{Discussion}
\par Formerly, there was a certain misunderstanding by some authors concerning the phenomenological finite-size scaling hypothesis. It was thought 
that the variable $\frac{L}{\xi}$ governed the approach to 
bulk criticality: whenever $\frac{L}{\xi} \rightarrow \infty$ the $\epsilon$-expansion results shown for the above diagram could be 
trusted, whereas $\frac{L}{\xi} \leq 1$ would be the regime in which perturbation theory could no longer be reliable. In the modern approach to 
finite-size criticality proposed in the present work, this scaling variable looses its meaning 
as discussed in the  $ABC$ and $PBC$ cases. In fact, we are going to show next that even in the massive case for $DBC$ and $NBC$, 
the regions $\frac{L}{\xi} \sim 1$ and $\frac{L}{\xi} < 1$  are also available to scrutiny within the present technique.
\par How small $\frac{L}{\xi}$ (or how large $\tilde{r}$) can be without invalidating the $\epsilon$-expansion results? Setting $\mu=\xi^{-1}$, 
the variable can be rewritten as $\tilde{r}= \frac{\pi \xi}{L}$, so that the limit $L \rightarrow 0$ corresponds to 
$\tilde{r} \rightarrow \infty$. Before we start our discussion, identify the dimensional single pole $(\propto \epsilon^{-1})$ with the 
$ln \bigl(\frac{\Lambda}{\mu} \bigr) \equiv ln(\Lambda \xi)$ in a theory regularized with a cutoff, which will be helpful to us in what follows. 
\par The dimensional crossover condition for $NBC$ implies that the linear term in $\tilde{r}$ is greater or equal to that coming from the pole 
in $\epsilon$. (We can extract further information by identifying the ultraviolet cutoff with the inverse of the 
lattice constant $a$ as $\Lambda \sim \frac{1}{a}$.) In other words, $L \leq \frac{2 \pi \xi}{3 ln(\frac{\xi}{a})}$. 
\par Away from the critical temperature $\xi$ is finite and typically a multiple of the lattice parameter. Take $\xi=10a$. In that case,  
$L \leq 9.1 a$. There is a narrow region $9.1a <L <10a$ where $\frac{L}{\xi} <1$ and the perturbative expansion is still valid. The number of 
parallel plates is $n= \frac{L}{a} + 1$, where $n$ is an integer. Then, for $n \leq 10.1$ the $\epsilon$-expansion results break down. Since 
$n$ has to be integer, this condition implies that for $n \leq 11$ the perturbative expansion is invalid. However, it is valid (at least 
formally) for $n>12$. Hereafter, anytime we mention the region of validity of the $\epsilon$-expansion for $L$ in a certain range, it is 
implicit that this variation interval on $L$ obeys the criterion $\frac{L}{\xi} <1$. Now take $\xi=100a$, which implies $L \leq 45.5a$, $n \leq 47$. There is a wider region $45.5a< L < 100a$ which does not invalidate perturbation theory. For $\xi=10^{3}a$, the breaking 
conditions are $L \leq 303.2a$ (or $n \leq 305$) and the window of validity of perturbation theory is contained in the range $303.2a < L < 10^3a$.
By allowing $\xi=10^{4}a$ leads to the condition(s) $L \leq 2274a$(, $n \leq 2275$) and a wider range ($2274a < L < 10000a$) exists which does 
not invalidate perturbation theory, and so on.   
\par In $PBC$ from Ref. \cite{BL}, the linear term was equal 
to $\frac{r}{2}$, where $r=\frac{2 \pi \xi}{L}$. Using exactly the same argument as above in $NBC$, the dimensional crossover condition reads 
now  $L \leq \frac{\pi \xi}{ln(\frac{\xi}{a})}$. When $\xi=10a$ implies $L \leq 13.64 a$ and $n \leq 15$. For that value of the correlation 
length, perturbation theory is only valid for $\frac{L}{\xi} >1$. The choice $\xi=100a$ turns out to produce the breaking conditions 
$L \leq 68.21 a$ and $n\leq 70$, with a region $68.21 a < L < 100a$ where $\frac{L}{\xi} <1$ does not invalidate $\epsilon$-expansion 
results. Taking $\xi=10^{3}a$, perturbation theory is invalid for  $L \leq 454.8a$ or $n \leq 456$. The window of validity 
is $454.8a < L < 1000a$. Finally, the choice $\xi=10^{4}a$ leads to the dimensional 
crossover conditions $L \leq 3410.9a$ and $n \leq 3412$, with a wider region of validity, namely, $3410.9 a < L < 10000a$. Comparing with $NBC$ 
results, the bulk critical behavior requires a smaller number of plates for $NBC$ than in $PBC$ and a thinner film geometry. Since fewer plates 
are required in $NBC$ to keep the bulk critical behavior, from the energetic viewpoint it is easier to provoke a bulk-surface transition in 
$NBC$ than in $PBC$. Therefore, our approach explains why nature chooses $NBC$ as the prototype for the bulk-surface transition, since this 
information is encoded in the finite-size correction to the bulk behavior.
\par We turn our attention to the dimensional crossover regime for $DBC$. The 
logarithmic divergence in $\tilde{r}\rightarrow \infty$ in $DBC$ can be understood similarly: regularizing the theory with a cutoff 
$\Lambda$ such that $\frac{1}{\epsilon} \sim ln\Bigl(\frac{\Lambda}{\mu}\Bigr)$. The correction will become as big as 
the dimensional pole whenever $L \leq \pi a \sim 3.14 a$, which is independent of the bulk correlation length $\xi$. Using the same line 
of thought for $ABC$ yields $L \leq 2 \pi a \sim 6.28a$ for the collapse of the bulk critical behavior description in perturbation theory. Both 
results are independent of the (fixed, finite) bulk correlation length $\xi$. For different fixed 
bulk correlation lengths one needs $n \leq 5$ ($n \leq 8$ )for $DBC$ ($ABC$) by demanding integer values for $n$. For $n > 6$ ($n > 9$) 
the bulk critical behavior dominates the finite-size corrections and there is a real experimental possibility of construction of nanodevices for 
$DBC$ ($ABC$). 
\par For instance, by slicing a thick material presenting bulk critical behavior in thin films satisfying the above conditions, one 
could vary the temperature to different values from the bulk critical temperature (variation of the correlation length) and use neutron 
scattering experiments to see whether the bulk behavior varies for large enough values of $n$. If it does not, the boundary conditions on the 
limiting plates are either $DBC$ or $ABC$ and we have a practical way to determine the boundary conditions in a certain material. If it does, 
the boundary condition should be either $NBC$ or $PBC$ and the number $n$ should be much bigger than in the previous situation.   
\par The description of the dimensional crossover regime in the massless case is also worthwhile. It is important to mention that finite 
values of $L$ already implies $\frac{L}{\xi} \rightarrow 0$ since $\xi=\infty$ in the massless (critical temperature) formulation. The 
important 
quantity here is the external momenta scale which generates the renormalization flow, although it has no obvious physical interpretation like 
the connection of the mass with the correlation length. Nevertheless, we can vary it and see what happens with the correction. 
\par For $NBC$, the linear term in the massless case is $\frac{2 \pi \hat{r}}{3}$, where $\hat{r}=\frac{\tilde{\sigma}}{\kappa}$ 
($\sigma = \frac{\pi}{L}$) as can be verified from Eq. (\ref{121}). What is left here is to perform the variation of 
$\kappa$ in units of the inverse multiple of the lattice parameter. Following the same trend we find the critical value $L \leq\frac{2 \pi^{2}}
{3 \kappa ln\bigl(\frac{1}{\kappa a}\bigr)}$. For instance, we can set $\kappa^{-1}=10a$, $\kappa^{-1}=100a$, 
$\kappa^{-1}=10^{3}a$, and $\kappa^{-1}=10^{4}a$. Obviously, the picture does not change. All the critical values of $L$ below which the system 
undergoes the dimensional crossover gets a factor of $\pi$ with respect to those from the massive theory. For instance, $\kappa^{-1}=10a$ 
implies that $L \leq 28.6$ ($n \leq 30$); for  $\kappa^{-1}=100a$, one has $L \leq 142.9$ ($n \leq 144$); when $\kappa^{-1}=10^{3}a$, we obtain 
$L \leq 952.5$ ($n \leq 954$); etc.     
\par For $PBC$, a comparison involves the correction to the four-point function which is given by the second, third, fourth and fifth terms 
from Eq. (61) from Ref. \cite{BL} by taking $\tau=0$. Of course, we have to reconstruct that expression (taken at $\kappa=1$) by multiplying all 
terms proportional to $L$ by $\kappa$. Then, the second term gives the logarithmic divergence on $r=\frac{\sigma}{\kappa}$ 
($\sigma = \frac{2 \pi}{L}$) while the third gives the linear divergence on $r$ and the other terms are not important in what follows. Altogether, 
the aforementioned correction for small values of $L$ has the linear divergence as the dominant term which invalidates perturbation expansion 
whenever $L \leq\frac{\pi^{2}}{\kappa ln\bigl(\frac{1}{\kappa a}\bigr)}$. It is not necessary to go into the details, to see that several values 
of $\kappa$ in the same range as before produce critical values of $L$ which are bigger by a factor of $\pi$ in comparison with the analogous 
situation in the massive theory. The reader can check that the same feature takes place to $NBC$ when compared to the massless case. This should 
be not surprising since the fluctuations are enhanced in the massless regime which requires thicker slabs to guarantee the bulk criticality.
\par Perhaps the most interesting result regards $DBC$ and $ABC$ which {\it have the same critical value for} $L$ {\it in both massive and massless regimes}. This is so because in the massless and massive regimes the logarithm term for the variables in both situations does not change its 
coefficients. The full unveiling of this invariance remains to be investigated.
\par In closing this discussion, if these inequalities are not satisfied, the new situation correspond to new critical behavior not belonging to 
the usual bulk universality class, since the smallness of this variable disturbs the 
system in an uncontrolable fashion driving it to a ``dimensional crossover''. Note, however, that the regions 
$\frac{L}{\xi}\sim 1$ and $\frac{L}{\xi}<1$ are away from the dimensional crossover regime since they are not too close 
to zero. They can be safely described within the present modern finite-size scaling approach.

\section{Concluding Remarks}
\par In this work we have developed a momentum space method to calculate critical exponents for a critical system whose order parameter 
is defined on a layered geometry with Dirichlet and Neumann boundary conditions imposed on the limiting (parallel) plates at $z=0,L$ with 
$z$ characterizing the finite-size direction perpendicular to the $(d-1)$-dimensional (hyper)planes of infinite extent along all its linear 
dimensions. The main features introduced are: {\it i)} the tensors needed to construct arbitrary loop diagrams for $1PI$ vertex parts through 
the utilization of the exponential representation for the basis eigenfunctions with or 
without insertion of composite operators which can be renormalized multiplicatively; {\it ii)} the representation of the finite-size 
correction written in terms of a sum of Bessel functions instead of that involving integrals without elementary primitive; {\it iii)} 
massless framework for the computation of critical indices.
\par There are many similarities with the $PBC$ and $ABC$ treated recently \cite{BL}. For instance, the unification of the results for the 
boundary conditions, except that for $DBC$ this only occurs as long as the external quasi-momentum are set to nonzero values. (For $PBC$ and $ABC$ this unification was obtained using 
vanishing external quasi-momentum.) This restriction causes no loss of generality, since it 
is possible to show that $NBC$ can also be formulated with zero external quasi-momenta and the results do not change. We leave this topic 
for a future work. The region of validity of the $\epsilon$-expansion is consistent either in the massless or in the massive formalism. In 
that region finite-size scaling is valid, the dominant critical behavior is the one from the bulk with respect to the critical exponents 
even though the finite-size corrections might appear in other universal quantities, like amplitude ratios \cite{Leiteetal1,Leiteetal2}. From 
the dimensional crossover regime viewpoint where the $\epsilon$-expansion results are no longer applicable when $L$ gets smaller and 
smaller, $DBC$ and $ABC$ diverge logarithmically with $L$ whereas $NBC$ and $PBC$ diverge like $L^{-1}$. This confirms a previous claim by 
Nemirovsky and Freed in their Green's function framework for the massive case. 
\par When the $\epsilon$-expansion ceases to be valid, those 
finite-size corrections become bigger than the poles in $\epsilon$ (representing logarithmic divergences of the bulk theory). In that 
case, the correction for $NBC$  $\propto L^{-1}$ in the limit $L \rightarrow 0$ dominates over the term $lnL$ 
(which is one piece of the total contribution). In the present framework, the onset of the dimensional crossover is directly connected to 
surface effects but not in the way previously imagined, {\it i.e.}, the passage from $(N,d)$ to the $(N,d-1)$ universality class: the 
interaction between the plates provoke a more complicated effect transliterated in those $L$-dependent terms. In $NBC$ the dimensional 
crossover regime can be identified with the bulk-surface transition region although the identification 
is not complete since a proper description of this region should probably include external surface fields as well \cite{Di,Nami1,Nami2}. 
In other words, the dimensional crossover regime marks the transmutation of finite-size contributions into surface effects.
\par From the technical point of view, there are other similarities among ($DBC$, $NBC$) and ($ABC$, $PBC$). Our utilization of the basis 
functions in terms of the exponentials for ($DBC$, $NBC$) makes it possible to use results from ($ABC$, $PBC$) analysis like performing 
summations in the range ($-\infty,\infty$), since the functional form of these results are invariant in spite of minor 
modifications in the finite size parameter $r$ ($\tilde{r},\hat{r}$, etc.). Besides, the terms which have been discarded along the way 
for being more regular than those kept in all stages of the process revealed themselves identical in their functional form for all of these 
boundary conditions, with differences obviously in the massless and massive approaches. 
\par But there are also totally different aspects when comparing ($DBC$, $NBC$) and ($ABC$, $PBC$). The latter are boundary conditions 
whose terms do not break translation invariance. The former actually do have translation invariance breaking terms. Previously, the common 
belief about translation invariance breaking was necessarily attached to surface contributions. These could stem from two origins, 
namely, if either surface fields {\it i)} are not allowed to begin with and these contributions are corrections to finite-size effects or 
{\it ii)} are permitted and the subject goes beyond the finite-size problem itself \cite{B,Di}. According to our study here the description of 
finite-size systems subject to $DBC$ or $NBC$ can be understood entirely out of fields within the volume between the plates without necessity 
of referring to surface fields. Plus, the breaking of translation invariance in our treatment has nothing to do with surface effects, provided 
the critical system is kept away from the dimensional crossover regime. This is highly desirable in comparison with the deconfined situation: 
the approach to bulk criticality in the region $0 < \frac{L}{\xi} \leq \infty$ clearly leads to a novel modern finite-size scaling paradigm 
not arising from externally imposed surface fields. The structure unveiled in the present work demonstrates that translation invariance 
breaking comes from the ``nondiagonal terms''. These terms actually make it difficult to get rid of mass insertions in order to have a smaller 
number of diagrams in the computation of critical exponents, for instance. Our prescription to eliminate the mass insertion graphs without 
invoking external surface fields or any sort of fields other than that representing the confined bulk system might be worthy to take the 
subject to another level of understanding. 
\par One interesting aspect is to pursue other renormalization schemes like minimal subtraction for massless and massive fields for the 
boundary conditions just presented. The discussion for massless fields subject to $ABC$ and $PBC$ already appeared in Ref. \cite{BL}. A 
thorough discussion for all boundary conditions utilizing massless and massive fields is still lacking. It would be nice to see whether 
setting the external quasi-momenta at fixed values should be sufficient to renormalize the massive theory \cite{CLJMP}. 
\par It is tempting to employ the machinery just developed to treat the case of anisotropic $m$-fold Lifshitz type competing 
systems \cite{L1,L2}. A simple follow-up idea is to take the finite direction perpendicular to the $m$(-dimensional subspace) competing axes 
and investigate whether the critical exponents are affected by the introduction of this new ingredient. Amplitude ratios such as that from 
the susceptibility (see Ref. \cite{L3} for the $m=1$ case) and specific heat \cite{L4} could also be tackled within the technique introduced 
here in order to figure out whether the finiteness modify them in comparison with those from the bulk. To extend this topic to its full 
generality, we could use it in the investigation of finite-size corrections of arbitrary anisotropic competing systems \cite{L5,L6} and take 
the finite-size direction perpendicular to all types of competing axes inherent to the problem. The treatment of amplitude ratios for 
generic competing systems, for instance, those discussed in Refs. \cite{L7,L8} poses no obstacle in principle and can be investigated too.   
\par There are many applications in the critical phenomena context which could be unraveled 
utilizing the results contained in our present results. For instance, further understanding of this topic would lead to the fabrication of new 
devices involving thin films of materials displaying bulk critical behavior. The importance of our work in guiding this enterprise to 
experimentalists is the indication of how small $L$ can be without spoiling the bulk critical properties, which is certainly important in the 
nanotechnology scenario. 
\par Since we have shown that the critical exponents for a noncompeting system in a parallel plate geometry with one direction of finite 
extent is identical of those from the infinite system we could try to find an exact perturbative analytical solution to the uniaxial 
($m=1$) Lifshitz critical exponents. In fact, if we allow the finite-size to be along the competing axis the problematic quartic integral 
would get transformed into a summation which is easier to perform at least in principle. Taking this achievement for simple enough boundary 
conditions like $PBC$ would automatically yield the critical exponents of the bulk systems for the uniaxial case. Can this be possible? The 
results just obtained will help to make sure whether this idea is feasible in the near future.

\section{Acknowledgments}
MVSS would like to thank CAPES grant number 76640 and CNPq grant number 141912/2012-0 for financial support. JBSJ acknowledges support 
from CNPq, grant number 142220/2007-8. MML thanks CNPq grant number 232352/2014-3 for partial financial support.

\appendix
\section{Massive integrals in dimensional 
regularization}
\par In the text we have omitted some steps concerning the one-loop integral $I_{2}(k,i,\tilde{r})$ in order to get Eq. (\ref{78}) 
due to its similarity with $PBC$ and $ABC$ integrals already worked out in Ref. \cite{BL}. We commence by deriving explicitly Eq. (\ref{78})
along with the nondiagonal integral $\tilde{I}_{2}(k,i,j,\tilde{\sigma},\mu)$. We shall plug the renormalized mass $\mu$ in all diagrams to be 
discussed for the reasons discussed in the main text.
\par Let us get started with Eq. (\ref{26a}). We factor out the renormalized mass $\mu$ and utilize utilize a Feynman 
parameter to obtain
\begin{eqnarray}\label{A1}
&& I_{2}^{(\tau)} (k, i; \tilde{\sigma}, \mu) = \tilde{r} \mu^{-\epsilon} 
\sum_{j=-\infty}^{\infty} \int_{0}^{1} dx \int d^{d-1}p \nonumber\\
&& \;\; \times\;\;\frac{1}{\Bigl[p^{2} + 2xkp + xk^{2} 
+ \tilde{r}^{2}[(j+ ix)^{2} + x(1-x)i^{2}] + 1 \Bigr]^{2}}.
\end{eqnarray}
Dimensional regularization is usually expressed in the Feynman's integrals evaluations through the identity   
(see Ref.\cite{amit}) 
\begin{equation}\label{A2}
\int \frac {d^{d}q}{(q^{2} + 2 k.q + \mu^{2})^{\alpha}} =
\frac{1}{2} \frac{\Gamma(\frac{d}{2}) \Gamma(\alpha - \frac{d}{2}) (\mu^{2} - k^{2})^{\frac{d}{2} - \alpha}}{\Gamma(\alpha)} 
S_{d},
\end{equation} 
where $S_{d}$ is the area of the $d$-dimensional unit sphere. Replacing this into the previous expression leads us to
\begin{eqnarray}\label{A3}
I_{2}^{(\tau)} (k, i; \tilde{\sigma}, \mu) &=& \tilde{r} \mu^{-\epsilon} 
\frac{1}{2} S_{d-1} \Gamma(\frac{d-1}{2}) \Gamma(2 - \frac{(d-1)}{2}) \nonumber \\
&& \times \int_{0}^{1} dx \sum_{j=-\infty}^{\infty} 
[x(1-x) (k^{2} + i^{2} \tilde{r}^{2})+ \tilde{r}^{2} (j + ix)^{2} + 1]^{\frac{d-1}{2} - 2}.
\end{eqnarray}
The summation can be performed using the generalized thermal function identity \cite{BF} 
\begin{eqnarray}\label{A4}
&D_{\alpha}(a,b)= \overset{\infty}{\underset{n=-\infty}{\sum}} [(n+a)^{2} + b^{2}]^{-\alpha} 
\nonumber\\
& = \frac{\sqrt{\pi}}{\Gamma(\alpha)}\;\left[\frac{\Gamma(\alpha - \frac{1}{2})}{b^{2\alpha -1}} + f_{\alpha}(a,b)\right], 
\end{eqnarray}
where 
\begin{equation}\label{A5} 
f_{\alpha}(a,b)= 4 \sum_{m=1}^{\infty} cos(2\pi ma)
\left(\frac{\pi m}{b}\right)^{\alpha - \frac{1}{2}} K_{\alpha - \frac{1}{2}}(2\pi mb),
\end{equation} 
and $K_{\nu}(x)$ is the modified Bessel function of the second kind. By identifying 
$a(x)= ix$, $b(x)=\tilde{r}^{-1}\sqrt{(k^{2} + \tilde{r}^{2}i^{2})x(1-x) +1}$ 
and performing the continuation $\epsilon=4-d$, we find 
\begin{eqnarray}\label{A6}
I_{2}^{(\tau)} (k, i; \tilde{\sigma}, \mu) &=& \mu^{-\epsilon} 
\frac{1}{2} S_{d-1} \Gamma(\frac{d-1}{2}) \sqrt{\pi}\; \int_{0}^{1} dx \nonumber 
\\
&& \times \;\; \left[\Gamma(\frac{\epsilon}{2}) 
[x(1-x) (k^{2} + i^{2} \tilde{r}^{2})+ 1]^{-\frac{\epsilon}{2}} 
+ f_{\frac{1}{2} + \frac{\epsilon}{2}}(a,b)\right]. 
\end{eqnarray}
Transforming the argument of the $\Gamma$-function with the recipe
$\sqrt{\pi}\Gamma(\frac{d-1}{2})S_{d-1}= \Gamma(\frac{d}{2}) S_{d}$ 
along with the expansion in $\epsilon=4-d$, we get to
\begin{eqnarray}\label{A7}
I_{2}^{(\tau)} (k, i; \tilde{\sigma}, \mu) &=& S_{d}\mu^{-\epsilon} \Bigl[
\frac{1}{\epsilon}\Bigl(1 - \frac{\epsilon}{2}\Bigr)
\times\;\; \int_{0}^{1} dx [x(1-x) (k^{2} + i^{2} \tilde{r}^{2})+ 1]^{-\frac{\epsilon}{2}} \nonumber\\
&&  +  \frac{1}{2} \tilde{r}^{-\epsilon} \Gamma\Bigl(2-\frac{\epsilon}{2}\Bigr)\int_{0}^{1} dx f_{\frac{1}{2} + \frac{\epsilon}{2}}\Bigl(ix, 
\tilde{r}^{-1}\sqrt{x(1-x) (k^{2} + i^{2} \tilde{r}^{2})+ 1}\Bigr)\Bigr]. 
\end{eqnarray}
After dividing by $S_{d}$ this object is precisely $I_{2}(k,i;\tilde{r})$ from Eq. (\ref{75}) (see also Eq. (\ref{76})). 
Setting $k=0$ and neglecting $O(\epsilon)$ contributions leads to the following equation
\begin{eqnarray}\label{A8}
I_{2}^{(\tau)} (k=0, i; \tilde{r}) &=& \mu^{-\epsilon} 
\frac{1}{\epsilon} \Bigl[1 - \frac{\epsilon}{2} - \frac{\epsilon}{2}\int_{0}^{1} dx ln[x(1-x) i^{2} \tilde{r}^{2}+ 1]  
+  \frac{\epsilon}{2} \tilde{F}_{0}(0,i;\tilde{r}) \Bigr]. 
\end{eqnarray}
\par We also need the integral $\tilde{I}_{2}(k,i,j;\tilde{\sigma},\mu)$. Factorizing the mass and using a Feynman parameter 
in this integral from Eq. (\ref{26b}) one finds
\begin{eqnarray}\label{A9}
&& \tilde{I}_{2}^{(\tau)} (k, i,j; \tilde{\sigma}, \mu) = \tilde{r} \mu^{-\epsilon} 
\sum_{j=-\infty}^{\infty} \int_{0}^{1} dx \int d^{d-1}q \nonumber\\
&& \;\; \times\;\;\frac{1}{\Bigl[q^{2} + 2xkq + xk^{2} 
+ \tilde{r}^{2}[(i^{2} x + (1-x)j^{2}] + 1 \Bigr]^{2}}.
\end{eqnarray}
Integrating over $q$ as before, using $\sqrt{\pi}\Gamma(\frac{d-1}{2})S_{d-1}= \Gamma(\frac{d}{2}) S_{d}$ 
and performing the continuation $\epsilon=4-d$, we can rewrite this object as
\begin{eqnarray}\label{A10}
\tilde{I}_{2}^{(\tau)} (k, i,j; \tilde{\sigma}, \mu) &=& S_{d} \frac{\tilde{r} \mu^{-\epsilon}}{2} 
\int_{0}^{1} dx [x(1-x)k^{2} + [x i^{2} + (1-x) j^{2}] \tilde{r}^{2})+ 1]^{-\frac{1}{2}}, 
\end{eqnarray}
that results in Eq. (\ref{77}) after dividing the last expression by $S_{d}$. In particular, the 
combination which appears to produce the one-loop diagram with all external quasi-momenta set to $i$ is given by
\begin{eqnarray}\label{A11}
&& \parbox{12mm}{\includegraphics[scale=0.4]{fig13DN.eps}}\quad\;= \tilde{\sigma}\frac{(N+8)}{9}
\Big[I_{2}(k,2i;\tilde{r}) + 2I_{2}(k,0;\tilde{r}) + 2 \tau \tilde{I}_{2}(k,0,0;\tilde{r}) 
+ 2 \tau \tilde{I}_{2}(k,2i,0;\tilde{r}) \nonumber\\
&& \qquad+ 4 \tilde{I}_{2}(k,i,i;\tilde{r})\Bigr]. 
\end{eqnarray}
Utilizing the results above mentioned evaluated at $k=0$ yields Eqs. (\ref{79}). This completes our first task.
\par Let us consider two- and three-loop integrals. We preclude the dimensional crossover region for too small values of $L$ which spoils 
the $\epsilon$-expansion analysis. 
\par The easiest two-loop contribution for the four-point vertex function is given by the diagram which consists of the 
following combination of integrals
\begin{eqnarray}\label{A12}
&& \parbox{10mm}{\includegraphics[scale=0.3]{fig14DN.eps}}\qquad\quad = \tilde{\sigma} \frac{(N^{2}+6N+20)}{27} 
\Bigl[I_{2}^{2}(k,2i;\tilde{\sigma},\mu) + 2I_{2}^{2}(k,0;\tilde{\sigma},\mu) + 4 \tau I_{2}(k,2i;\tilde{\sigma},\mu) 
\nonumber\\
&& \;\;\times \tilde{I}_{2}(k,0,2i;\tilde{\sigma},\mu) + 4 \tau I_{2}(k,0;\tilde{\sigma},\mu) 
\tilde{I}_{2}(k,0,0;\tilde{\sigma},\mu) + 4 \tilde{I}_{2}(k,i,i; \tilde{\sigma},\mu)[I_{2}(k,2i;\tilde{\sigma},\mu) \nonumber\\
&& + I_{2}(k,0;\tilde{\sigma},\mu)] + 2 \overset{\infty}{\underset{j=-\infty}{\sum}}[\tilde{I}_{2}^{2}(k,j,j+2i; \tilde{\sigma},\mu) 
+ \tilde{I}_{2}^{2}(k,j,j;\tilde{\sigma},\mu) + \tau \tilde{I}_{2}(k,j,j+2i;\tilde{\sigma},\mu) \nonumber\\
&& \;\;\times \tilde{I}_{2}(k,j+i,j+i;\tilde{\sigma},\mu) + \tau \tilde{I}_{2}(k,j,j;\tilde{\sigma},\mu) \tilde{I}_{2}(k,j-i,j+i;\tilde{\sigma},\mu)] \Bigr],
\end{eqnarray}  
computed at zero external momenta. Since the contributions inside the summation in the above expression is regular 
($O(\epsilon^{0})$), we can neglect them. It is then easy to show that
\begin{eqnarray}\label{A13}
\parbox{10mm}{\includegraphics[scale=0.3]{fig14DN.eps}}\qquad\quad = 3\tilde{\sigma} \frac{(N^{2}+6N+20)}{27} 
\mu^{-2 \epsilon}\Bigr[\frac{1}{\epsilon^{2}} \Bigl(1- \epsilon + 2 \epsilon \zeta^{(\tau)}(i;r) \Bigr)\Bigl].
\end{eqnarray}
\par The nontrivial two-loop diagram contributing to the four-point vertex function with arbitrary external momenta was 
derived in the main text and corresponds to Eq. (\ref{59}). The integral 
$\hat{I}_{4}(P,k_{3}, i,j,k,\tilde{\sigma},\mu)$ (where $P=k_{1}+k_{2}$) is regular and does not contribute to the singular 
part of the diagram. At the symmetric point it reads
\begin{eqnarray}\label{A14}
&& \parbox{10mm}{\includegraphics[scale=0.32]{fig8DN.eps}}\quad =\frac{(5N+22)}{27} \tilde{\sigma} \Bigr[I_{4}(0,2i,i) + 
2I_{4}(0,0,i) + 3 \tilde{I}_{4}(0,i,i,0)\nonumber\\ 
&& + 2 \tau \tilde{I}_{4}(0,0,0,i) + \tau \tilde{I}_{4}(0,0,2i,i) 
+ \tau \tilde{I}_{4}(0,2i,0,i) + \tilde{I}_{4}(0,i,i,2i) + O(\hat{I}_{4})\Bigr]
\end{eqnarray}  
The diagonal terms are composed by the integral $I_{4}(0,i,j)$ which can be written as 
\begin{equation}\label{A15}
I_{4}^{(\tau)}(0,i,j) = \tilde{\sigma} \sum_{j_{1}=-\infty}^{\infty} 
\int d^{d-1}q 
\frac{I_{2}^{(\tau)}(q,j_{1}-j;\tilde{\sigma},\mu)}{[q^{2} + \tilde{\sigma^{2}}(j_{1}-i)^{2}+ \mu^{2}][q^{2} 
+ \tilde{\sigma^{2}} j_{1}^{2}+ \mu^{2}]}. 
\end{equation}
When we factorize $\mu$ we find
\begin{eqnarray}\label{A16}
&& I_{4}(0,i,j)= \tilde{r} \mu^{-2 \epsilon} \frac{1}{\epsilon} (1-\frac{\epsilon}{2})
\int_{0}^{1} dx \sum_{j_{1}=-\infty}^{\infty} \int d^{d-1}q 
\frac{1}{[q^{2} + \tilde{r}^{2}(j_{1} - i)^{2}+ 1][q^{2} 
+ \tilde{r}^{2} j_{1}^{2}+ 1]}\nonumber\\
&&\;\times \frac{1}{[(q^{2} + \tilde{r}^{2}(j_{1}-j)^{2})x(1-x) + 1]^{\frac{\epsilon}{2}}} 
+ \tilde{r} \mu^{-2 \epsilon} \frac{1}{2}\tilde{F}_{\frac{\epsilon}{2}, 2}^{(\tau)}(0,i,j) S_{d},
\end{eqnarray}
where 
\begin{eqnarray}\label{A17}
&& \tilde{F}_{\frac{\epsilon}{2},2}(0,i,j)= \sum_{j_{1}=-\infty}^{\infty} 
\int d^{d-1}q 
\frac{\tilde{F}_{\frac{\epsilon}{2}}^{(\tau)}(q,j_{1}-j)}{[q^{2} + \tilde{r}^{2} (j_{1}-i)^{2}+ 1][q^{2} 
+ \tilde{r}^{2} j_{1}^{2}+ 1]}. 
\end{eqnarray}
The singular terms in $I_{4}$ come from the high-momentum region of the momentum integrations, since they correspond to 
ultraviolet divergences. Focusing our attention in the last term, it will contribute to the singular part if and only if 
the object $\tilde{F}_{\frac{\epsilon}{2}}^{(\tau)}(q,j_{1}-j)$ is proportional to $q^{p}, p\geq 0$ in the limit 
$q \rightarrow \infty$. We can discard it in the computation of the singularities if we can prove that it is proportional 
to $q^{p}$ for $p<0$ in the ultraviolet region. Indeed we can neglect this term as follows.
\par First, write it in the form
\begin{eqnarray}\label{A18}
&& \tilde{F}^{(\tau)}_{\frac{\epsilon}{2}}(q,j;\tilde{r})= 4 \tilde{r}^{-\epsilon} 
\int_{0}^{1} dx \sum_{m=1}^{\infty} cos[2\pi mjx)]\Bigl(\frac{\pi m}
{\tilde{r}^{-1} \sqrt{x(1-x)(q^{2} + \tilde{r}^{2} j^{2}) +1}}\Bigr)^{\frac{\epsilon}{2}} \nonumber\\
&& \;\;\times\;K_{\frac{\epsilon}{2}}\Bigl(2\pi m \tilde{r}^{-1} \sqrt{x(1-x)(q^{2} + \tilde{r}^{2} j^{2}) +1}\Bigr).
\end{eqnarray}
It follows trivially that 
\begin{eqnarray}\label{A19}
&& \tilde{f}^{' (\tau)}_{\frac{\epsilon}{2}}(q,j;\tilde{r})= 4 \tilde{r}^{-\epsilon} 
\int_{0}^{1} dx \sum_{m=1}^{\infty} \Bigl(\frac{\pi m}
{\tilde{r}^{-1} \sqrt{x(1-x)(q^{2} + \tilde{r}^{2} j^{2}) +1}}\Bigr)^{\frac{\epsilon}{2}} \nonumber\\
&& \;\;\times\;K_{\frac{\epsilon}{2}}\Bigl(2\pi m \tilde{r}^{-1} \sqrt{x(1-x)(q^{2} + \tilde{r}^{2} j^{2}) +1}\Bigr) > \tilde{F}^{(\tau)}_{\frac{\epsilon}{2}}(q,j;\tilde{r}).
\end{eqnarray}
We analyze the latter henceforth. Since the integrand is symmetric around $x=\frac{1}{2}$ we can write 
\begin{eqnarray}\label{A20}
&& \tilde{f}^{' (\tau)}_{\frac{\epsilon}{2}}(q,j;\tilde{r})= 8 \tilde{r}^{-\epsilon} 
\int_{0}^{\frac{1}{2}} dx \sum_{m=1}^{\infty} \Bigl(\frac{\pi m}
{\sigma^{-1} \sqrt{x(1-x)(q^{2}+\tilde{r}^{2}j^{2})+1}}\Bigr)^{\frac{\epsilon}{2}} \nonumber\\
&& \;\;\times\;K_{\frac{\epsilon}{2}}\Bigl(2\pi m \tilde{r}^{-1} 
\sqrt{x(1-x)(q^{2}+\tilde{r}^{2} j^{2}) +1}\Bigr).
\end{eqnarray}
When $q \rightarrow \infty$, we can take a small real parameter 
$\lambda<<1$ with the property  $\lambda q^{2} \rightarrow \infty$. We split the integral into two pieces: in 
the first one we integrate in the interval $(0,\lambda)$ neglecting the term $x^{2}$ in the integrand, whereas 
in the interval $(\lambda,\frac{1}{2})$ we replace the Bessel function by its asymtoptic form. We then write
\begin{eqnarray}\label{A21}
&& \underset{q \rightarrow \infty}{lim}\tilde{f}^{' (\tau)}_{\frac{\epsilon}{2}}(q,j;\tilde{r})= 8 \tilde{r}^{-\epsilon} 
\sum_{m=1}^{\infty}\Bigl[\int_{0}^{\lambda} dx  
\Bigl(\frac{\pi m}
{\tilde{r}^{-1} \sqrt{x(q^{2}+\tilde{r}^{2} j^{2})+1}}\Bigr)^{\frac{\epsilon}{2}} \nonumber\\
&& \;\;\times\;K_{\frac{\epsilon}{2}}\Bigl(2\pi m \tilde{r}^{-1} 
\sqrt{x(q^{2}+\tilde{r}^{2} j^{2})+1}\Bigr) + \int_{\lambda}^{\frac{1}{2}} dx  
\Bigl(\frac{\pi m}
{\tilde{r}^{-1} \sqrt{x(1-x)(q^{2}+\tilde{r}^{2} j^{2})+1}}\Bigr)^{\frac{\epsilon}{2}} \times \nonumber\\
&& \sqrt{\frac{\pi}{4 \pi m \tilde{r}^{-1} \sqrt{x(1-x)(q^{2}+\tilde{r}^{2} j^{2})+1}}} exp\Bigl(-2\pi m 
\tilde{r}^{-1} \sqrt{x(1-x)(q^{2}+\tilde{r}^{2} j^{2})+1}\Bigr)\Bigr].
\end{eqnarray}
The second term goes to zero exponentially and can be disregarded. After replacing the change of 
variables $y=1+x(q^{2} + \tilde{r}^{2} j^{2})$ in the integral, we find
\begin{eqnarray}\label{A22}
&& \underset{q \rightarrow \infty}{lim}\tilde{f}^{' (\tau)}_{\frac{\epsilon}{2}}(q,j;\tilde{r})= 
\frac{8\tilde{r}^{-\epsilon}} {q^{2}+\tilde{r}^{2} j^{2}} 
\sum_{m=1}^{\infty} (\pi m)^{\frac{\epsilon}{2}} 
\int_{1}^{\infty} dy  y^{-\frac{\epsilon}{4}} K_{\frac{\epsilon}{2}}(2\pi m \tilde{r}^{-1} \sqrt{y}).
\end{eqnarray}
We also know that \cite{GR}
\begin{equation}\label{A23}
\int_{1}^{\infty} dx x^{-\frac{\nu}{2}}(x-1)^{\mu-1}K_{\nu}(a\sqrt{x}) = 
\Gamma(\mu)2^{\mu}a^{-\mu}K_{\nu-\mu}(a).
\end{equation}
Considering the region outside the dimensional crossover region where $\tilde{r}<\infty$ implies the following 
asymptotic expression expression 
\begin{equation}\label{A24}
\underset{q \rightarrow \infty}{lim}\tilde{f}^{' (\tau)}_{\frac{\epsilon}{2}}(q,j;\tilde{r})= \frac{8\tilde{r}^{1-\epsilon}}
{q^{2}} 
\sum_{m=1}^{\infty} (\pi m)^{\frac{\epsilon}{2}-1} 
 K_{\frac{\epsilon}{2}-1}(2\pi m \tilde{r}^{-1}).
\end{equation}
Since $\tilde{f}^{' (\tau)}_{\frac{\epsilon}{2}}(q,j;\tilde{r})(>\tilde{F}^{(\tau)}_{\frac{\epsilon}{2}}(q,j;\tilde{r})$ is 
regular in $\epsilon$, in the ultraviolet region $\tilde{F}^{(\tau)}_{\frac{\epsilon}{2}}(q,j;\tilde{r})$ and the 
aforementioned integral involving it are both regular and we do not have to worry about its contribution to the singular 
part of the integral $I_{4}(0,i,j)$. 
\par We can write the remaining expression in the form  
\begin{eqnarray}\label{A25}
&& I_{4}^{(\tau)}(0,i,j) = \tilde{r} \mu^{-2\epsilon} \frac{1}{\epsilon}\Bigl(1-\frac{\epsilon}{2}\Bigr)
\int_{0}^{1} dx[x(1-x)]^{\frac{-\epsilon}{2}} \sum_{j_{1}=-\infty}^{\infty} 
\int d^{d-1}q \frac{1}{[q^{2} + \tilde{r}^{2} (j_{1}-i)^{2}+ 1]}\nonumber\\
&&\;\times \frac{1}{[q^{2} + \tilde{r}^{2} j_{1}^{2}+ 1][(q^{2} + \tilde{r}^{2} (j_{1}-j)^{2}) + \frac{1}{x(1-x)}]^{\frac{\epsilon}{2}}}.
\end{eqnarray}
\par We introduce the Feynman parameter $z$ to rewrite last equation as
\begin{eqnarray}\label{A26}
&& I_{4}^{(\tau)}(0,i,j) = \tilde{r} \mu^{-2 \epsilon}
\frac{1}{\epsilon}\Bigl(1-\frac{\epsilon}{2}\Bigr)
\int_{0}^{1} dx[x(1-x)]^{\frac{-\epsilon}{2}}\sum_{j_{1}=-\infty}^{\infty} 
\int_{0}^{1} dz \nonumber\\
&& \int \frac{d^{d-1}q}{[q^{2} + \tilde{r}^{2}j_{1}^{2} + 1 + \tilde{r}^{2}i(i-2j_{1})z]^{2}[(q^{2} + \tilde{r}^{2} (j_{1}-j)^{2}) + \frac{1}{x(1-x)}]^{\frac{\epsilon}{2}}},
\end{eqnarray}
followed by another $y$ to melt the two denominators into a single one. We then obtain:
\begin{eqnarray}\label{A27}
&& I_{4}^{(\tau)}(0,i,j) = \tilde{r}\mu^{-2 \epsilon}
\frac{\Gamma(2+\frac{\epsilon}{2})}{\Gamma(\frac{\epsilon}{2})\epsilon} \Bigl(1-\frac{\epsilon}{2}\Bigr)
\int_{0}^{1} dx[x(1-x)]^{\frac{-\epsilon}{2}}\sum_{j_{1}=-\infty}^{\infty} 
\int_{0}^{1} dz \int_{0}^{1} dy y(1-y)^{\frac{\epsilon}{2}-1}\nonumber\\
&& \int \frac{d^{d-1}q}{[q^{2} + \tilde{r}^{2}j_{1}^{2} + \tilde{r}^{2}[i(i-2j_{1})yz + (1-y)j(j-2j_{1}) + y 
+ \frac{1-y}{x(1-x)}]^{2+\frac{\epsilon}{2}}}.
\end{eqnarray}
Employing Eq. (\ref{A2}) to perform the momentum integration as well as our recipe to transform 
the unit sphere area $S_{d-1}$ into $S_{d}$, one can show that 
\begin{eqnarray}\label{A28}
&& I_{4}^{(\tau)}(0,i,j) = \frac{\tilde{r}\mu^{-2 \epsilon}S_{d}}{2\sqrt{\pi}} 
\frac{\Gamma(2-\frac{\epsilon}{2}) \Gamma(\frac{1}{2}+\epsilon)}{\Gamma(\frac{\epsilon}{2})\epsilon} \Bigl(1-\frac{\epsilon}{2}\Bigr)
\int_{0}^{1} dx[x(1-x)]^{\frac{-\epsilon}{2}} 
\int_{0}^{1} dy y(1-y)^{\frac{\epsilon}{2}-1}\nonumber\\ 
&& \sum_{j_{1}=-\infty}^{\infty} \Bigl[\tilde{r}^{2}[j_{1}^{2} + i^{2}yz + (1-y)j^{2} -2j_{1}(iyz + (1-y)j)] + y 
+ \frac{1-y}{x(1-x)}\Bigr]^{-\frac{1}{2}-\epsilon}.
\end{eqnarray}
By evaluating the summation using the generalized thermal function upon the identifications $a=-iyz -(1-y)j$, 
$b=\sqrt{i^{2}yz(1-yz)+ j^{2}y(1-y)+\tilde{r}^{-2}\Bigl(y + \frac{1-y}{x(1-x)}\Bigr)}$ and absorbing the factor 
$S_{d}$, that integral reads 
\begin{eqnarray}\label{A29}
&& I_{4}^{(\tau)}(0,i,j) = \frac{\tilde{r}^{-2 \epsilon} \mu^{-2 \epsilon}}{2\epsilon} \Bigl(1-\frac{\epsilon}{2}\Bigr)
 \frac{\Gamma(2-\frac{\epsilon}{2})}
{\Gamma(\frac{\epsilon}{2})}\int_{0}^{1} dx[x(1-x)]^{\frac{-\epsilon}{2}}
\int_{0}^{1} dy y(1-y)^{\frac{\epsilon}{2}-1}\nonumber\\
&&\;\;\times \Biggl[\Gamma(\epsilon)\Bigl(i^{2}yz(1-yz) + j^{2}y(1-y) +\tilde{r}^{-2} \Bigl(y+ \frac{1-y}{x(1-x)}\Bigr)\Bigr)^{-\epsilon} \nonumber\\
&& + f_{\frac{1}{2}+ \epsilon}\Biggl(-iyz-(1-y)j,\sqrt{i^{2}yz(1-yz)+ j^{2}y(1-y)+\tilde{r}^{-2}\Bigl(y + \frac{1-y}{x(1-x)}\Bigr)}\Biggr)\Biggr].
\end{eqnarray}
The integral over $y$ presents a pole in $y=1$. We then evaluate it at 
$y=1$ \cite{amit}, which is not only a lot easier but maintains the essential pole 
contribution of interest. Expanding in 
$\epsilon$ and getting rid of $O(\epsilon^{0})$ terms, we finally obtain
\begin{equation}\label{A30}
I_{4}^{(\tau)} (0,i,j)= \mu^{-2 \epsilon} \frac{1}{2 \epsilon^{2}}
\Bigl(1-\frac{\epsilon}{2} - \epsilon \int_{0}^{1} dz ln[\tilde{r}^{2}i^{2}z(1-z) + 1] + \epsilon \tilde{F}_{0}(0,i;\tilde{r})\Bigr).
\end{equation}
We still have to compute the nondiagonal contribution coming from $\tilde{I}_{4}(0,i,j,l)$. It can be expressed in terms of 
the one-loop integral $I_{2}(k,i,\tilde{\sigma},\mu)$ as 
\begin{equation}\label{A31}  
\tilde{I}_{4}(0,i,j,l)=\tilde{\sigma} 
\int d^{d-1}q 
\frac{I_{2}^{(\tau)}(q,l,\tilde{\sigma},\mu)}{[q^{2} + \tilde{\sigma^{2}}j^{2}+ \mu^{2}][q^{2} 
+ \tilde{\sigma^{2}}i^{2}+ \mu^{2}]}. 
\end{equation}
Extracting the mass from the integrand we get to
\begin{equation}\label{A32}
\tilde{I}_{4}(0,i,j,l)=\tilde{r}\mu^{-2 \epsilon} 
\int d^{d-1}q 
\frac{I_{2}^{(\tau)}(q,l;\tilde{r})}{[q^{2} + \tilde{r}^{2}j^{2}+ 1][q^{2} 
+ \tilde{r}^{2}i^{2}+ 1]}. 
\end{equation}
Note that the integrand in the last expression is symmetric by the change $i \rightarrow j$. This integral differs from $I_{4}(0,i,j)$ because 
{\it i)} the summation over $j_{1}$ is missing, {\it ii)} in the one-loop subdiagram there appears only a external 
quasi-momentum $l$, {\it iii)} each propagator in the ``external bubble'' are attached to external quasi-momentum. The 
computation is entirely analogous with the above computation for $I_{4}(0,i,j)$. Picking out only the singular term and 
putting aside the regular terms we are led to
\begin{equation}\label{A33}
\tilde{I}_{4}(0,i,j,l)= \frac{\tilde{r}\mu^{-2 \epsilon}}{2\epsilon} 
\int_{0}^{1} dx[\tilde{r}^{2}(i^{2} x + (1-x)j^{2})+1]^{-\frac{1}{2}}.  
\end{equation}  
Now, all we have to do is to take into account the combinations of the integrals just computed. The nontrivial 
two-loop diagram of the four-point function has the following result:
\begin{eqnarray}\label{A34}
\parbox{10mm}{\includegraphics[scale=0.4]{fig15DN.eps}}\quad = 3\tilde{\sigma} \frac{(5N+22)}{27} \mu^{-2\epsilon}\Bigl\{\frac{1}{2\epsilon^{2}}\Bigl(1-\frac{\epsilon}{2} + 2\epsilon 
\zeta^{(\tau)}(i,\tilde{r})\Bigr)\Bigr\}.
\end{eqnarray}   
\par The two-point vertex part two-loop diagram which interests us is
\begin{eqnarray}\label{A35}
&& \parbox{10mm}{\includegraphics[scale=0.3]{fig16DN.eps}} \quad = \left(\frac{N+2}{3}\right)[I_{3}(k,i,\tilde{\sigma},\mu) 
+ 3 \tilde{I}_{3}(k,i,0,\tilde{\sigma},\mu)+ 3 \tau \tilde{I}_{3}(k,0,i,\tilde{\sigma},\mu)].
\end{eqnarray}
We actually need the derivative of this diagram computed at zero external momenta. Let us first consider 
$I_{3}((k,i,\tilde{\sigma},\mu)$. Factorizing the mass, it can be written as follows
\begin{eqnarray}\label{A36}
&& I_{3}(k,i,\tilde{\sigma},\mu) = \tilde{r} \mu^{2-\epsilon} \sum_{j_{1}=-\infty}^{\infty} 
\int \frac{d^{d-1}q_{1} I_{2}(q_{1}+k, j_{1}+i; \tilde{r})}
{\left(q_{1}^{2} + \tilde{r}^{2}j_{1}^2 + 1 \right)}.
\end{eqnarray}
The solution of the subdiagram is then substituted into this expression and yields
\begin{eqnarray}\label{A37}
&& I_{3}(k,i,\tilde{\sigma},\mu) = \mu^{2- 2 \epsilon}\frac{1}{\epsilon}\Biggl\{\Bigl(1-\frac{\epsilon}{2}\Bigr) 
\tilde{r} \int_{0}^{1} dx \sum_{j_{1}=-\infty}^{\infty} \int \frac{d^{d-1}q_{1}}
{[q_{1}^{2} + \tilde{r}^{2}j_{1}^2 + 1][(q_{1}=k)^{2} + + \tilde{r}^{2}(j_{1}+i)^2 + 1]} \nonumber\\
&& + \frac{\epsilon}{2} \tilde{F}_{\frac{\epsilon}{2},1}^{(\tau)}(q_{1}+k, j_{1}+i;\tilde{r})\Biggr\},
\end{eqnarray}
where $\tilde{F}_{\frac{\epsilon}{2},1}^{(\tau)}(q_{1}+k, j_{1}+i;\tilde{r})$ was defined in Eq. (\ref{90a}). The remaining of the computation 
is very similar to what was worked out for $PBC$ and $ABC$; the reader is advised to consult Ref. (\cite{BL}) for further details. The 
derivative at zero external momentum can 
be shown to be given by
\begin{eqnarray}\label{A38}
\frac{\partial I_{3}(k,i;\tilde{r})}{\partial k^{2}}\Bigl|_{k^{2}=0} = -\frac{\mu^{- 2 \epsilon}}{8\epsilon}\Bigl[1 - \frac{\epsilon}{4}
+ \epsilon \tilde{W}_{0}(i;\tilde{r})\Bigr], 
\end{eqnarray}
where $\tilde{W}_{0}(i;\tilde{r})= \tilde{G}_{0}(i;\tilde{r}) + \tilde{H}_{0}(i;\tilde{r}) - 4 \tilde{F}_{0}^{' (\tau)}(i;\tilde{r})$,
\begin{subequations}\label{A39}
\begin{eqnarray}
 \tilde{G}_{0}(i;\tilde{r})&=& -\frac{1}{2} - 2\int_{0}^{1} \int_{0}^{1}dx dy (1-y)ln\Bigl[y(1-y)i^{2} + (1-y)\tilde{r}^{-2} 
+ \frac{y\tilde{r}^{-2}}{x(1-x)}\Bigr],\\ 
 \tilde{H}_{0}(i;r) &=& 2 \int_{0}^{1} \int_{0}^{1}dx dy (1-y)f_{\frac{1}{2}}(iy,\sqrt{y(1-y)i^{2} + (1-y)\tilde{r}^{-2} 
+ \frac{y\tilde{r}^{-2}}{x(1-x)}}),
\end{eqnarray}
\end{subequations}
with $\tilde{F}_{0}^{' (\tau)}(i;\tilde{r})$ defined in Eq. (\ref{90b}). We have to compute 
$\frac{\partial}{\partial k^{2}}\tilde{I}_{3}(k,i,j;\tilde{\sigma},\mu)\Bigl|_{k^{2}=0}$. First, we scale out the mass 
to write it as
\begin{eqnarray}\label{A40}
&& \tilde{I}_{3}(k,i,j;\tilde{\sigma},\mu) = \tilde{r} \mu^{2-\epsilon}
\int \frac{d^{d-1}q I_{2}(q+k, i; \tilde{r})}
{\left(q^{2} + \tilde{r}^{2}j^2 + 1 \right)} 
\end{eqnarray}
From the solution of the subdiagram we have the following intermediate result
\begin{eqnarray}\label{A41}
&& \tilde{I}_{3}(k,i,j;\tilde{r}) = \frac{\mu^{2-2\epsilon}}{\epsilon}\Biggl[\tilde{r}\Bigl(1-\frac{\epsilon}{2}\Bigr)
\int_{0}^{1} dx [x(1-x)]^{-\frac{\epsilon}{2}}
\int \frac{d^{d-1}q I_{2}(q+k, i; \tilde{r})}
{\Bigl(q^{2} + \tilde{r}^{2}j^2 + 1\Bigr) \Bigl[(q+k)^{2} + \tilde{r}^{2}i^2 + 1\Bigr]^{\frac{\epsilon}{2}}} \nonumber\\
&& + \frac{\epsilon}{2} \tilde{r} \int \frac{d^{d-1}q \tilde{F}_{\frac{\epsilon}{2}}^{(\tau)}(k,i;\tilde{r})}{(q^{2} + \tilde{r}^{2}j^2 + 1)}\Biggr].
\end{eqnarray}
Let us define the integral
\begin{eqnarray}\label{A42}
&& \tilde{i}_{3}=\int \frac{d^{d-1}q I_{2}(q+k, i; \tilde{r})}
{\Bigl(q^{2} + \tilde{r}^{2}j^2 + 1\Bigr) \Bigl[(q+k)^{2} + \tilde{r}^{2}i^2 + 1\Bigr]^{\frac{\epsilon}{2}}}.
 \end{eqnarray}
Introduce a parameter $y$ in order to write it in the form
\begin{eqnarray}\label{A43}
\tilde{i}_{3}&=&\int_{0}^{1}\int \frac{\Gamma(1+\frac{\epsilon}{2})\;d^{d-1}q \;dy \;y^{\frac{\epsilon}{2} -1}}{\Gamma(\frac{\epsilon}{2}) \Bigl[q^{2} + 2kqy + k^{2}y + \tilde{r}^{2}(j^{2}(1-y)+ i^{2}y) + 1-y + \frac{y}{x(1-x)}\Bigr]^{1 + \frac{\epsilon}{2}}}.
\end{eqnarray}
Next integrate over $q$ and transform $S_{d-1}$ into $S_{d}$ using the prescription already stated. When we use the definition 
Eq. (\ref{90c}) together with our labor on $\tilde{i}_{3}$ leads us to conclude that
\begin{eqnarray}\label{A44}
&& \tilde{I}_{3}(k,i,j;\tilde{r}) = \frac{\mu^{2-2\epsilon}}{\epsilon}\Biggl[\frac{\tilde{r} \Gamma(2-\frac{\epsilon}{2})
\Gamma(-\frac{1}{2}+\epsilon)}{\sqrt{\pi} \Gamma(\frac{\epsilon}{2})}\Bigl(1-\frac{\epsilon}{2}\Bigr)\int_{0}^{1} dx 
[x(1-x)]^{-\frac{\epsilon}{2}} \int dy  y^{\frac{\epsilon}{2} -1} \nonumber\\
&& \Bigl(k^{2}y(1-y)+ \tilde{r}^{2}(j^{2}(1-y)+i^{2}y) + 1-y 
+ \frac{y}{x(1-x)}\Bigr)^{\frac{1}{2} - \epsilon} 
+ \frac{\epsilon}{2} \mathcal{F}_{\frac{\epsilon}{2},1}^{(\tau)}(k,i,j;\tilde{r})\Biggr].
\end{eqnarray}
Taking the derivative at zero external momentum, performing the expansion in $\epsilon$, discarding $O(\epsilon)$ 
contributions and using the definition Eq. (\ref{90d}) we get the desired expression, namely
\begin{eqnarray}\label{A45}
&& \frac{\partial \tilde{I}_{3}(k,i,j;r)}{\partial k^{2}}\Bigl|_{k^{2}=0} = - \mu^{- 2 \epsilon}\Bigl[\frac{\tilde{r}}{4}
\int_{0}^{1}dx \int_{0}^{1} dy (1-y)\Bigl(\tilde{r}^{2}(j^{2}(1-y)+i^{2}y) + 1-y 
+ \frac{y}{x(1-x)}\Bigr)^{-\frac{1}{2}}\nonumber\\
&&  -\frac{1}{2} \mathcal{F}_{0}^{' (\tau)}(i,j;\tilde{r})\Bigr]
\end{eqnarray}
Plugging the previous expression in the combinations appearing in the two-loop graph, we get to an expression which goes 
beyond the result from the simple $PBC$ and $ABC$, or in other words
\begin{subequations}\label{A46}
\begin{eqnarray}
&& \frac{\partial}{\partial k^{2}}\left(\parbox{10mm}{\includegraphics[scale=0.3]{fig16DN.eps}}\quad\right)\Bigl|_{k^{2}=0}\quad =-\frac{\mu^{-2\epsilon} (N+2)}{24\epsilon}\Bigl[1-\frac{\epsilon}{4} + \epsilon \tilde{W}^{(\tau)}(i,\tilde{r})\Bigr],\\
&& \tilde{W}(i,\tilde{r})= \tilde{W}_{0}(i;\tilde{r}) + 6\tilde{r} \Biggl[\int_{0}^{1}dx \int_{0}^{1} dy (1-y)\Bigl(\tilde{r}^{2} i^{2}y + 1-y 
+ \frac{y}{x(1-x)}\Bigr)^{-\frac{1}{2}} + \tau \int_{0}^{1}dx \int_{0}^{1} dy \nonumber\\
&& \times(1-y)\Bigl(\tilde{r}^{2}i^{2}(1-y) + 1-y 
+ \frac{y}{x(1-x)}\Bigr)^{-\frac{1}{2}}\Biggr]  -12 \Bigl[\mathcal{F}_{0}^{' (\tau)}(i,0;\tilde{r})
+ \tau \mathcal{F}_{0}^{' (\tau)}(0,i;\tilde{r})\Bigr].
\end{eqnarray}
\end{subequations}
The three-loop diagram for the two-point function has a similar systematics. We just need to compute a combination of the objects
$\frac{\partial I_{5}(k,i,\tilde{\sigma},\mu)}{\partial k^{2}}\Bigl|_{k^{2}=0}$, $\frac{\partial \tilde{I}_{5}(k,i,0,\tilde{\sigma},\mu)}{\partial k^{2}}|_{k^{2}=0}$ and $\frac{\partial \tilde{I}_{5}(k,0,i,\tilde{\sigma},\mu)}{\partial k^{2}}|_{k^{2}=0}$. The first one can be checked to be
given by the following expression:
\begin{equation}\label{A47}
\frac{\partial I_{5}(k,i,\tilde{\sigma},\mu)}{\partial k^{2}}\Bigl|_{k^{2}=0} = \mu^{-3\epsilon} \Bigl(-\frac{1}{6 \epsilon^{2}}\Bigr) 
\Bigl[ 1 - \frac{\epsilon}{4} + \frac{3W_{0}^{(\tau)}(i,\tilde{r})}{2}\Bigr],
\end{equation}
where $W_{0}^{(\tau)}= -\frac{1}{2} - 2\int_{0}^{1} \int_{0}^{1}dx dy (1-y)ln\Bigl[y(1-y)i^{2} + (1-y)\tilde{r}^{-2} 
+ \frac{y\tilde{r}^{-2}}{x(1-x)}\Bigr] + 2 \int_{0}^{1} \int_{0}^{1}dx dy (1-y)f_{\frac{1}{2}}(iy,\sqrt{y(1-y)i^{2} + (1-y)\tilde{r}^{-2} 
+ \frac{y\tilde{r}^{-2}}{x(1-x)}}) - 4\tilde{F}'_{0}(i;\tilde{r})$. This is precisely equal to the contribution from $PBC$ in Ref. \cite{BL}. We 
shall discuss the computation of the nondiagonal contribution. First, notice that 
$\tilde{I}_{5}(k,i,j,\tilde{\sigma}, \mu)$ can be written in the form
\begin{equation}\label{A48}
\tilde{I}_{5}(k,i,j,\tilde{\sigma}, \mu)= \mu^{2-3\epsilon} \tilde{r} \int \frac{d^{d-1}q I_{2}^{2}(q+k,j,\tilde{r})}{q^{2} + \tilde{r}^{2}i^{2} +1}.
\end{equation} 
Replacing the value of the subdiagram already computed previously (with $S_{d}$ already divided), we find
\begin{eqnarray}\label{A49}
&& \tilde{I}_{5}(k,i,j,\tilde{\sigma}, \mu)= \mu^{2-3\epsilon}\frac{1}{\epsilon^{2}}\Biggl\{ \tilde{r}\int_{0}^{1} dx 
\int \frac{d^{d-1}q}{[q^{2} + \tilde{r}^{2}i^{2} +1]\Bigl[(q+k)^{2} + \tilde{r}^{2}j^{2} + \frac{1}{x(1-x)}\Bigr]^{\epsilon}} \nonumber\\
&& + \tilde{r}\epsilon \int \frac{d^{d-1}q F_{\frac{\epsilon}{2}}(q+k,j,\tilde{r})}{q^{2} + \tilde{r}^{2}i^{2} +1} \Bigr\}. 
\end{eqnarray}
Now utilize another Feynman parameter, integrate over $q$ (transform the product $S_{d-1}\Gamma(\frac{d-1}{2})$ into $S_{d}\Gamma(\frac{d}{2})$ as 
described in the text) and divide the result by $S_{d}$. Expand the remaining $\Gamma$ functions in $\epsilon$ and neglect $O(\epsilon^{0})$. 
Next, taking the derivative at zero external momenta we find
\begin{eqnarray}\label{A50}
&& \frac{\partial \tilde{I}_{5}(k,i,j,\tilde{\sigma},\mu)}{\partial k^{2}}\Bigl|_{k^{2}=0} = \mu^{-3\epsilon} \Bigr(-\frac{1}{2 \epsilon}\Bigr) \Bigl\{\tilde{r} 
\int_{0}^{1} dy (1-y) \int_{0}^{1} dx \Bigr[\tilde{r}^{2}(i^{2}(1-y)+j^{2}y) + 1-y \nonumber\\
&& + \frac{y}{x(1-x)}\Bigr]^{-\frac{1}{2}} - 2 \mathcal{F}_{0}^{' (\tau)}(i,j,\tilde{r}) \Bigr\},
\end{eqnarray}
where $\mathcal{F'}_{0}^{(\tau)}(i,j,\tilde{r})$ was defined in Eq. (\ref{90d}). The other terms involve the integration of products of the type 
$I_{2}\tilde{I}_{2}$ and $\tilde{I}_{2}\tilde{I}_{2}$. The former can be shown to be $O(\epsilon^{0})$ whereas the former are $O(\epsilon)$. 
Both terms which contribute a total of ten integrals in the computation of the three-loop graph of the two-point function are all regular and 
can be neglected in the evaluation of the singular part of the diagram.  
\par With the resources furnished in this Appendix, 
the reader is invited to check that the following expression holds
\begin{eqnarray}\label{A51}
&& \frac{\partial}{\partial k^{2}}\left(\parbox{10mm}{\includegraphics[scale=0.3]{fig17DN.eps}}\quad\right)_{k^{2}=0}\quad =-\frac{\mu^{-3\epsilon}(N+2)(N+8)}{162 \epsilon^{2}}\Bigl[1-\frac{\epsilon}{4} + \frac{3\epsilon}{2} \tilde{W}^{(\tau)}(i;\tilde{r})\Bigr].
\end{eqnarray}

\end{document}